\documentclass{article}
\usepackage[utf8]{inputenc}
\usepackage{fancyhea}
\usepackage{bm}       %    enables bold math symbols  e.g.  \bm{\gamma}
\usepackage{graphicx}
\usepackage{hyperref}      % hypertext links %%ARXIV
\usepackage{wasysym}
\usepackage{cite}  % bibtex myarticle
\usepackage{xcolor}

\newcommand{\nc}{\newcommand}  

\def\beq{\begin{equation}}
\def\eeq#1{\label{#1}\end{equation}}
\def\eeqn{\end{equation}}

\newenvironment{Eqnarray}%
   {\arraycolsep 0.14em\begin{eqnarray}}{\end{eqnarray}}
\def\beqa{\begin{Eqnarray}}
\def\eeqa#1{\label{#1}\end{Eqnarray}}
\def\eeqan{\end{Eqnarray}}

\nc{\ra}{\rightarrow}  
\nc{\slsh}{\slash\hspace*{-0.22cm}}
\def\Re{{\cal R \mskip-4mu \lower.1ex \hbox{\it e}\,}}
\def\Im{{\cal I \mskip-5mu \lower.1ex \hbox{\it m}\,}}

\nc{\vev}[1]{ \left\langle {#1} \right\rangle }
\nc{\bra}[1]{ \langle {#1} | }
\nc{\ket}[1]{ | {#1} \rangle }
\nc{\fb}{\,{\rm fb}^{-1}}
\nc{\ev}{{\rm eV}}
\nc{\kev}{{\rm keV}}
\nc{\Mev}{{\rm MeV}}
\nc{\gev}{{\rm GeV}}
\nc{\tev}{{\rm TeV}}
\nc{\mev}{{\rm MeV}}

\def\del{\partial}
\def\Dslash{\not{\hbox{\kern-4pt $D$}}}
\def\dslash{\not{\hbox{\kern-2pt $\del$}}}
\def\pslash{\not{\hbox{\kern-2pt $p$}}}
\def\ETmiss{ \not{\hbox{\kern-4pt $E$}}_T }

\def\msb{{\bar{\ssstyle M \kern -1pt S}}}

\usepackage{hyperref}
\usepackage{latexsym}
\usepackage{xspace}
\usepackage{relsize}
\usepackage{subfigure}
\usepackage[section]{placeins}

\usepackage[figuresright]{rotating}

\begin{document}
\pagenumbering{none}
\def\bibname{References}

\raggedbottom

\pagenumbering{arabic}

\parindent=0pt
\parskip=8pt
%\setlength{\evensidemargin}{0pt}
%\setlength{\oddsidemargin}{0pt}
%\setlength{\marginparsep}{0.0in}
%\setlength{\marginparwidth}{0.0in}
%\marginparpush=0pt

% The content begins here

%\renewcommand{\chapname}{chap:intro_}
%\renewcommand{\chapterdir}{.}
%\renewcommand{\arraystretch}{1.25}
%\addtolength{\arraycolsep}{-3pt}

%------

%\title{2017 Community Planning Report \\ on Light Dark Matter Studies}
%\maketitle
%\tableofcontents

\centerline{\large PTOLEMY:  A Proposal for Thermal Relic Detection of Massive Neutrinos}
\centerline{\large and Directional Detection of MeV Dark Matter}
%\vspace*{35pt}
%\centerline{Proposal submitted by:  }
%\centerline{}
%\vspace*{-20pt}
\noindent
%Authors:
E.~Baracchini$^3$,
M.G.~Betti$^{11}$,
M.~Biasotti$^{5}$,
A.~Bosc\'a$^{16}$,
F.~Calle$^{16}$,
J.~Carabe-Lopez$^{14}$,
G.~Cavoto$^{10,11}$,
C.~Chang$^{22,23}$,
A.G.~Cocco$^{7}$,
A.P.~Colijn$^{13}$,
J.~Conrad$^{18}$,
N.~D'Ambrosio$^{2}$,
P.F.~de~Salas$^{17}$,
M.~Faverzani$^{6}$,
A.~Ferella$^{18}$, 
E.~Ferri$^{6}$,
P.~Garcia-Abia$^{14}$,
G.~Garcia~Gomez-Tejedor$^{15}$,
S.~Gariazzo$^{17}$,
F.~Gatti$^{5}$,
C.~Gentile$^{25}$,
A.~Giachero$^{6}$,
J.~Gudmundsson$^{18}$,
Y.~Hochberg$^{1}$, 
Y.~Kahn$^{26}$,
M.~Lisanti$^{26}$,
C.~Mancini-Terracciano$^{10}$,
G.~Mangano$^{7}$,
L.E.~Marcucci$^{9}$,
C.~Mariani$^{11}$,
J.~Mart\'inez$^{16}$,
G.~Mazzitelli$^{4}$,
M.~Messina$^{20}$,
A.~Molinero-Vela$^{14}$,
E.~Monticone$^{12}$,
A.~Nucciotti$^{6}$,
F.~Pandolfi$^{10}$,
S.~Pastor$^{17}$,
J.~Pedr\'os$^{16}$,
C.~P\'erez~de~los~Heros$^{19}$,
O.~Pisanti$^{7,8}$,
A.~Polosa$^{10,11}$,
A.~Puiu$^{6}$,
M.~Rajteri$^{12}$,
R.~Santorelli$^{14}$,
K.~Schaeffner$^{3}$,
C.G.~Tully$^{26}$,
Y.~Raitses$^{25}$,
N.~Rossi$^{10}$,
F.~Zhao$^{26}$,
K.M.~Zurek$^{21,22}$

%\vspace*{5pt}
$^{1}$Racah Institute of Physics, Hebrew University of Jerusalem, Jerusalem, Israel\\
$^{2}$INFN Laboratori Nazionali del Gran Sasso, L'Aquila, Italy\\
$^{3}$Gran Sasso Science Institute (GSSI), L'Aquila, Italy\\
$^{4}$INFN Laboratori Nazionali di Frascati, Frascati, Italy\\
$^{5}$Universit\`a degli Studi di Genova e INFN Sezione di Genova, Genova, Italy \\
$^{6}$Universit\`a degli Studi di Milano-Bicocca e INFN Sezione di Milano-Bicocca, Milano, Italy\\
$^{7}$INFN Sezione di Napoli, Napoli, Italy \\
$^{8}$Universit\`a degli Studi di Napoli Federico II, Napoli, Italy\\
$^{9}$Universit\`a degli Studi di Pisa e INFN Sezione di Pisa, Pisa, Italy\\
$^{10}$INFN Sezione di Roma, Roma, Italy \\
$^{11}$Universit\`a degli Studi di Roma La Sapienza, Roma, Italy\\
$^{12}$Istituto Nazionale di Ricerca Metrologica (INRiM), Torino, Italy\\
$^{13}$Nationaal instituut voor subatomaire fysica (NIKHEF), Amsterdam, Netherlands\\
$^{14}$Centro de Investigaciones Energ\'eticas, Medioambientales y Tecnol\'ogicas (CIEMAT),  Madrid, Spain\\
$^{15}$Consejo Superior de Investigaciones Cientificas (CSIC), Madrid, Spain\\
$^{16}$Universidad Polit\'ecnica de Madrid, Madrid, Spain\\
$^{17}$Instituto de F\'isica Corpuscular (CSIC-Univ. de Val\`{e}ncia), Valencia, Spain\\
$^{18}$Stockholm University, Stockholm, Sweden\\
$^{19}$Uppsala University, Uppsala, Sweden\\
$^{20}$New York University Abu Dhabi, Abu Dhabi, UAE \\
$^{21}$Lawrence Berkeley National Laboratory, University of California, Berkeley, CA, USA \\
$^{22}$Department of Physics, University of California, Berkeley, CA, USA \\
$^{23}$Argonne National Laboratory, Chicago, IL, USA \\
$^{24}$Kavli Institute for Cosmological Physics, University of Chicago, Chicago, IL, USA \\
$^{25}$Princeton Plasma Physics Laboratory, Princeton, NJ, USA \\
$^{26}$Department of Physics, Princeton University, Princeton, NJ, USA \\
%$^4$Savannah River National Laboratory, Aiken, SC, USA \\

%\vspace*{10pt}
%\noindent
%and ATLAS, CMS, ILC, CLIC Collaborations

\vspace*{8pt}
\centerline{Submitted to the LNGS Scientific Committee on March 19$^{\rm th}$, 2018} %\today}
\vspace*{12pt}
\centerline{\bf Abstract}
% \vspace*{-5pt}
%This report summarizes the work of the Cosmic Frontier Light Dark Matter
%working group of the 2017 Community Planning Study.  We
%identify the key elements of a light dark matter physics program and
%document the physics potential of future experiments.

We propose to achieve the proof-of-principle of the PTOLEMY project to directly detect the Cosmic Neutrino Background (CNB). 
Each of the technological challenges described in~\cite{betts2013development,Cocco2007Probing}
%$^[$\footnote{Betts, et. al, 2013. ``Development of a Relic Neutrino Detection Experiment at PTOLEMY: Princeton Tritium Observatory for Light, Early-Universe, Massive-Neutrino Yield'', \url{http://arxiv.org/abs/1307.4738}.}$^,$\footnote{
%A. G. Cocco, G. Mangano, M. Messina, ``Probing low energy neutrino backgrounds with neutrino capture on beta decaying nuclei.''  JCAP. (2007) 0706:015.  \url{http://dx.doi.org/10.1088/1742-6596/110/8/082014}.}$^]$
will be targeted and hopefully solved by the use of the latest experimental developments and profiting from the low background environment provided by the LNGS underground site.
%\footnote{D. Mei, A. Hime, ``Muon-induced background study for underground laboratories,'' Phys. Rev. D73 (2006) 053004.}.
The first phase will focus on the graphene technology for a tritium target and the demonstration of TES microcalorimetry
with an energy resolution of better than 0.05~eV for low energy electrons.  These technologies will be evaluated using the PTOLEMY prototype, proposed for underground installation, using precision HV controls to step down the kinematic energy of endpoint electrons to match the calorimeter dynamic range and rate capabilities.  The second phase will produce a novel implementation of the EM filter that is scalable to the full target size and which demonstrates intrinsic triggering capability for selecting endpoint electrons.
Concurrent with the CNB program, we plan to exploit and develop the unique properties of graphene to implement an intermediate program for direct directional detection of MeV dark matter~\cite{hochberg,cavoto}.
%\footnote{Hochberg, et. al, 2016. ``Directional Detection of Dark Matter with 2D Targets", \url{http://doi.org/10.1016/j.physletb.2017.06.051}.  Cavoto, Luchetta, Polosa, ``Sub-GeV Dark Matter Detection with Electron Recoils in Carbon Nanotubes", \url{http://doi.org/10.1016/j.physletb.2017.11.064}.}.  
This program will evaluate the radio-purity and scalability of the graphene fabrication process with the goal of using recently identified ultra-high radio-purity CO$_2$ sources.
The direct detection of the CNB is a snapshot of early universe dynamics recorded by the thermal relic neutrino yield 
taken at a time that predates the epochs of Big Bang Nucleosynthesis, the Cosmic Microwave Background and the recession of galaxies (Hubble Expansion).  Big Bang neutrinos are believed to have a central role in the evolution of the Universe and a direct measurement with PTOLEMY will unequivocally establish the extent to which these predictions match present-day neutrino densities.

%High radio-purity wafer-level fabrication\footnote{Low background contamination lithography has been demonstrated, see for example ``Cryogenic Dark Matter Search detector fabrication process and recent improvements" by Jastram et. al, 2015.  NIM A: 772:14-25.}, ultra-low ratio $^{14}$C/C graphene growth\footnote{Litherland et. al, 2005.  ``Low-level $^{14}$C measurements and Accelerator Mass Spectrometry'' in AIP Conference Proceedings, vol. 785, p. 48. \url{http://dx.doi.org/10.1063/1.2060452}.}, a cryogenic fiducialized volume, the coincidence of the FET-to-FET trajectories of electron recoils and 3km w.e. overburden\footnote{D. Mei, A. Hime, ``Muon-induced background study for underground laboratories,'' Phys. Rev. D73 (2006) 053004.} provide the conditions for a low background observatory of MeV dark matter interactions.  
%The evaluation of the G$^3$ active target and low background methods are an important step for the PTOLEMY project and will enable concurrent demonstration of the resolutions required for detection of the cosmic neutrino background\footnote{A. G. Cocco, G. Mangano, M. Messina, ``Probing low energy neutrino backgrounds with neutrino capture on beta decaying nuclei.''  JCAP. (2007) 0706:015.  \url{http://dx.doi.org/10.1088/1742-6596/110/8/082014}.}.
%PTOLEMY-G$^3$ is the only experiment with direct directional detection capability for MeV dark matter and has a projected detection sensitivity that exceeds an equivalent mass target of low noise (5~$e^-$ threshold) germanium cryogenic detectors.

%\vspace*{10pt}

%\pagenumbering{arabic}

\tableofcontents

\clearpage

\section*{Overview}
\addcontentsline{toc}{section}{Overview}

The basic concept for the detection of the Cosmic Neutrino Background (CNB) was laid out in the original paper by Steven Weinberg~\cite{WeinbergRelic} in 1962 and further refined by the work of Cocco, Mangano and Messina~\cite{Cocco2007Probing} in 2007 in view of the finite neutrino mass discovered by oscillation experiments.

An experimental realization of this concept was proposed based on PTOLEMY~\cite{betts2013development} in 2013; since then, many of those points that were considered as major obstacles toward the realization of the detector have been overcome. To mention the most important ones: high radio-pure graphene for the substrate of a tritium target appears now as a viable choice, single electron RF detection has been successfully achieved and Transition Edge Sensors (TES) with extremely good energy resolution may be employed to discriminate the few signal events from the overwhelming background. Notwithstanding this, each and every aspect of a future experiment aiming at the Cosmological Relic Neutrino detection has to be tested and validated for the specific case; in this respect PTOLEMY is presently conceived as a set of technological demonstrators that have to converge toward a detector design. 
We can divide this process in mainly three phases:
proof-of-principle demonstrator, scalable prototype realization and tests, full detector construction.
PTOLEMY proof-of-principle phase at the Laboratori Nazionali del Gran Sasso is thus the first of those fundamental steps.

\section*{PTOLEMY proof-of-principle at LNGS}
\addcontentsline{toc}{section}{PTOLEMY proof-of-principle at LNGS}

The proposed work plan for the activities to be carried out in a three- to five-year period in order to complete the PTOLEMY phase I can be summarized as follows:
\begin{itemize}
\item[1)]
validate graphene as target substrate, measure hydrogen/tritium bond characteristics, measure electron-graphene interaction properties,
\item[2)]
achieve an electron energy measurement resolution of 0.05 eV to separate the CNB signal from the $\beta$-decay spectrum, couple detector with spectrometer magnetic field,
\item[3)]
commission the prototype with high stability HV and (single) electron gun,
\item[4)]
operate the prototype and measure the background rate in the CNB signal region (using a pure graphene tritium-free target),
\item[5)] further reduce background by implementing high radio-pure graphene -- eventually exploiting a concurrent program in MeV dark matter searches,
\item[6)]
implement RF antenna for triggering on single electrons in coincidence with an energy measurement, and
\item[7)]
design and simulate a scalable target mass setup with high acceptance kinematic filtering.
\end{itemize}
In order to address these issues a set of Working Packages (WP) has been setup and 
descriptions of the work plans are presented in this document, including current estimates in terms of funds, workforce and timescale for the primary WPs.
%an estimate in terms of \textcolor{red}{funds, workforce and timescale} 
%is reported for each of the WP in the next sections of this document. 
Substantial progress has already been made within the Collaboration on points 1) and 2) with world's best records in hydrogenation of graphene (Princeton) and $\Delta E_{\rm FWHM}$ = 0.12 eV energy resolution for 0.8 eV IR photons at 300 mK (INRiM, Torino). As will be described in detail in the WP sections, the use of an underground low background site is mandatory in order to successfully perform points 3) to 6). Finally, point 7) may be performed in parallel, profiting of the outcome of the previous ones; this represents also the starting point for the second phase of the PTOLEMY project.

Key operations to be performed at LNGS underground site are summarized here in time-order; a more detailed schedule may be found in Fig.~\ref{fig:schedule}:
\begin{itemize}
\item move the equipment and the PTOLEMY prototype from Princeton to LNGS
\item prepare and setup an underground area ( + above-ground service space )
\item commission the prototype as-is
\item realize high precision HV system
\item commission hi-res TES
\item run the prototype with new HV + high precision e-gun + hi-res TES
\item run the prototype with new HV + graphene + hi-res TES
\item commission high radio-purity graphene target
\item realize and test the RF system
\item run the prototype with new HV + high radio-purity graphene + hi-res TES
\end{itemize}

\section*{The PTOLEMY prototype}
\addcontentsline{toc}{section}{PTOLEMY prototype}

\begin{figure*}[h!]
\begin{center}
      \includegraphics[scale=0.1]{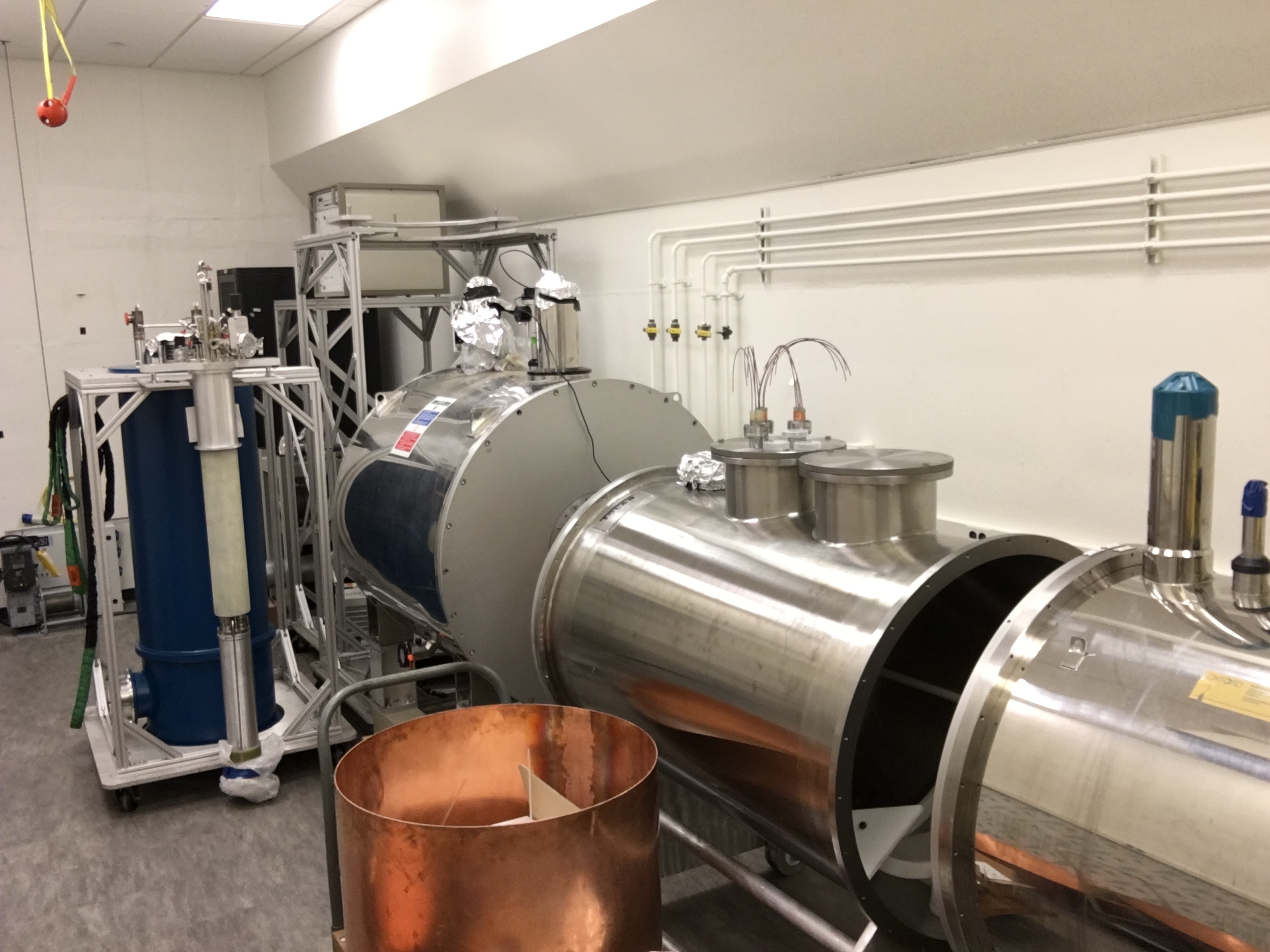}
\vspace*{10pt}
   \caption{
PTOLEMY prototype located in Jadwin Hall Physics Department, Princeton University (9 June 2017) showing the
readout rack (far wall), Kelvinox MX400 dilution refrigerator (gas handler and pumps), 4.7~T Oxford Instruments 200/300 superconducting magnet, central vacuum tank (copper electrodes and HV feed-throughs) and 7.05~T Oxford
Instruments 300/183 superconducting magnet.
 }
\label{fig:PTOLEMYJadwinHall}
\end{center}
%\vspace*{-15pt}
\end{figure*}

The PTOLEMY prototype schema is shown in Fig.~\ref{fig:PTOLEMYLAYOUT} and a picture of the equipment in its current location in the Princeton University Physics Department site is shown in Fig.~\ref{fig:PTOLEMYJadwinHall}. The detector has been funded by the Simons Foundation, Princeton University and the John Templeton Foundation for a total 1~M\$ capital investment.

The PTOLEMY prototype consists of the following components:
\begin{itemize}
\item[$-$] Dilution Refrigerator (Kelvinox MX400 - Oxford Instruments)
\item[$-$] StarCryo Precision X-Ray TES Calorimeter and SQUID readout system
\item[$-$] 1.9~T Oxford Instruments 85/310 horizontal bore superconducting magnet
\item[$-$] 7.05~T Oxford Instruments 300/183 horizontal bore superconducting magnet
\item[$-$] Central vacuum tank hosting a 9-segment high precision HV electrostatic filter
\item[$-$] Spellman 210-30R HV Supply
\item[$-$] Oerlikon Leybold TurboVAC 450 iX (160CF) and ScrollVAC SC15D pumping system.
\end{itemize}
The lowest base temperature of the Dilution Refrigerator is 7~mK and up to 400~mW of cooling power is provided at 100~mK.  A custom cryostat with a sample space exceeding a volume of 10$^3$~cm$^3$ and with a vacuum path connecting to a horizontal port matching the vertical height of the horizontal bore magnets is present in order to match the StarCryo calorimeter with the electromagnetic filter. The MAC-E filter is designed to achieve 1\% energy resolution from the filtering process of adiabatic collimation through a pair of superconducting solenoidal magnets and a retarding electric field~\cite{Beamson1980Collimating,Lobashev1985Method,Picard1992Solenoid}.
%A novel HV reference system developed at LNGS and based on precision voltages held on capacitors with field mill sensing.

%These components are shown in figure~\ref{fig:PTOLEMYJadwinHall} in their current location in the Princeton University Physics Department. 
%The procedures for the shipment, installation and operation for the
%commercial equipment from Oxford Instruments are all documented in manuals.

\section*{Time schedule}
\addcontentsline{toc}{section}{Time schedule}

\begin{figure*}[p]
\begin{center}
      \includegraphics[angle=90,scale=0.46]{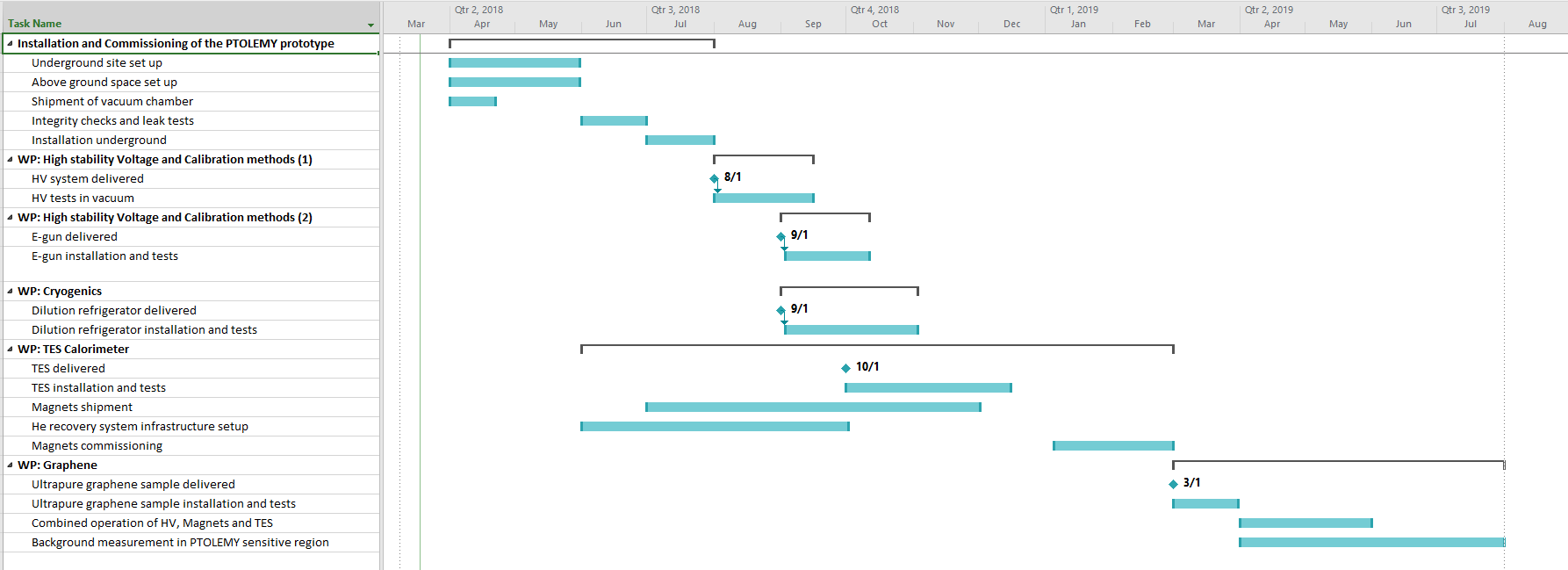}
\vspace*{10pt}
   \caption{Time schedule of the PTOLEMY prototype installation and operation.}
\label{fig:schedule}
\end{center}
%\vspace*{-15pt}
\end{figure*}

%\textcolor{red}{To be revised}
Subject to a positive decision and coordination process with LNGS, the first step is to ship the PTOLEMY prototype to the lab and to ready the system for underground installation.  An approved local contractor would be hired to design and equip the designated underground area with the services, described below, needed to operate the prototype.
Installation and measurement schedules are shown in Fig~\ref{fig:schedule}.
The work breakdown structure is described in detail in the following sections.
LNGS will serve as a center to lead an already substantial and growing international
research program to detect the Cosmic Neutrino Background and to achieve new
experimental capabilities for graphene-based targets with ultra-high radio-purity 
Carbon-12 sources.

As indicated in Fig~\ref{fig:schedule}, the activities on the LNGS site in the second and third quarters of 2018 will focus on the setup of the underground space for the PTOLEMY prototype and the receiving of prototype equipment in a surface hall for transport underground.  The HV system and electron gun calibration systems, developed at LNGS, will be installed on electrodes in the central vacuum tank in quarters three and four of 2018. We anticipate publishing the results of this new HV system by the end of 2018.

The installation of the cryogenic system for the dilution refrigerator(DR) will be the next step in expanding the underground setup and will be coordinated with other groups at LNGS with the goal of making shared use of the helium recovery system.  The delivery of the DR and underground installation are foreseen to be completed by the end of 2018. The DR is pre-wired for the StarCryo TES cryogenic calorimeter with on-detector and off-detector SQUID amplifiers connected through three thermal stages to room temperature electronics. At this time the ongoing R\&D on the TES cryogenic calorimeter at INRiM, Genova and Milano will move to LNGS to integrate into the PTOLEMY prototype.  We expect the results of the calorimeters to already be in the publication process based on this R\&D.  The first PTOLEMY prototype results with the cryogenic calorimeter will be on evaluating the energy resolution of electrons produced with the high-precision electron gun and transported through the electrodes of the HV system.  With this system we will evaluate the energy resolution for keV electrons that are stepped down to match the dynamic range of the cryogenic calorimeter.  This will be a first result of this type and an exciting result for PTOLEMY.  We expect this result in the first quarter of 2019.  

The commissioning of the dewar will also mark the start of testing for MeV dark matter targets for their radio-purity by targets produced by the CIEMAT, Rome and Princeton groups.  
This phase begins in the second quarter of 2019.  We will begin evaluation of the $^{14}$C backgrounds of a standard graphene target source across the keV energy range.  Once the $^{14}$C spectrum is understood, the PTOLEMY prototype will begin evaluating very small admixtures of tritium source (nanogram quantities) into a fully hydrogenated graphene target to provide the first results on molecular smearing from the graphene substrate.

The magnet spectrometer system is essential for evaluating isotropically emitted electrons from a $\beta$-decay source.  The magnets will be delivered in the second quarter of 2018.  The setup of the magnets underground is expected to extend through the first quarter of 2019.  This positions the magnetic field operation to begin following the first results on the TES cryogenic calorimeter described above and will continue through the remainder of 2019.  During this time, the spectrum of backgrounds will be evaluated with the PTOLEMY prototype.  We anticipate integrating single electron RF detection methods into the PTOLEMY prototype following the program of background measurements.  The background suppression factor from the RF electron identification provides important input on the scalability of the neutrino telescope design.  We expect an active period of full simulation and data comparisons to fully describe the PTOLEMY prototype and to extrapolate background measurements to larger target masses.
At the end of 2019, we anticipate launching the program of MeV dark matter searches and to collect data through 2020.

After the first three years of PTOLEMY prototype operation, a second phase of neutrino telescope evaluation will be proposed based on what has been learned from the first phase described above.  The development of a scalable design will be conducted in parallel with the PTOLEMY prototype tests and will lead to a new configuration that is constructed from individual spectrometers whose services will be compatible with the PTOLEMY prototype for evaluation.  The deployment of a full scale telescope array will depend on the prototype performance and background requirements developed at the LNGS.

\section*{Space and service requirements to the lab}
\addcontentsline{toc}{section}{Space and service requirements to the lab}

In order to setup and operate the PTOLEMY prototype an adequate space is required both on surface laboratory and underground.

\subsection*{Surface Lab requirements}

About 50 to 60 square meters would be needed on surface labs in order to test and verify detector components shipped from Princeton. This area could also be used to host the assembly and test of mechanical and electronic parts to be installed on the detector according to the measurements schedule as shown in the previous chapters.  The magnets can be moved on paddle jacks with a weight capacity of 850~kg over a 1~m span.  The refrigerator is mounted on a stand with coasters and can be rolled on a flat surface.  The total weight of the stand, dewar and refrigerator with cryogens at full capacity is 365~kg.  Overhead access (provided by winch or crane) to 6 meters is required to lift the (175~kg) dewar and place it on the stand.  The stand can be moved into an adjacent hall to gain overhead access.  The refrigerator insert fits inside the dewar and must be lowered vertically into the dewar from a height of approximately 6 meters.
The weight and height of these components are described below.

\subsection*{Underground Lab requirements}

\begin{figure*}[h!]
\begin{center}
      \includegraphics[width=0.9\textwidth]{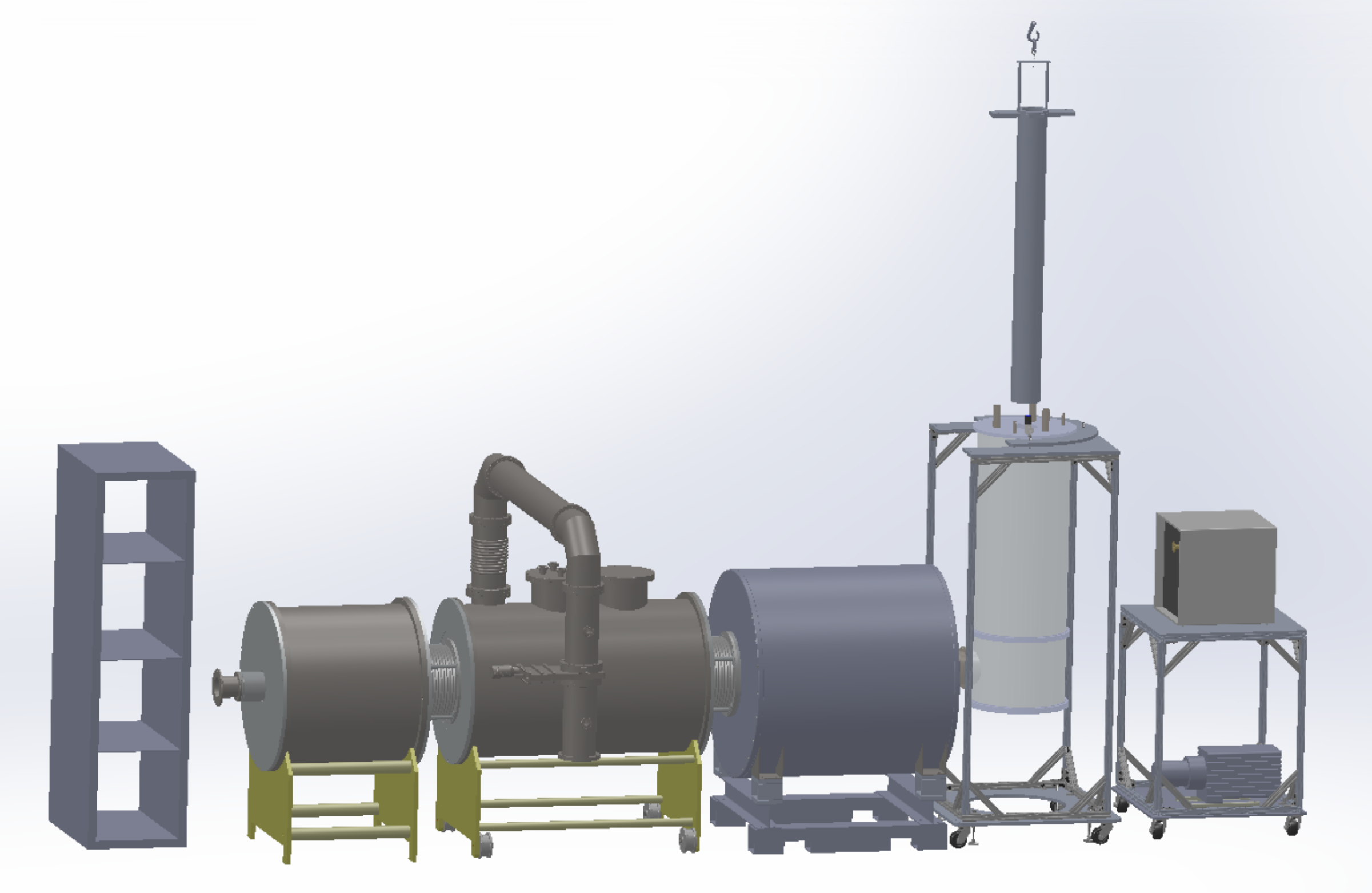}
   \caption{
3D CAD image of PTOLEMY Layout. From left to right: target region and first superconducting magnet, central vacuum chamber, second superconducting magnet, Dilution Refrigerator hosting the TES calorimeter.
 }
\label{fig:PTOLEMYLAYOUT}
\end{center}
\vspace*{-15pt}
\end{figure*}

\begin{figure*}[h!]
\begin{center}
      \includegraphics[width=0.9\textwidth]{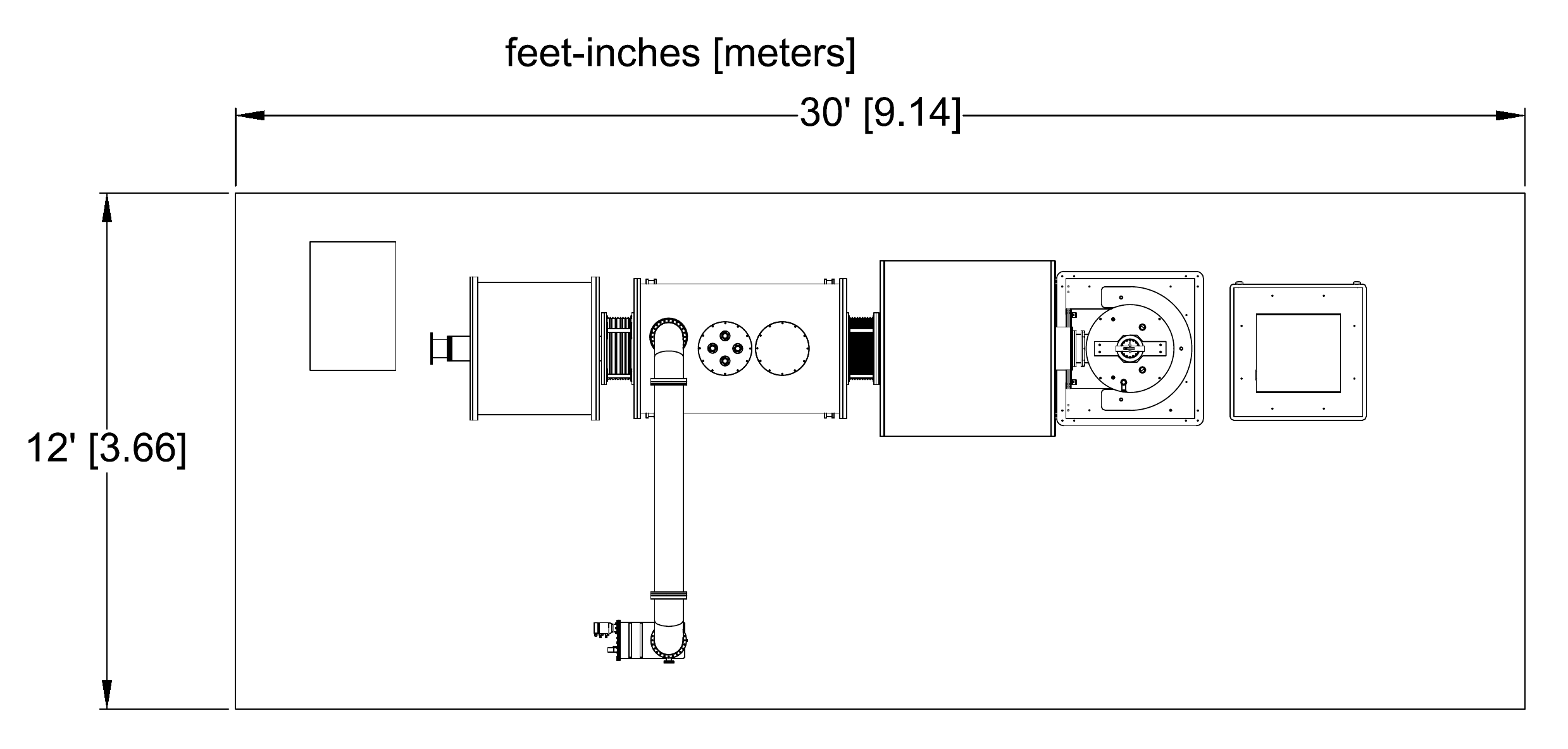}\\
      \includegraphics[width=0.9\textwidth]{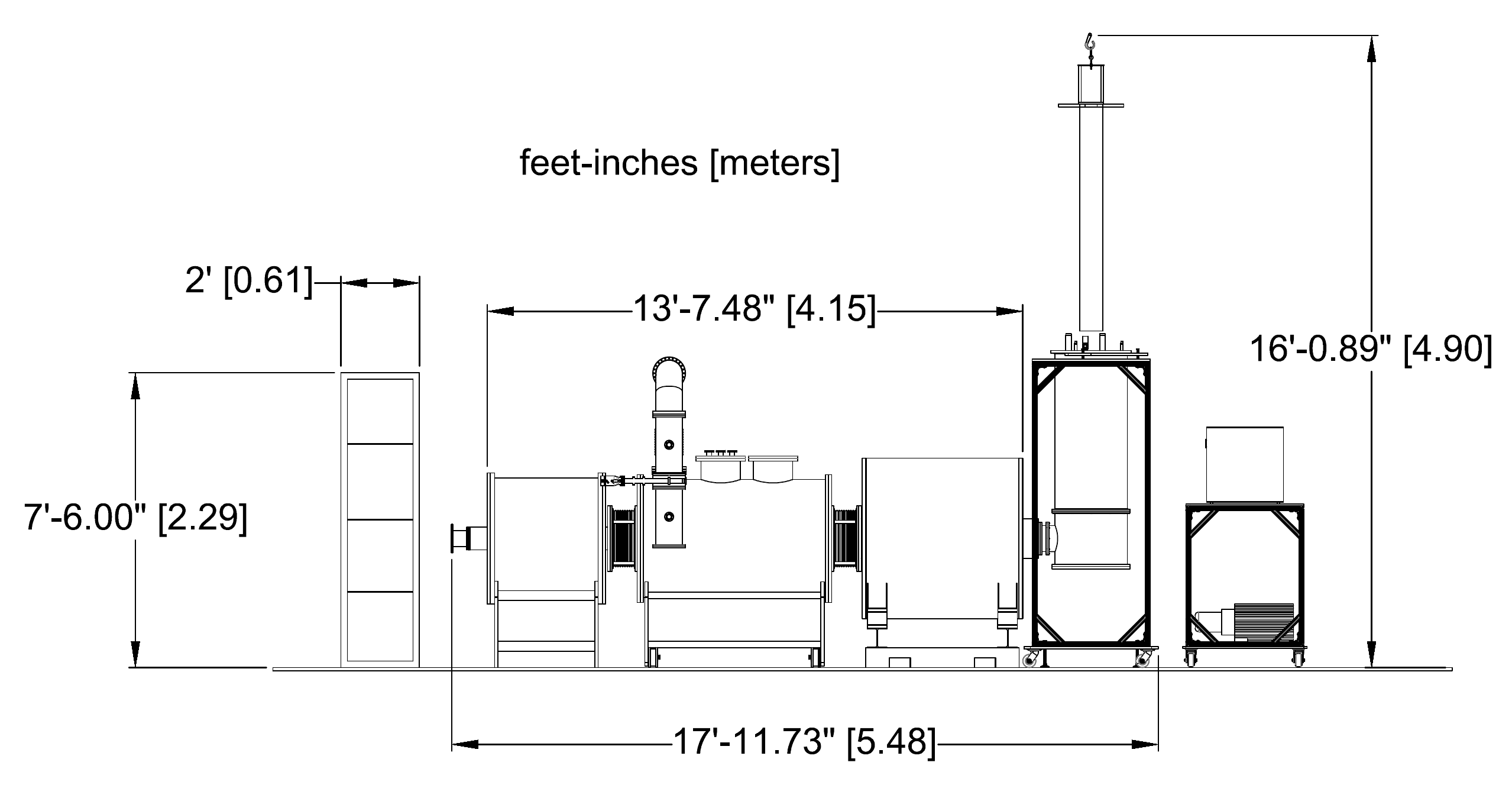}
   \caption{
(\emph{top}) Floor plan of the PTOLEMY setup. (\emph{bottom}) Vertical height of the PTOLEMY setup.
 }
\label{fig:PTOLEMYFloor}
\end{center}
\vspace*{-15pt}
\end{figure*}

Underground space requirements are driven by the actual footprint of the PTOLEMY prototype as shown in Fig.~\ref{fig:PTOLEMYLAYOUT}. The space needed in order to adequately commission and operate the detector is of about 11 meters length by 7 meters width; the extra width with respect to the footprint is needed in order to accommodate data acquisition and control site at an appropriate distance from the superconducting magnets. Required height clearance is of about 6 meters due to the insertion of the refrigerator in the dewar, as shown in Fig.~\ref{fig:PTOLEMYFloor} (bottom); this operation is expected to be infrequent and may possibly be done in a neighboring hall, if needed.

An overhead winch or crane access to place the dewar (175~kg) on the stand and the refrigerator insert (40~kg) into the dewar is required.  As the stand can move on coasters, the overhead access can be provided in a neighboring hall.  Max crane 
required weight is 225~kg.

\noindent Service requirements are summarized as follows:

\begin{itemize}
\item[] \underline{Liquid Helium Requirements} (expansion factor 740) \\ 
Oxford Instrument 300/183 magnet: \\
\hspace*{20pt} 2 liters of liquid helium per day (1.4~cc/min = 1.0~lpm) \\
\hspace*{20pt} Refill once per two weeks (64 liter capacity: 35 liter refill volume) \\
Oxford Instrument 85/310 magnet: \\
\hspace*{20pt} 1.2 liters of liquid helium per day (0.8~cc/min = 0.6~lpm) \\
\hspace*{20pt} Refill once per four weeks (64 liter capacity: 35 liter refill volume) \\
\hspace*{20pt} Volume for cooldown: 150 liters \\
Oxford Instrument Kelvinox MX400 Refrigerator: \\
\hspace*{20pt} 3.5 liters of liquid helium (1k pot) per day (2.4~cc/min = 1.8~lpm) \\
{\bf Total Daily LHe:} 6.7 liters liquid helium per day (operation)
\end{itemize}

\begin{itemize}
\item[] \underline{Liquid Nitrogen Requirements:} (expansion factor 680) \\
Oxford Instrument 300/183 magnet: \\
\hspace*{20pt} 7.5 liters of liquid nitrogen per day (5~cc/min = 3.4~lpm) \\
\hspace*{20pt} Refill once per two weeks (158 liter capacity: 134 liters refill volume) \\
Oxford Instrument 85/310 magnet: \\
\hspace*{20pt} 15 liters of liquid nitrogen per day (10~cc/min = 6.8~lpm) \\
\hspace*{20pt} Refill once per week (158 liter capacity: 134 liters refill volume) \\
\hspace*{20pt} Volume for cooldown: 300 liters \\
Oxford Instrument Kelvinox MX400 Refrigerator: \\
\hspace*{20pt} 15 liters of liquid nitrogen per day (10~cc/min = 6.8~lpm) \\
{\bf Total Daily LN2:} 37.5 liters liquid nitrogen per day (operation)
\end{itemize}

\begin{itemize}
\item[] \underline{Crane operation:} \\
Oxford Instrument 300/183 magnet: \\
\hspace*{20pt} Magnet and Cryostat Weight: 850~kg \\
\hspace*{20pt} Cryogens: 140~kg \\
\hspace*{20pt} {\bf Total Weight:} 990~kg \\
\hspace*{20pt} Height: 2600~mm \\
Oxford Instrument 85/310 magnet: \\
\hspace*{20pt} Magnet and Cryostat Weight: 777~kg \\
\hspace*{20pt} Cryogens: 143~kg \\
\hspace*{20pt} {\bf Total Weight:} 920~kg \\
\hspace*{20pt} Height: 2600~mm \\
Oxford Instrument Kelvinox MX400 Refrigerator: \\
\hspace*{20pt} Insert Weight: 40~kg \\
\hspace*{20pt} Dewar Weight: 175~kg \\
\hspace*{20pt} Cryogens: 150~kg \\
\hspace*{20pt} {\bf Total Weight:} 365~kg \\
\hspace*{20pt} Height (access during installation): 4900~mm \\
\hspace*{20pt} Height: 3000~mm
\end{itemize}

\begin{itemize}
\item[] \underline{Electrical requirements:} \\
Oxford Instruments Relay Box (4400~W) for Kelvinox MX400: \\
\hspace*{20pt} 3-Phase 208~V, 30~Amps \\
\hspace*{20pt} Powered by Relay Box: He3 PUMP 3-Phase 208~V, 2200~W \\
\hspace*{20pt} Powered by Relay Box: ROOTS PUMP 3-Phase 208~V, 1500~W \\
\hspace*{20pt} Powered by Relay Box: He4 PUMP 1-Phase 208~V, 550~W \\
Oxford Instruments IGH Gas Handler (250~W) for Kelvinox MX400: \\
\hspace*{20pt} 1-Phase 200/230~V, 10~Amps \\
Oxford Instruments IPS Magnet Power Supply (only to energize magnets): \\
\hspace*{20pt} 1-Phase 208~V, 20~Amps \\
Pumping System: \\
\hspace*{20pt} Oerlikon Turbo Integra 100/240V, 403~W \\
\hspace*{20pt} Oil-Free Scroll Nidec 100/240V, 400~W \\
Spellman 210-30R HV Supply: \\
\hspace*{20pt} 1-Phase 230~V, 2.5~Amps \\
Readout Rack (19 inch width): \\
\hspace*{20pt} 1-Phase 208-V UPS, 30~Amps (98~kg, 5U)\\
\hspace*{20pt} Powered by UPS: Control PC (400W, 4U), LCD Monitor (2U)
\end{itemize}

\clearpage

\section*{Work Breakdown Structure of the PTOLEMY project}

%For each WP the following items must be included: 
%\begin{itemize}
%\item[-] R\&D description (where applicable)
%\item[-] Resources required for prototype run
%\item[-] Tentative timetable for the final detector
%\item[-] Cost and manpower estimate 
%\end{itemize}

\begin{figure*}[h!]
\begin{center}
      \includegraphics[width=0.85\linewidth]{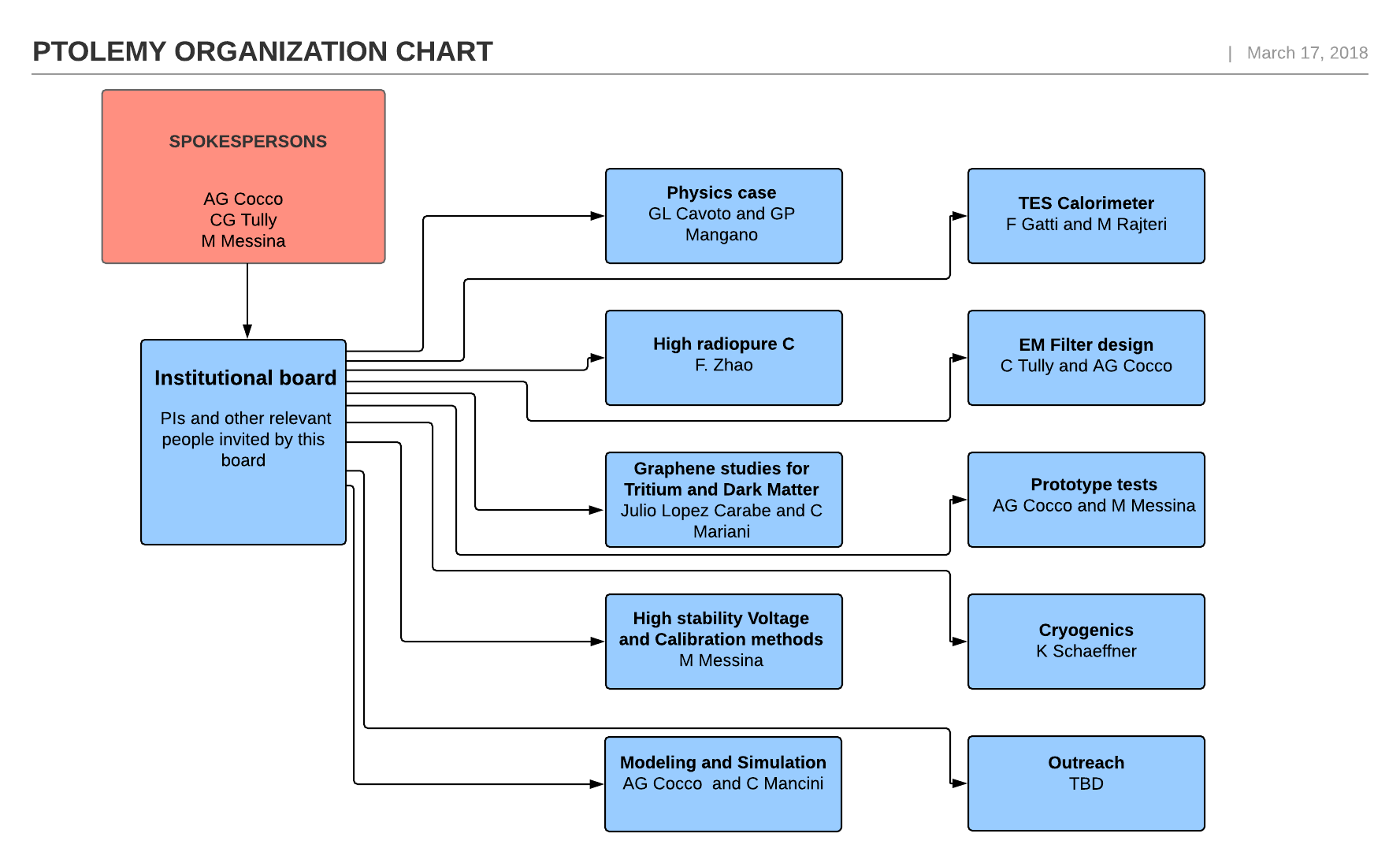}
   \caption{Organizational chart of the PTOLEMY project.
 }
\label{fig:PTOLEMY_OrgChart}
\end{center}
%\vspace*{-15pt}
\end{figure*}

The organizational structure of the PTOLEMY project is shown in Fig~\ref{fig:PTOLEMY_OrgChart}.  The project has a work breakdown structure (WBS) whereby ten parallel efforts (Physics case, High radiopure C, Graphene studies for Tritium and Dark Matter, High stability Voltage and Calibration methods, Modeling and Simulation, TES Calorimeter, EM Filter design, Prototype tests, Cryogenics, Outreach) are coordinated by one or two individuals from the institutes with membership in the institutional board (IB).  The institutional board will form a constitution and conduct regular elections for the IB chair and deputy pending the decision of the LNGS.  The spokespeople report to the institutional board and coordinate the project across the work packages.

\section{Physics cases}
%\noindent \large {\bf Coordinators: G.P.~Mangano and G.L.~Cavoto} \normalsize
%\begin{itemize}
%\item[-] relic (three flavor cases and inverted and normal hierarchy and polarized target)
%\item[-] Dark Matter (planar and nano-tube)
%\item[-] Sterile neutrino
%\end{itemize}
%please insert text here

The PTOLEMY project aims to develop a scalable design for a Cosmic Neutrino Telescope, the first of its kind and the only telescope conceived that can look directly at the image encoded in neutrino density fluctuations of the Universe in the first second after the Big Bang.  The scope of work for the next three years is to complete the design of the Cosmic Neutrino Telescope and to validate with direct measurement that the non-neutrino backgrounds are below the expected signal from the Big Bang.  An array of telescopes of this design will reach discovery sensitivity for the Cosmic Neutrino Background.  The number and deployment of these telescopes around the world will depend on the next phase of PTOLEMY developments described in this proposal.

A complementary physics program based on high radio-purity graphene targets is proposed to further the research into backgrounds originating from interactions in the graphene target and to explore directional detection sensitivity for MeV dark matter searches~\cite{hochberg,cavoto}.

%Apart from the ultimate goal of studying the cosmological neutrino background (CNB),
%the physics case of the PTOLEMY experiment proposal is quite wide, spanning from
%MeV directional detection of Dark Matter (DM)~\cite{hochberg,cavoto} candidates, to neutrino mass scale measurement in the sub-eV range, as well as to bounds to sterile neutrino states in both the eV mass range, for their cosmological implications see e.g.~\cite{mirizzi}, and keV~\cite{adhikari}. 
%In fact,
%eV neutrinos which mix with active states have been suggested since the LSND results
%to solve some anomalies in neutrino oscillation experiments (short baseline data, reactor
%anomaly and Gallex anomaly). On the other hand neutrino states with a mass in the keV
%range are excellent candidates of warm DM~\cite{boyarsky}.

\subsection*{Cosmic Neutrino Background}

The Universe has expanded by a factor of over one billion between the present-day and the early thermal epoch known as the neutrino decoupling.  We observe this dynamics in many forms:  the recession of galaxies (Hubble Expansion), the dim afterglow of the hot plasma epoch (Cosmic Microwave Background) and the abundances of light elements (Big Bang Nucleosynthesis).  The epoch of neutrino decoupling produced a fourth pillar of confirmation – the Cosmic Neutrino Background (CNB). These early universe relics have cooled under the expansion of the Universe and are sensed indirectly through the action of their diminishing thermal velocities on large-scale structure formation. Experimental advances both in the understanding of massive neutrino physics and in techniques of high sensitivity instrumentation have opened up new opportunities to directly detect the CNB, an achievement which would profoundly confront and extend the sensitivity of precision cosmology data.

Exciting new ideas for late-time inflationary and non-inflationary predictions provide new opportunities to search for non-standard anisotropies in the cosmic neutrinos.
A new design for incorporating angular resolution into the telescope has been recently introduced, making use of expert knowledge of polarized tritium cross section calculations~\cite{marcucci2012chiral,chang2017nuclear,lisanti2014measuring} from the Pisa group. 
Similarly, the direct measurements per neutrino mass of the CNB from a future Cosmic Neutrino Telescope will provide the only test of cosmic neutrino physics at a precision comparable to the new frontier of CMB measurements.  We are also very enthusiastic about the central role of further collaboration with center for computational astrophysics to bring together exact numerical predictions about the Universe across a rapidly growing field with new and rich challenges to confront an ever expanding gamut of direct measurements. 

\begin{figure*}[h!]
\begin{center}
      \includegraphics[width=0.9\textwidth]{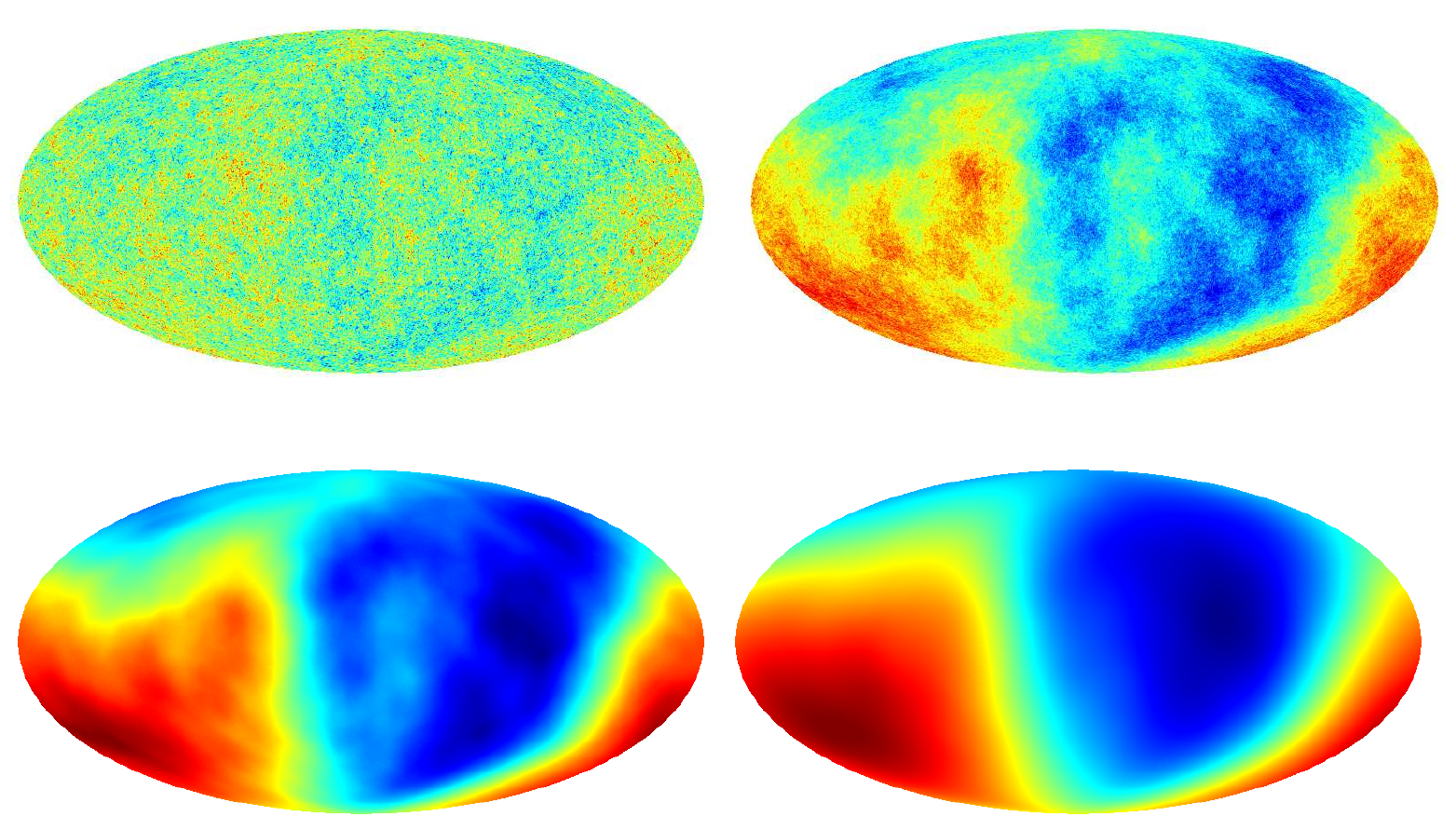}
   \caption{
With a polarized tritium target, an angular dependence of the neutrino capture cross section introduces the possibility of mapping the all sky density of cosmic neutrinos~\cite{brandbyge2009grid,lisanti2014measuring}.  The lightest neutrinos (top-left, 10$^{-5}$eV, top-right, 10$^{-3}$eV, bottom-left, 10$^{-2}$, bottom-right, 10$^{-1}$eV) will probe late-time inflationary density fluctuations and the heaviest neutrinos will measure the gravitational instability during long-scale structure formation following the transition from relativistic to non-relativistic kinetic energies.
 }
\label{fig:neutrinoanisotropies}
\end{center}
\vspace*{-15pt}
\end{figure*}

The three species of neutrinos will undergo different levels of gravitational instability from the period of neutrino decoupling to the present day~\ref{fig:neutrinoanisotropies}.  The heaviest neutrinos are predicted to have relatively large dipolar distributions in density that span the full sky.  The lighter neutrinos, depending on the mass hierarchy and absolute mass scale in the neutrino sector, will elucidate more detail about late-time inflation density fluctuations~\cite{brandbyge2009grid}.  
%The lightest neutrino may also have an important impact on the tritium endpoint spectrum, as proposed in original paper by Weinberg'~\cite{WeinbergRelic}.  
These effects are to be quantified with full simulation of the PTOLEMY prototype with measured resolutions and performances.

PTOLEMY is based on the detection of the CNB by the process of neutrino capture on tritium~\cite{WeinbergRelic}
\begin{equation}
\nu_e + {\rm ^3H} \rightarrow {\rm ^3He} + e^- .
\end{equation}
Tritium has been chosen among other target candidates because of its availability, lifetime
of 12.3 years, high neutrino capture cross section and 
low $Q$-value~\cite{Cocco2007Probing}.
The smoking gun signature of a relic neutrino capture is the peak in the electron spectrum, shifted $\sim 2 m_\nu$ above the $\beta$-decay endpoint.
Because flavor neutrino eigenstates are a composition of mass eigenstates with different
masses, relic neutrinos quickly decohere into those, on a time scale less than one
Hubble time~\cite{weiler}. Therefore, the capture rate of relic neutrinos by their absorption in tritium
\begin{equation}
\Gamma_{CNB} = \sum_{i=1}^{N_\nu} \Gamma_i ,
\end{equation}
must be computed from the capture rates of the neutrino mass eigenstates $\Gamma_i$.
Following Ref.~\cite{long} it can be obtained
\begin{equation}
\Gamma_i = \bar{\sigma} N_T |U_{ei}|^2 f_{c,i} n_0 .
\end{equation}
where $\bar{\sigma}$ is the cross section for neutrino capture, $N_T$ the number of tritium nuclei,
$U_{ei}$ is the mixing of each mass eigenstate with the electron flavor, and the factor $f_{c,i} n_0$ is the number density of the $i^{th}$ mass eigenstate relic neutrino, where $f_{c,i}$ is the clustering factor, i.e. the factor that gives the overdensity of such particles due to the gravitational attraction of our galaxy~\cite{ringwald,desalas2,zhang},
and
\begin{equation}
n_0 \frac{ 3 \zeta(3)}{4 \pi^2} T^3_{\nu,0} = 56~{\rm cm}^{-3}
\end{equation}
is the number density without clustering, per neutrino and degree of freedom.  It is
obtained from a Fermi-Dirac distribution with a temperature $T_{\nu,0}=0.168$~meV, which
corresponds to the temperature of a relativistic gas of fermions with the same distribution
as that of the CNB today.

Because of the finite energy resolution, the main background to this process comes from
the most energetic electrons of the $\beta$-decay of tritium, since they can be measured with
energies larger than the $\beta$-decay endpoint.  To estimate the rate of such background,
we need to account for the $\beta$-decay spectrum~\cite{masood}.
Now we consider the experimental energy resolution $\Delta$, and introduce a smearing in
the electron spectrum by convolving both, the CNB and the $\beta$-decay spectra, with a
Gaussian of full width at half maximum (FWHM) equal to the energy resolution $\Delta$. 
This relates with the standard deviation as
\begin{equation}
\sigma = \Delta/\sqrt{8 \ln 2} .
\end{equation}

\begin{figure*}[h!]
\begin{center}
	\includegraphics[width=0.9\textwidth]{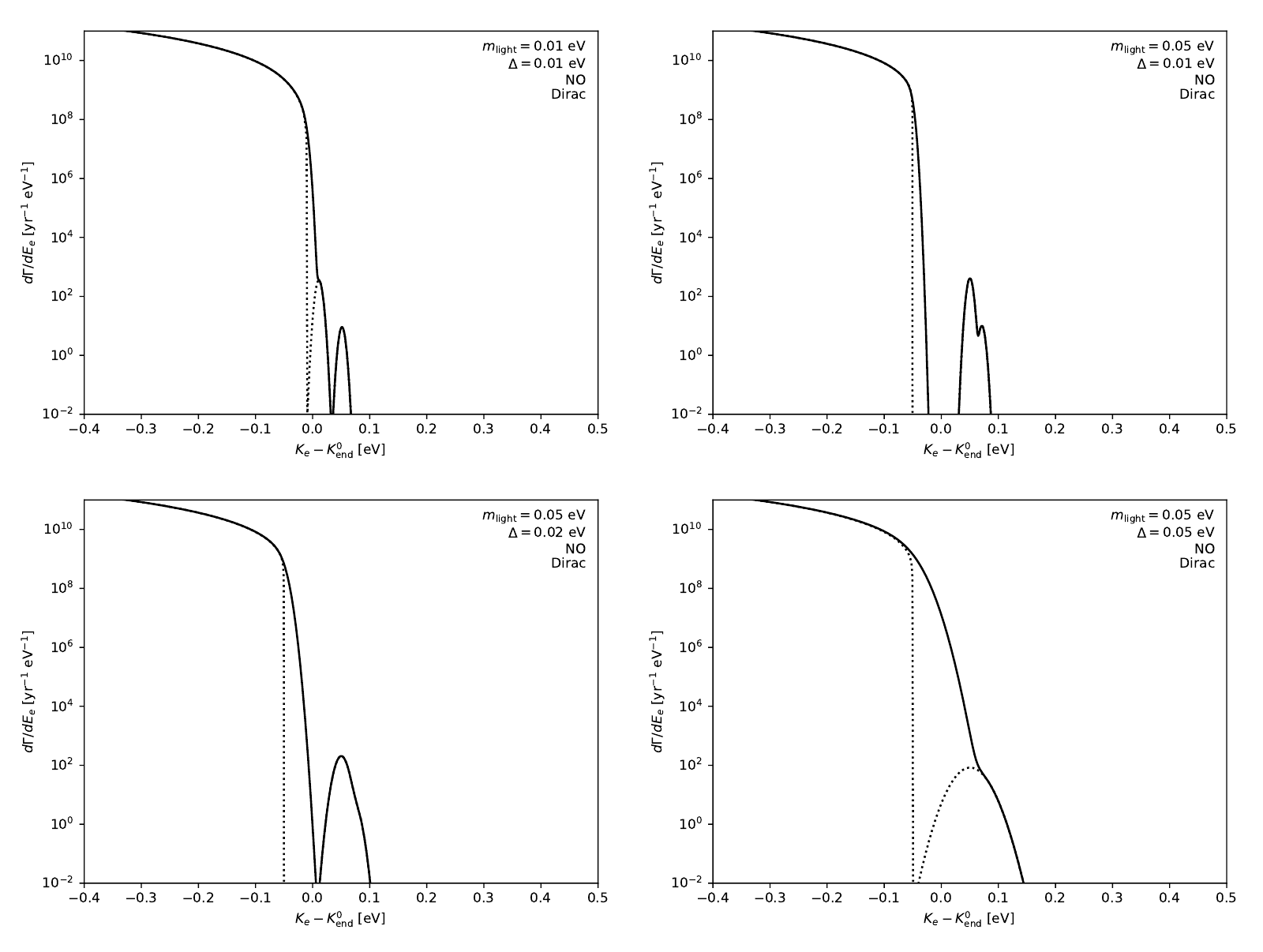}
\caption{
Expected event rates in a direct-detection experiment near the beta-decay endpoint
for different lightest neutrino masses and energy resolutions. Solid lines represent
the event rates convolved with a Gaussian envelope of FWHM equal to the assumed energy
resolution. Dotted lines show the same event rates without the convolution.
Normal ordering of neutrino masses and Dirac nature are assumed.
}
\label{fig:massesNO}
\end{center}
\end{figure*}

\begin{figure*}[h!]
\begin{center}
	\includegraphics[width=0.9\textwidth]{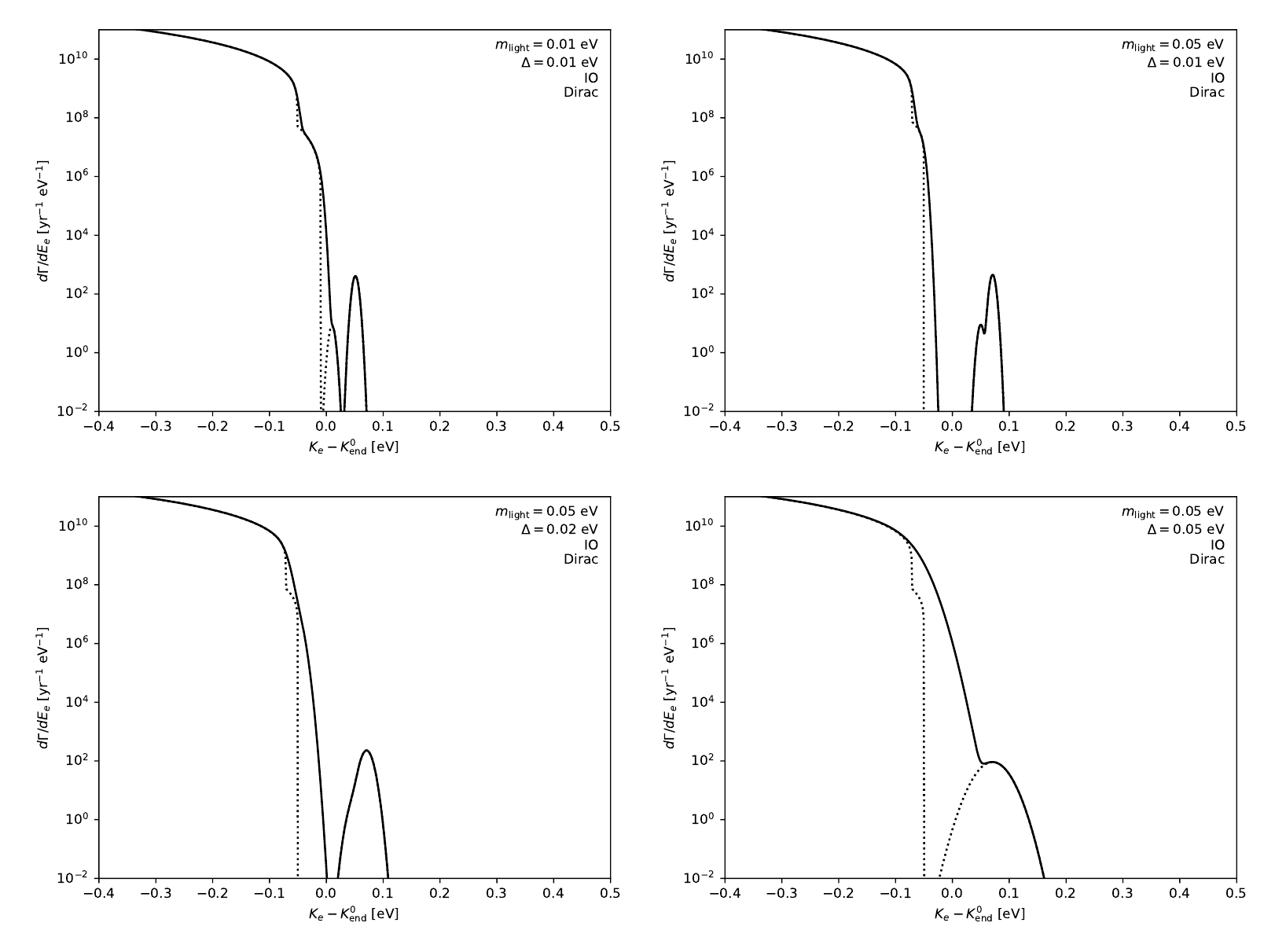}
\caption{
Same as Fig.~\ref{fig:massesNO} for inverted ordering of neutrino masses.
}
\label{fig:massesIO}
\end{center}
\end{figure*}

Figs.~\ref{fig:massesNO} and \ref{fig:massesIO} show the expected event rates at energies close to the $\beta$-decay endpoint.  Different neutrino masses and energy resolutions are assumed. From these figures we notice that those neutrinos whose masses are larger than the energy resolution can be more easily resolved from the $\beta$-decay background. Otherwise, their peaks are confused with the endpoint of the more prominent $\beta$-decay spectrum. In the inverted ordering scenario of neutrino masses (see Fig.~\ref{fig:massesIO}) we further appreciate the kink in the $\beta$-decay spectra produced in the transition when the neutrino capture through the two heavier states is kinematically allowed.
These figures emphasize the primary goals of the PTOLEMY prototype, energy resolution and low backgrounds.

The PTOLEMY prototype, operating at LNGS, will leverage the factor of one million in cosmogenic background reduction to probe with high sensitivity, the spectrum of backgrounds that appear in the energy range relevant to the CNB detection. The goal of operating underground is to demonstrate that the high radio-pure graphene target, cryogenic calorimeter and precision high-voltage system and electromagnetic filters can achieve the low background requirements needed to proceed to a full implementation of the cosmic neutrino experiment.  A relatively small target system proven to have high radio-purity can be scaled to a large area system with known background levels.  The cosmic neutrino telescopes will not need to be located deep underground, but we need to start in a low background environment to understand what backgrounds and background-induced instabilities are relevant.

\subsection*{Directional Detection of MeV Dark Matter}

The development goals of high radio-purity graphene targets are two-fold.  The first is to yield a low background target for the CNB measurement, and the second is for the deployment of high sensitivity detectors using the unique properties of graphene.
The proposal is to focus initially on the latter goal and to conduct a significant MeV dark matter search with novel graphene-based detectors~\cite{hochberg,cavoto}.
The count rate requirements are more stringent for MeV dark matter searches due to the broad energy spectrum of low energy recoil electrons, as described below.

There are two approaches to directional detection MeV dark matter(DM) searches that will be investigated with the PTOLEMY prototypes.  One of them, called PTOLEMY-G$^3$, self-instruments the graphene target at the level of single electron sensitivity.  This extraordinary level of sensitivity enables the detector to sense low levels of radio-impurities throughout the target volume without the need of a magnetic spectrometer.
The second is a carbon nanotube(CNT) detector, called PTOLEMY-CNT herein, that leverages the anisotropic scattering and absorption properties of tightly packed, aligned CNTs to determine the scattering direction of MeV dark matter.  The advantage of this approach is in the large scale increase in graphene mass that is achieved with this approach, while still maintaining sensitivity to the forward-backward scattering cross section asymmetry imposed by the direction of the dark matter wind.

\subsection*{PTOLEMY-G$^3$}

\begin{figure*}[h!]
\begin{center}
	\includegraphics[width=0.4\textwidth]{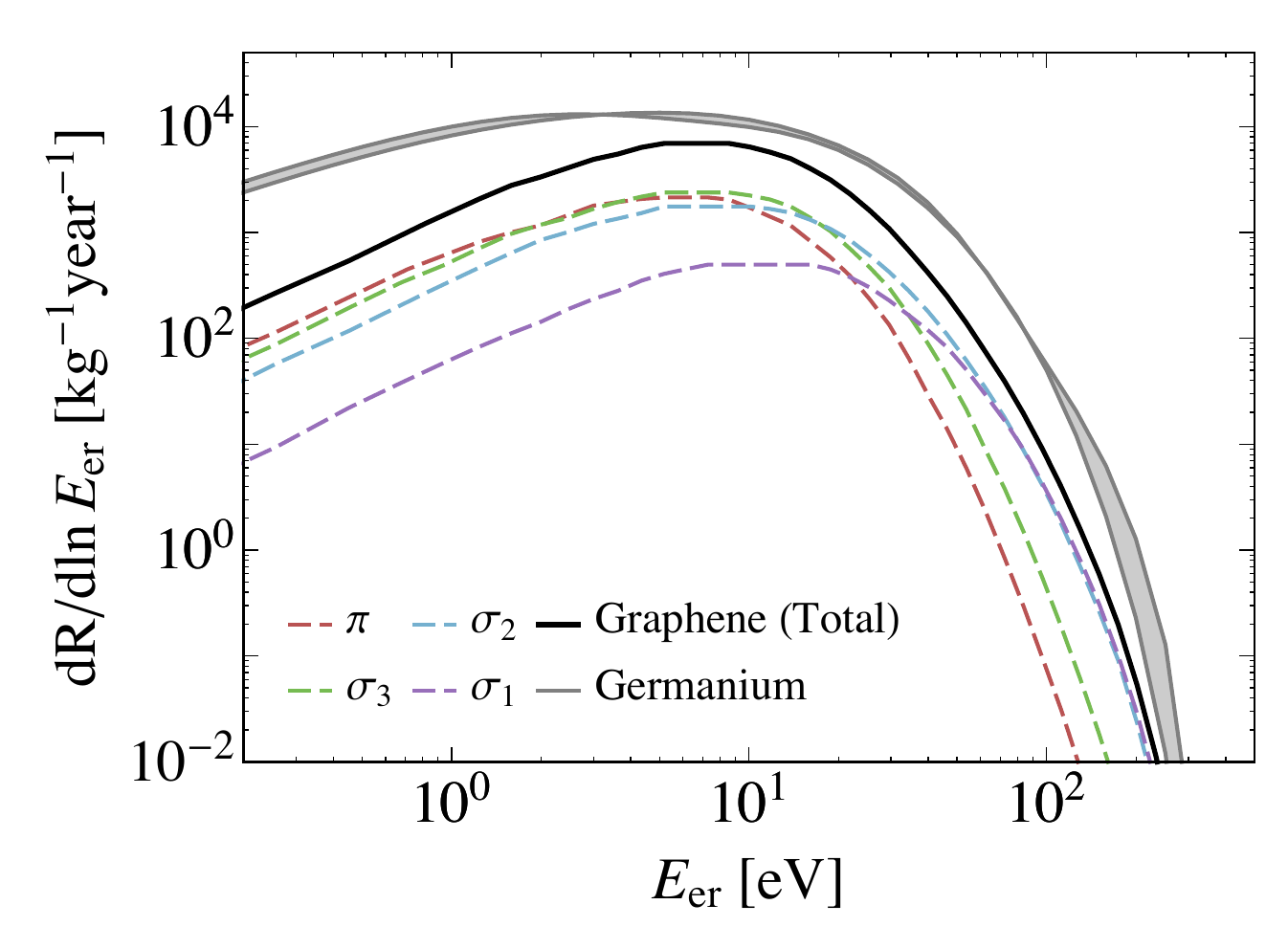}
	\includegraphics[width=0.4\textwidth]{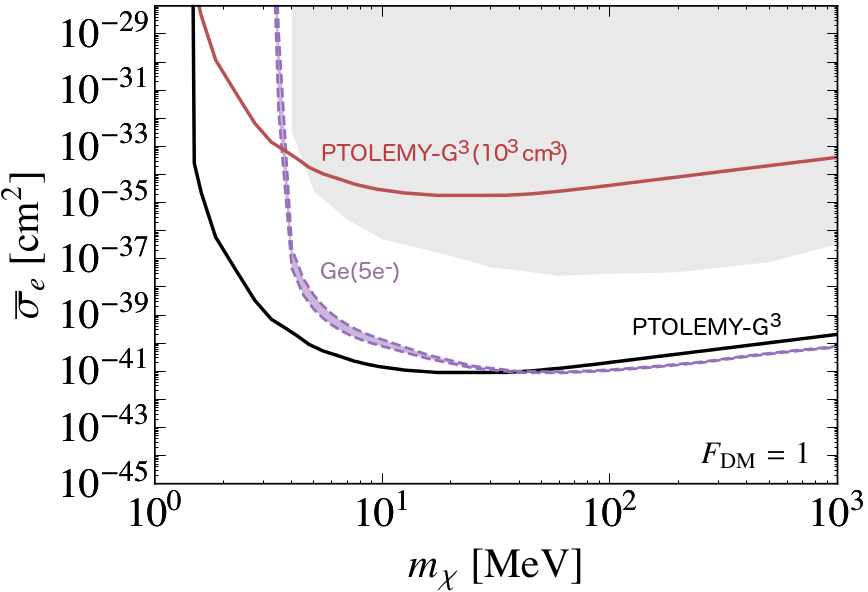}
\caption{(\emph{left}) Differential rate for a 100~MeV DM particle scattering off an electron in graphene is shown with the solid black line with $\bar{\sigma}_e = 10^{-37}$~cm$^2$ and $F_{\rm DM}(q) = 1$.  (\emph{right})  Expected background-free 95\% C.L. sensitivity for a graphene target with a 1-kg-year exposure (black).  A first experiment with a G$^3$ volume of 10$^3$~cm$^3$ (target surface of 10$^{4}$~cm$^2$) will search down to approximately $\bar{\sigma}_e = 10^{-33}$~cm$^2$ at 4~MeV.
}
\label{fig:diffR}
\end{center}
\end{figure*}

\begin{figure*}[h!]
\begin{center}
	\includegraphics[width=0.8\textwidth]{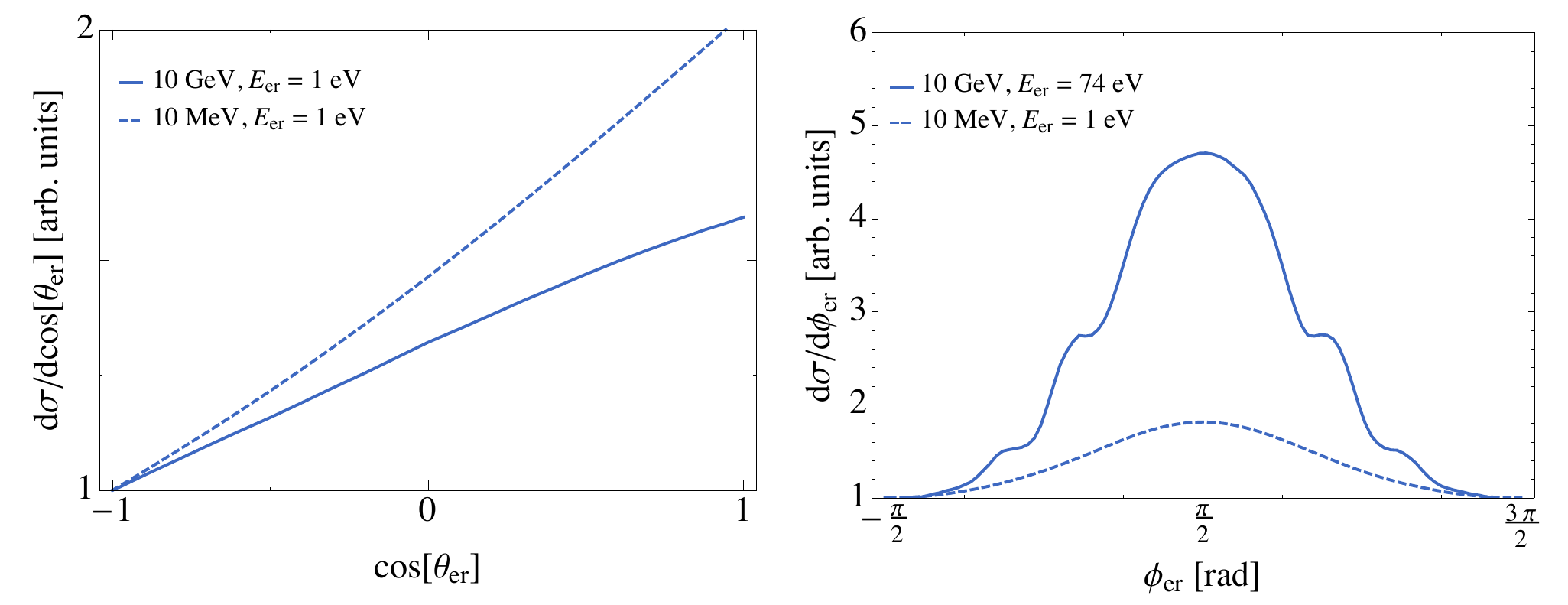}
\caption{Predicted angular distributions for DM masses 10~MeV (dashed) and 10~GeV (solid) in a DM stream with $v_{\rm stream} = 550$~km/s in the lab frame. ({\em left}) Polar distribution of the final-state electron when the stream is oriented perpendicular to the graphene plane and points along $\cos\theta =1$. ({\em right}) Azimuthal distribution of the final-state electron when the stream is oriented parallel to the graphene plane and points along $\phi = \pi/2$.  The outgoing electron direction is highly correlated with the initial DM direction.}
\label{fig:diffrac}
\end{center}
\end{figure*}

With a small-scale deployment of PTOLEMY-G$^3$~\cite{hochberg}, based on G-FET sensors, a fiducialized volume of 10$^3$~cm$^3$ consisting of 100 stacked 4-inch wafers will search down to approximately $\bar{\sigma}_e = 10^{-33}$~cm$^2$ for dark matter masses of 4~MeV in one year, uncovering a difficult blind spot inaccessible to current nuclear recoil experiments (see Fig.~\ref{fig:diffR}).  This new approach will open up for the first time direct directional detection of MeV dark matter (see Fig.~\ref{fig:diffrac}), a capability that no other light dark matter proposal has and which would be highly complementary to a detection, for example, in DAMIC or SENSEI.
The graphene target will follow high radio-purity wafer-level fabrication procedures~\cite{jastram2015cryogenic}.
The support structures will use materials that have achieved
high radio-purity~\cite{chen2017pandax}.

%\vspace*{-30pt}

\begin{figure*}[h!]
\begin{center}
      \includegraphics[scale=0.6]{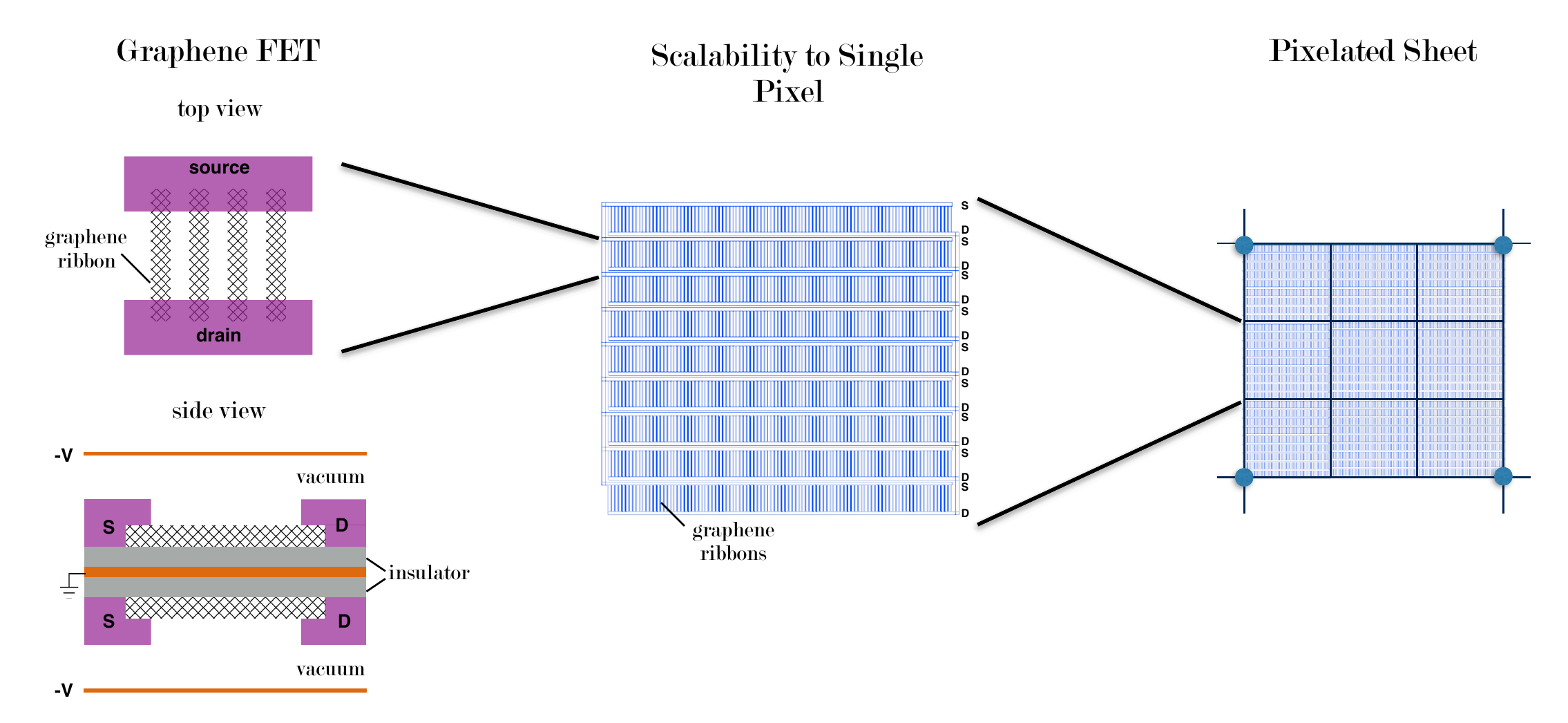}
   \includegraphics[scale=0.5]{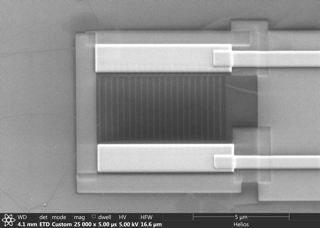}
   \includegraphics[scale=0.3]{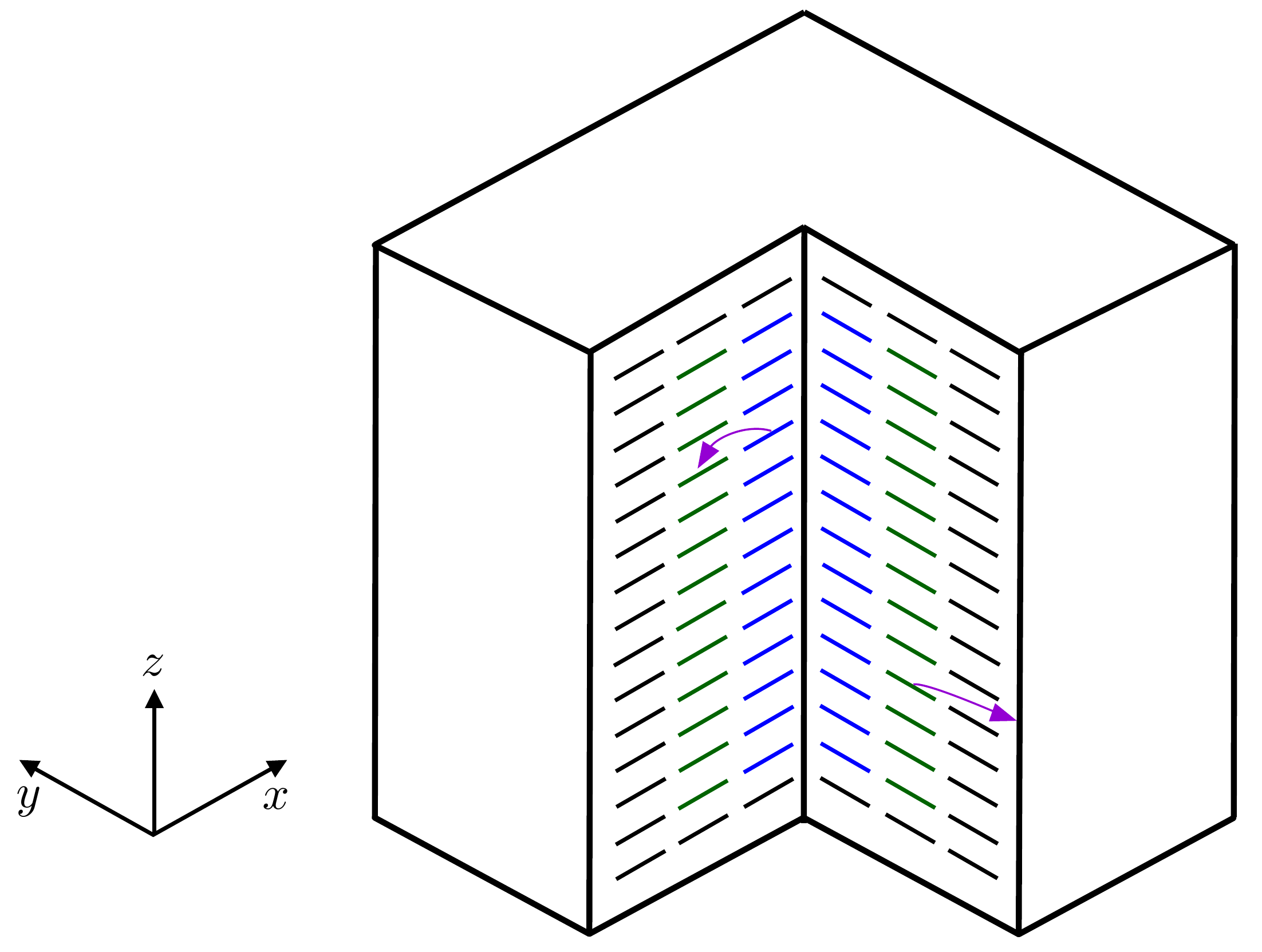}
   \vspace*{5pt}
   \caption{
 ({\em top}) The FET plane will be double-sided, separated by two insulating layers and a bottom gate electrode.  Top gate electrodes will provide the $\sim -100$~V needed to accelerate ejected electrons away from the electrodes and back towards the graphene planes.  Multiple graphene FETs can be arranged into a single pixel (center) with interdigitated source and drain and multiple pixels are arranged into sheets that are stacked together to form a cube structure and
 multiple cubes are assembled to form a fiducialized volume.
 ({\em left bottom}) Prototype graphene FET sensor made at Princeton University consists of a source and drain separated by a planar graphene layer segmented finely into ribbons. ({\em right bottom}) Cutaway view of a conceptual design for graphene directional detection. When an electron is ejected from a graphene sheet, it is deflected by an electric field, where electrons follow a ``FET-to-FET'' trajectory.
 }
\label{fig:conceptdesign}
\end{center}
%\vspace*{-15pt}
\end{figure*}

The G-FET sensor has a tunable meV band gap (see Fig.~\ref{fig:conceptdesign}), a full three orders of magnitude smaller than cryogenic germanium detectors.  This sensitivity is used to switch on and off the conductivity of the G-FET channel by 10 orders of magnitude in charge carriers in response to the gate voltage shift from a single scattered electron.  A narrow, vacuum-separated front-gate imposes kinematic discrimination on the maximum electron recoil energy, where low energy recoil electrons above the graphene work function follow FET-to-FET directional trajectories within layers of the fiducialized G$^3$ volume.
Each FET plane will be vacuum sealed on top and bottom during assembly.
The target will be kept at cryogenic temperatures and have no line-of-sight vacuum trajectories from the outer vacuum region to the sealed FET planes.  Residual gas backgrounds will be cryopumped to the outer boundaries of the fiducialized volume.
%\vspace*{-10pt}

%With an area per plane of $10^{6} \ {\rm cm}^2$, the overburden of cosmic-ray muon flux is an important concern for
%dead-time associated with a cosmic-ray veto.  The instrumented target is designed to have no more than a
%percent-level fill factor of support material, mostly epoxy or a similar material to support the graphene sheets at the corners as shown in Fig~\ref{fig:conceptdesign} (right).  The remainder of the target volume will be highly sensitive to charged
%particles entering the volume, and therefore the electric field regions that control the conductivity of the 
%graphene FETs, including the regions between the vacuum-separated top gate electrode and the graphene and 
%underneath the graphene with the insulator-separated bottom gate electrode, will be active regions for cosmic-ray vetos.  With an overburden of roughly 3~km or greater, the total flux of muons across the entire graphene target falls below $10^{-1} \ {\rm s}^{-1}$.  With a finite readout time of the FET planes, this rate would introduce less than 1\% of dead-time depending to a lesser extent on the size of the fiducialized volume used in the veto.

The fine segmentation of the G-FETs provide localization of backgrounds and the FET-to-FET coincidence further suppresses the background count rate.  The intrinsic $^{14}$C background from the graphene target will profit from a newly identified source of CO$_2$ that is estimated to be three orders of magnitude lower in $^{14}$C/$^{12}$C than achieved in Borexino.  This source was recently identified in the collection of low background underground Argon.  The AMS methods for verifying low-level
$^{14}$C/$^{12}$C are described in~\cite{litherland2005low}, and the
AMS facilities described in this paper are now located at the Lalonde AMS Lab at the University of Ottawa.  The fabrication process to implement high radio-pure CO$_2$ to grow the graphene target is described in the high radio-pure $^{12}$C WP.

\subsubsection*{Cost Estimate for PTOLEMY-G$^3$}

PTOLEMY-G$^3$ is ready for a first phase experiment.  Graphene sensor results are in progress and the existing PTOLEMY prototype has the volume and cooling capacity to host PTOLEMY-G$^3$ with a fiducialized volume of 10$^3$~cm$^3$.
Current, research-level production capacity of graphene exceeds 200~m$^2$ per year with prices of approximately \$1
per cm$^2$.  Substrate and graphene wafer costs are less than \$200 per wafer (100~mm dia.), but current research-level processing costs are an order of magnitude higher.  Wafer-level processing costs are \$2,000 per wafer (100~mm dia.) to implement the G-FET array structure and applying expected yields.  A 10$^3$~cm$^3$ fiducialized volume corresponds to a 100 wafer production run.  Each readout card equipment with low noise buffers and a Kintex-7 readout is estimated at \$800 with one readout card serving $10^6$ FETs per wafer.  This gives \$280,000 as the core construction cost for PTOLEMY-G$^3$.  Significant understanding of the wafer-level fabrication would be achieved with a pre-production of 10 wafers.  The dominant production cost, wafer-level processing, could be further reduced through collaborations with industry.  The collaboration with CIEMAT and Graphenea Inc. as part of PTOLEMY is detailed on section on the graphene work package.

\subsection*{PTOLEMY-CNT}

\begin{figure*}[h!]
\begin{center}
	\includegraphics[width=0.55\textwidth]{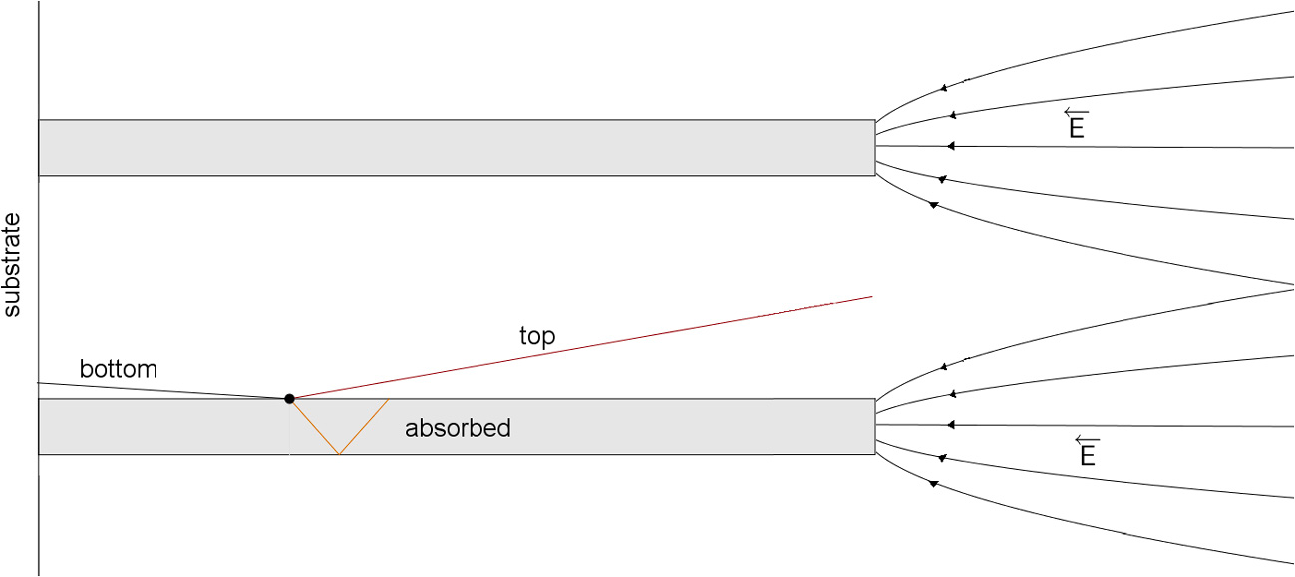} \ \ 
    \includegraphics[width=0.35\textwidth]{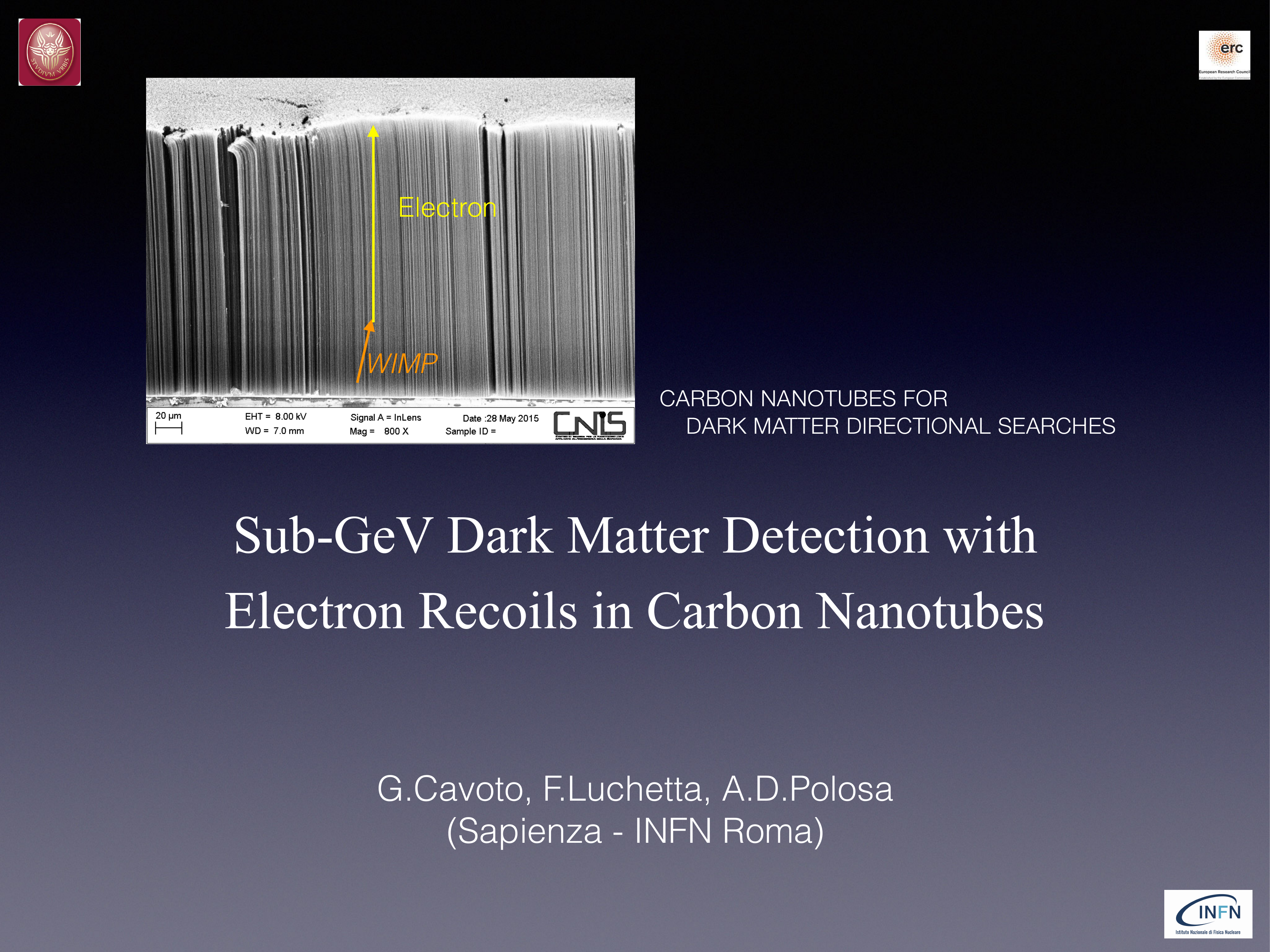}
\caption{(\emph{left}) Angular directionality from the CNT target depends on the orientation of the CNTs (shown here horizontally) relative to the dark matter wind. A voltage potential at the open-end of the CNTs accelerate the low energy recoil electrons into a single electron sensor.  (\emph{right}) SEM image of densely packed CNTs (oriented vertically).}
\label{fig:CNTdirectionality}
\end{center}
\end{figure*}

The physics reach for MeV dark matter and the principles of operation of the carbon nanotube(CNT) target are described in~\cite{cavoto}. 
The path of propagation of scattered low energy electrons through the CNT target is depicted schematically in Fig.~\ref{fig:CNTdirectionality}.  The directionality derives from the forward-backward anisotropy of the dark matter cross section stemming from the dark matter wind direction relative to the open-end of the CNT target orientation.
A low energy electron exiting the CNT target through the open-end will be accelerated into a single electron detector through an applied voltage potential.
Measurements of the graphene-electron interactions, including transmission, absorption and scattering, indicate that the electrons produce in the CNT target will have a high transparency to reach the end of the CNT target for direct detection of the electrons.
The characterization of these properties is the focus of the graphene work package.

\begin{figure*}[h!]
\begin{center}
	\includegraphics[width=0.45\textwidth,height=4cm]{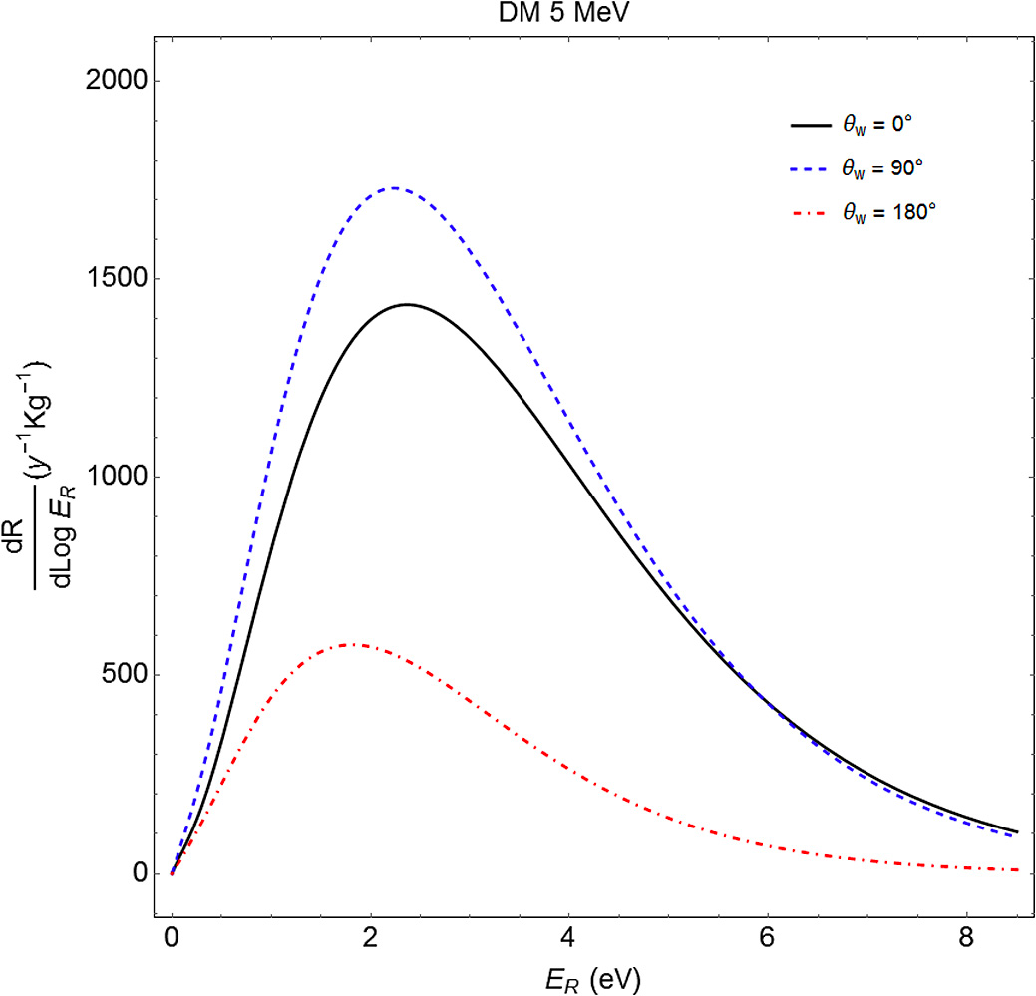}
	\includegraphics[width=0.5\textwidth,height=4cm]{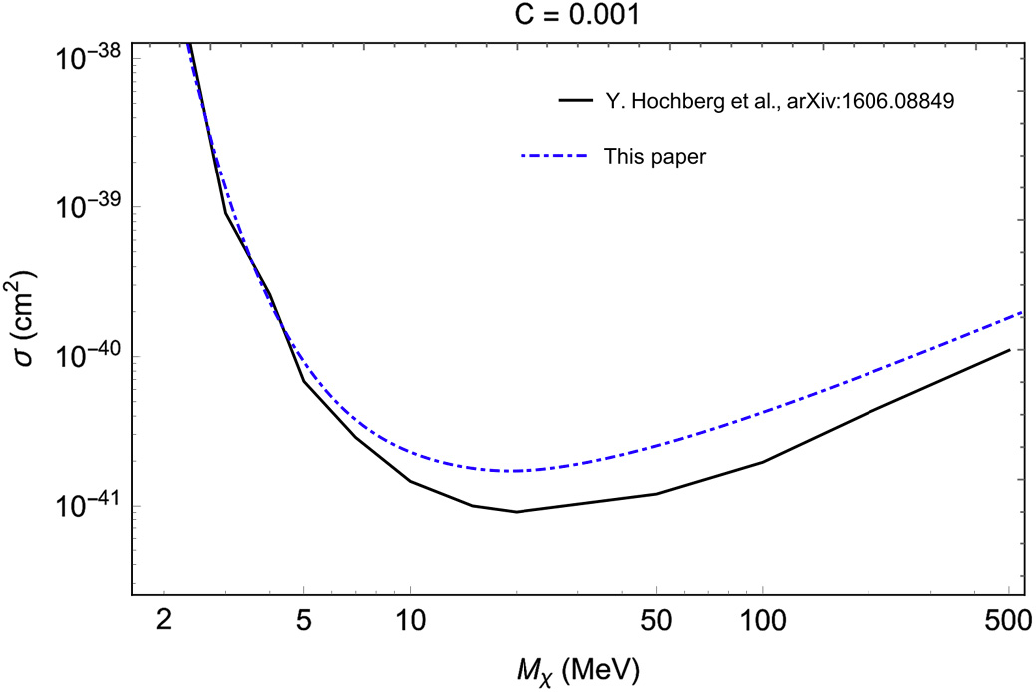}
\caption{(\emph{left}) Differential rate for a 5~MeV DM particle scattering off an electron in the CNT target is shown.  (\emph{right})  Expected background-free 95\% C.L. sensitivity for the CNT target with a 1-kg-year exposure.
}
\label{fig:CNTsearch}
\end{center}
\end{figure*}

Fig.~\ref{fig:CNTsearch} shows the recoil energy spectrum of electrons from 5~MeV dark matter scattering in the CNT.
The cross section sensitivity to MeV dark matter for a CNT mass exposure of $M \cdot t$(kg$\cdot$day)$\simeq 16$ reaches a sensitivity of $\bar{\sigma}_e = 10^{-37}$~cm$^2$ for dark matter masses of 5~MeV owing the relatively large target mass that can be achieved in the CNT configuration.  The physics reach for a 1-kg-year exposure is shown in Fig.~\ref{fig:CNTsearch}.

\clearpage

\section{High Radio-pure $^{12}$C}
\begin{figure*}[h!]
\begin{center}
      \includegraphics[width=0.75\textwidth]{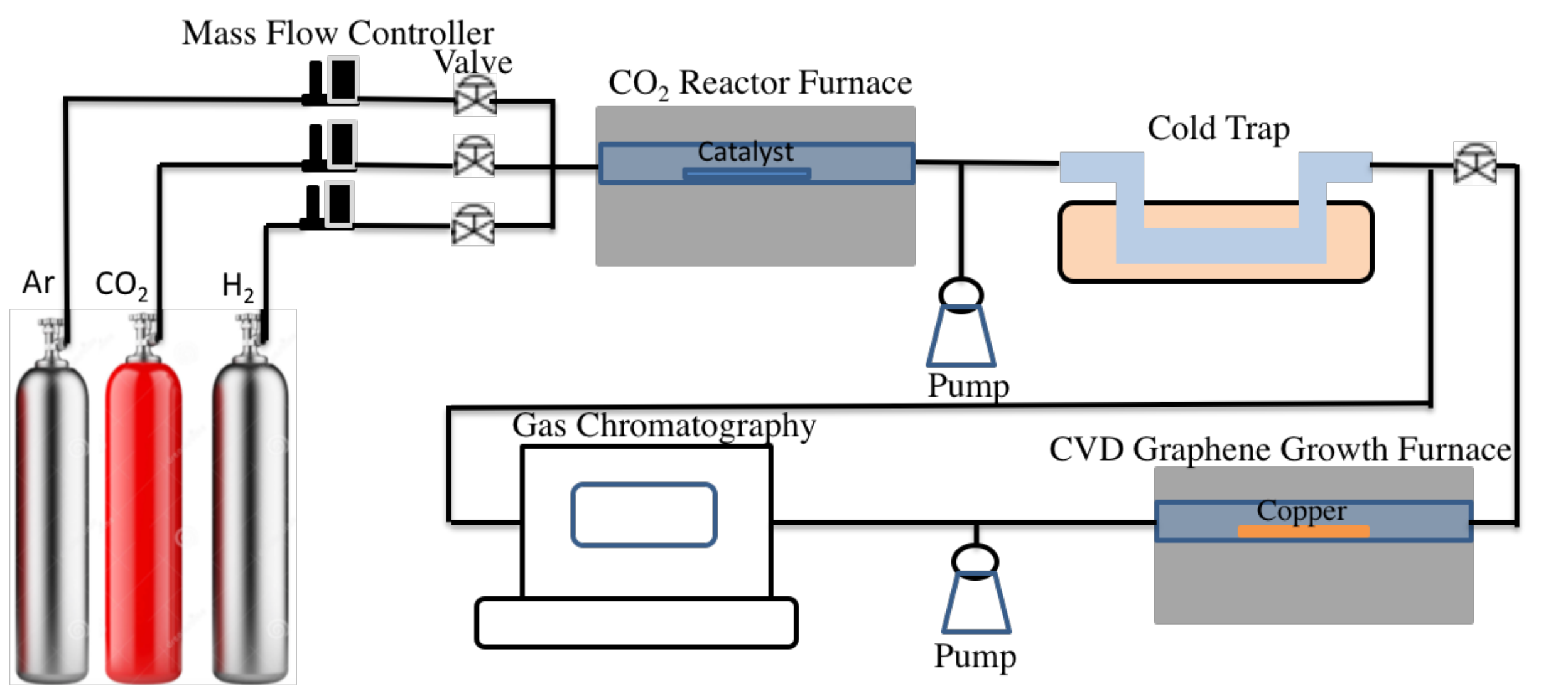}
   \caption{
High radio-purity Carbon-12 graphene growth system. High radio-purity Carbon-12 CO$_2$ with Ar and H$_2$ (as the carrier gases) is injected to a reactor furnace to covert to CH$_4$, H$_2$ and H$_2$O by catalysts. The flux rates are controlled by mass flow controllers. The cold trap is used for water removal. The gaseous products ratio is analyzed by a GC system. Single-layer graphene is grown on the copper foil in a furnace with the controlled gas ratio and flow rates. 
 }
\label{fig:radiopureC}
\end{center}
\end{figure*}

%\noindent \large {\bf Coordinator: F.~Zhao} \normalsize
The aim is to use a high radio-purity Carbon-12 source of CO$_2$ gas extracted from the Earth’s mantle to convert to CH$_4$ which is the carbon source of chemical vapor deposition(CVD) graphene growth. The electrochemical reduction of CO$_2$ to produce CH$_4$ is a promising route to produce clean fuel under ambient pressure and temperature by catalysts. The gaseous products of the reduction process contain CH$_4$, H$_2$ and H$_2$O and are detected by a gas chromatography(GC). CH$_4$ and H$_2$ are the effective carbon and hydrogen source for high quality single-layer graphene growth on a copper foil. The experiment setup is shown in Fig~\ref{fig:radiopureC}. After the conversion reaction, the gaseous products are run through a cold trap to remove the H$_2$O molecules.

The mechanism of the electrochemical reduction of CO$_2$ to CH$_4$ on catalysts has been discussed~\cite{radiopurec1,radiopurec2}. In this process, an adduct of CO$_2$-CO$_2^{\cdot -}$ is formed over catalysts. Then the free-energy pathway becomes thermodynamically downhill to transfer the second electron to form adsorbed CO$_{ads}$. In the third step, the CO$_{ads}$ can be desorbed or converted into CHO$_{ads}$ after accepting an electron and proton.  In the final step, the CHO$_{ads}$ can be transformed to CH$_4$ after accepting additional electrons and protons. The CHO$_{ads}$ is the key intermediate towards the breaking of the C-O band, leading to formation of CH$_4$.

Initial samples of CVD graphene will be produced with the setup described above and evaluated in the PTOLEMY prototype at LNGS.  Further collaboration into the fabrication of high radio-purity Carbon-12 veto systems will be discussed with LNGS groups.
Interest in these systems has been expressed by Frank Calaprice at Princeton University, Anna Pla-Dalmau at Fermilab and members of the DarkSide collaboration for potential use in plastic and liquid scintillators.

\clearpage

\section{Graphene studies for Tritium and Dark Matter}

\subsection*{Graphene}
Graphene is a unique material which offers a combination of properties that make it a key enabler for electronic applications, generating new products that cannot or are difficult to be obtained with current technologies or materials \cite{FER15}. Applications of graphene and carbon nanotubes (CNTs) have been identified and will be developed in this project: detection of relic neutrinos and measurements of dark matter.
In order to develop these experiments, five different tasks have been identified, namely: preparation of graphene, characterization of the material, target design, experiments of electron-graphene interaction, preparation of graphene field-effect transistors (GFET), and study of channeling in carbon nanotubes.

\subsection*{Preparation of graphene}
\begin{figure*}[h!]
\begin{center}
  \includegraphics[scale=0.3]{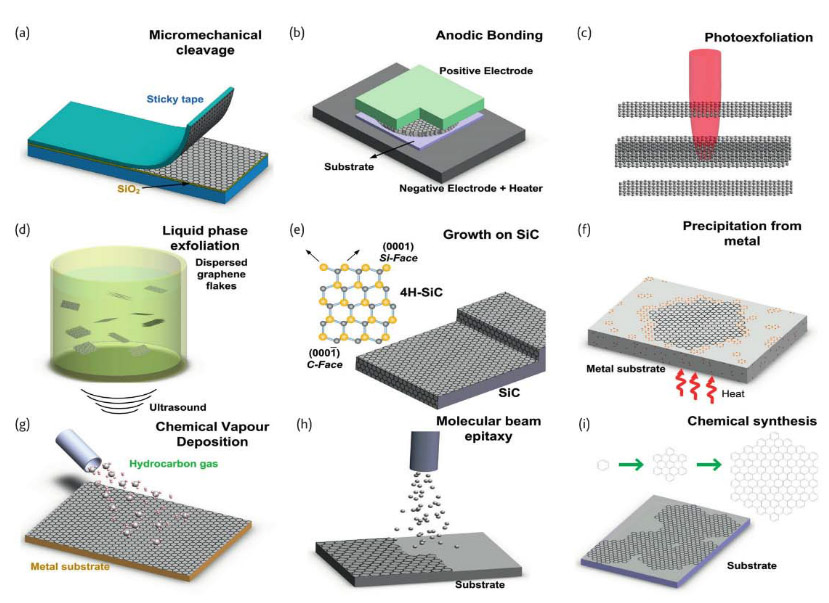}
   \vspace*{5pt}
   \caption{Illustration of the main production graphene techniques (From \cite{BON12}).}
\label{fig:graphenetechnique}
\end{center}
%\vspace*{-15pt}
\end{figure*}

There are several growth techniques of 2D graphene (see Fig.~\ref{fig:graphenetechnique}). Among them, micromechanical exfoliation from ultrapure graphite provides the highest quality material at the expense of quantity and uniformity. The tradeoff between quality and quantity, both required by this proposal, is chemical vapor deposition (CVD). A catalytic substrate, typically a copper foil of thin film, is used for CVD of graphene, which has to be removed after growth prior to device processing. An automatic growth transfer system will be used to deposit the electronic-grade graphene onto the final substrate, including semiconductor, dielectric or free-standing platforms.

\subsection*{CVD growth and optimization}
%{\it (from the beginning of the project, six months for first runs of optimized material, and continued)}

High electronic-quality graphene is required to achieve maximum device performance and reduce contamination 
and degradation arising from defects. In this regard, several parameters of the CVD growth have to be optimized. Figures of merit will be based in structural properties (defect density, domain size) as well as electrical performance (doping and mobility).

\subsection*{Transfer to arbitrary substrates}
%{\it(from the beginning of the project, and continued)}

\begin{figure*}[h!]
\begin{center}
  \includegraphics[scale=0.3]{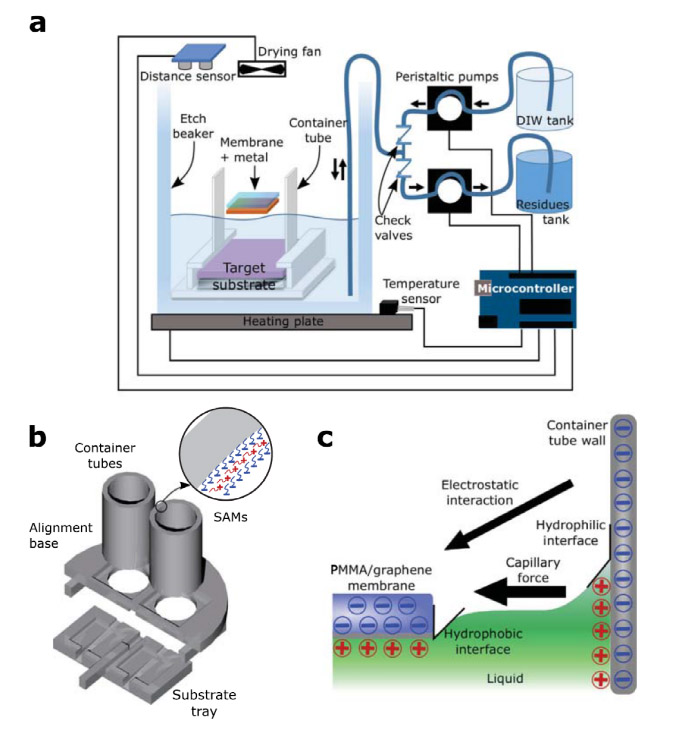}
   \vspace*{5pt}
   \caption{(a) Schematic of the automatic transfer system components and their connections. 
   A microcontroller communicates with all the sensors and actuators. (b) Holding structure 
   for the automatic transfer system. (c) Schematic of the centering mechanism of the membrane 
   between the graphene film and the self-assembled-monolayer treated inner wall of the tube. (From \cite{BOS16}).}
\label{fig:automatictransfer}
\end{center}
%\vspace*{-15pt}
\end{figure*}

In most applications,  the transfer of the CVD graphene sheet from the 
growth substrate to the target substrate is a critical step for the final device 
performance. Manual procedures are time consuming and depend on handling skills, whereas 
existing automatic roll-to-roll methods work well for flexible substrates but induce 
mechanical damage in rigid ones. A system that automatically transfers CVD graphene 
to an arbitrary target substrate was recently developed by members in the consortium \cite{BOS16} (Fig.~\ref{fig:automatictransfer}). 
The procedure, based on the all-fluidic manipulation of the graphene to avoid mechanical 
damage, strain and contamination, will be used from the beginning of the project. In addition, 
several improvements of the processing will be performed to enhance in all the quality, 
uniformity and yield of the automatically transferred graphene for the fabrication of GFET \cite{BOS15}.

\subsection*{Characterization of graphene and carbon nanotubes material}
%{\it(from the beginning of the project, and continued)}

A combination of different techniques to assess the structural, optical and electrical characteristics 
of graphene will be performed.  

\subsubsection*{Structural and electronic characterization}
This structural and electronic characterization includes:

\begin{itemize}
\item Atomic Force Microscopy (AFM);
\item Field-emission Scanning Electron Microscopy (FE-SEM), to determine the surface topography;
\item X-ray photoelectron spectroscopy (XPS), to measure the elemental composition (purity,  oxygen), $^{14}$C content, chemical-state of C and influence of defects and doping elements on the electronic state \cite{DIB17,MON17,MON16}; eventually, spatially-resolved XPS with synchrotron-radiation beamlines \cite{DIB17,DIB18}.
\end{itemize}

\subsubsection*{Optical: Raman and Nomarski}

Raman spectroscopy is the most routine technique to find out the structural quality, comparing the existence of peaks in the spectra and comparing their amplitudes and widths. 
Using Nomarski microscopy we will determine the domain size and relate that to the electronic properties.

\subsubsection*{Electrical (mobility, doping type and level)}\label{electricalmobility}
We will use a method for extracting relevant electrical parameters from GFET devices using a simple electrical characterization and a model fitting developed by the members of the consortium (\cite{BOS15}). With experimental data from the device output characteristics, the method allows to calculate parameters such as the mobility, the contact resistance, and the fixed charge. The application of this control method to GFET arrays also allows to correlate the quality with several issues during key fabrication steps such as the graphene growth and transfer, lithography and metalization processes, which will help to optimize the material fabrication.

\subsection*{Target design}

Initial graphene samples provided by suppliers will be loaded  with deuterium at CIEMAT, 
as the chemical behavior of deuterium is equal to tritium and samples without tritium are easier for handling.

In order to evaluate any effect associated to electron self irradiation consequence of beta emission absorption/desorption, experiments will be carried out during irradiation, in the beam line of the Van de Graaff electron accelerator at CIEMAT at different sample temperatures and deuterium pressures. It is expected that both absorption and desorption of tritium in graphene will be deeply modified by irradiation \cite{BAO18}.

Graphene samples will be modified by ion implantation and electron irradiation in order to increase the graphene 
capability to absorb tritium by the production of traps.  Then the samples will be deuterium loaded at different 
temperatures and deuterium pressures and the quantity absorbed will be evaluated by thermal desorption (TSD) 
techniques and SIMMS.

When deuterium loaded graphene samples will be characterized, tritium loaded graphene samples will be produced and tested.

\subsection*{Experiments on electron-graphene interactions}

The main objective of this section is to prove the electrical and transport properties of graphene by using collimated electron beam sources and related spectroscopic techniques.

%{\it(from the beginning of the project, and continued)}

\begin{figure*}[h!]
\begin{center}
  \includegraphics[scale=0.5]{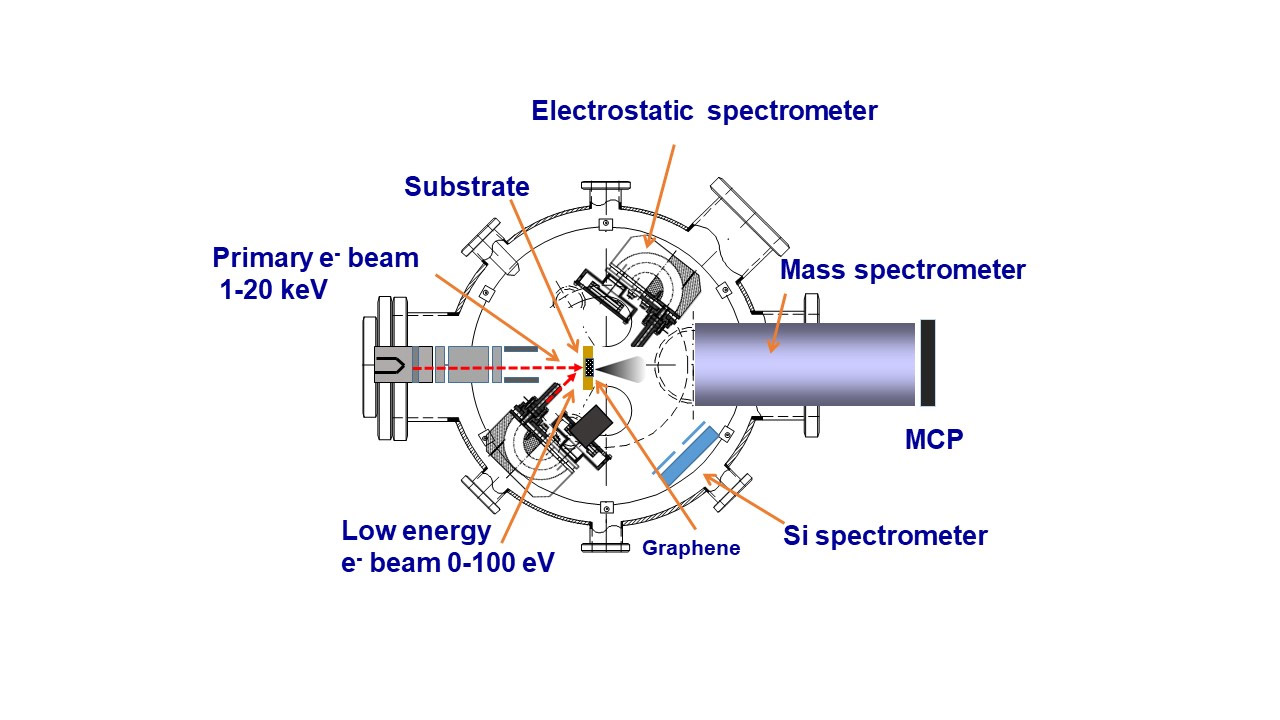}
   \vspace*{5pt}
   \caption{Schematic of the experiment on electron interactions with graphene structures  (see text for details).}
\label{fig:schematicoftheebeam}
\end{center}
%\vspace*{-15pt}
\end{figure*}

Several experimental techniques will be addressed, including:\\
Low (0-100~eV) and intermediate (1-20~keV) energy electron interactions with graphene. 
The schematics of the experiment is shown in Fig.~\ref{fig:schematicoftheebeam}.

The primary electron beam (typically 100~nA current) is generated by an emitting filament then collimated (1~mm diameter), 
accelerated (up to 1-20~keV), focused and deflected towards the sample holder where graphene, in different configurations (foil, nanotubes), is placed either in vacuum (UHV conditions) or deposited on different substrates (both electric and dielectric materials). After the interaction, the energy and angular distribution of generated low energy (0-50~eV) secondary electrons is analyzed with an electrostatic energy analyzer which can rotate around the sample. Primary electron transmitted and scattered in selected directions are detected and analyzed in energy with a pure Silicon spectrometer cooled with liquid nitrogen. A quadrupole mass spectrometer analyses the neutral fragmentation induced to the graphene sample by electron impact. Alternatively, charged fragments are also analyzed with a time of flight (TOF) mass spectrometer. In addition, to understand better the low energy electron interactions with graphene a 0-100~eV electron source, formed by an electron gun attached to an electrostatic hemispherical monochromator, generates 10~nA electron current beams with an energy resolution of about 50~meV. This low energy electron source will be used to study electron transport properties and induced dissociation by electron attachment to the graphene structures. 
As a complement of the experimental data, electron scattering cross sections with different geometrical configurations of graphene will be calculated for impact energies ranging from 0.1 to 20000~eV. The calculation method is based on our independent atom with screening correction additivity rule (IAM-SCAR) procedure which has been successfully applied in previous studies to different atomic and molecular structures.
Finally, experimental and theoretical interaction probabilities will be integrated into a Monte Carlo modeling procedure to simulate electron transport properties of different graphene structures and evaluate their response to electron irradiation.

\subsection*{Preparation of G-FETs}
%{\it(from month seven onwards)}

Modeling will be performed to calculate the band-gap and the required design for the nano-ribbon 
width and number. Graphene nano-ribbons (GNR) down-to-100-nm wide will be prepared by e-beam lithography 
and characterized structurally and optically. GFET arrays will be processed based on them, and electrical 
assessment of these devices will be performed following the procedures described above.
%in \ref{electricalmobility}.
We cannot process narrower GNR.

\subsection*{Channeling in carbon nanotubes}
%{\it (from first months onwards)}

Study of ion channeling in carbon nanotubes (CNTs) will be carried out by low-energy ion bombardment of highly-aligned multi-wall CNT (MWCNT).  

This study constitutes a proof of principle for CNTs as potential targets in novel detection scheme for dark matter particles \cite{CAP15,CAV16,cavoto}, where the intrinsic anisotropy of aligned CNTs could be exploited. 

Experiments:
\begin{itemize}
\item (\it{month} 1-6) Low-energy (0.2-5~keV) Ar+ ion-bombardment of highly-aligned MWCNT as a function of energy and flux; 
XPS of the C 1s core level and spatially resolved Raman analysis (investigation of penetration depth);
\item (\it{month 7-onwards}) Low-energy Ar+ ion-bombardment of highly-aligned MWCNT as a function of ion impinging angle with respect to the CNT axis;
\item spatially resolved Raman in the different angular configurations, study of ion penetration depth, defect generation;
\item XPS of the C 1s in the different angular configurations, with quantification of the different spectral fingerprints associated to the defects (sp$^2$, sp$^3$-like, $\pi$-excitation, dangling bond-related).
\end{itemize}

\clearpage

\section{High stability HV development and Calibration methods}

%\noindent \large {\bf Coordinator: M.~Messina} \normalsize
%\begin{itemize}
%\item[-] Voltage stability
%\item[-] Feed-through
%\item[-] high precision E-gun
%\item[-] sources
%\end{itemize}
A series of electrodes will be used to build the MAC-E filter, key element of the PTOLEMY concept, and to generate the finally retarding potential
to adjust the energy of the electrons before the active measuring device.
Thus, the voltage stability of those electrodes will directly affect
the absolute energy scale and the final energy resolution ($\sigma_E \sim 0.05$~eV).

For the reasons mentioned above a special High Voltage (HV) system
with unprecedented stability features and also with a 
peculiar voltage monitoring device needs to be developed. The latter is dictated by the fact that any 
resistive device would affect significantly the voltage we aim at measuring. 

It is worth pointing out that the PTOLEMY detector design is based on electrodes, generating a constant E-field, which do not provide any current but only voltages with high stability. For this reason, the HV generator can be a very large charge reservoir, i.e. a capacitor, that locks the voltage on the electrode with which is connected in parallel. 
In this case any possible noise generated by switching devices, always present in standard power supplies, is prevented.
Not to neglect are dense cosmic ray showers that can generate signal pulses on the measuring device.

Furthermore, long-time decay rate of voltages can be due to ions generated by cosmic rays in the surrounding atmosphere, Rn decay, internal leak of the capacitor and surface current on spacers in between points at different voltages. While the effect generated by ions produced by cosmic rays and Rn can be avoided by installing the setup in vacuum the others needs to be monitored with high resolution and eventually trigger the system to reset the voltage to the desired value. 

\begin{figure}[h!]
\begin{center}
      \includegraphics[scale=0.2]{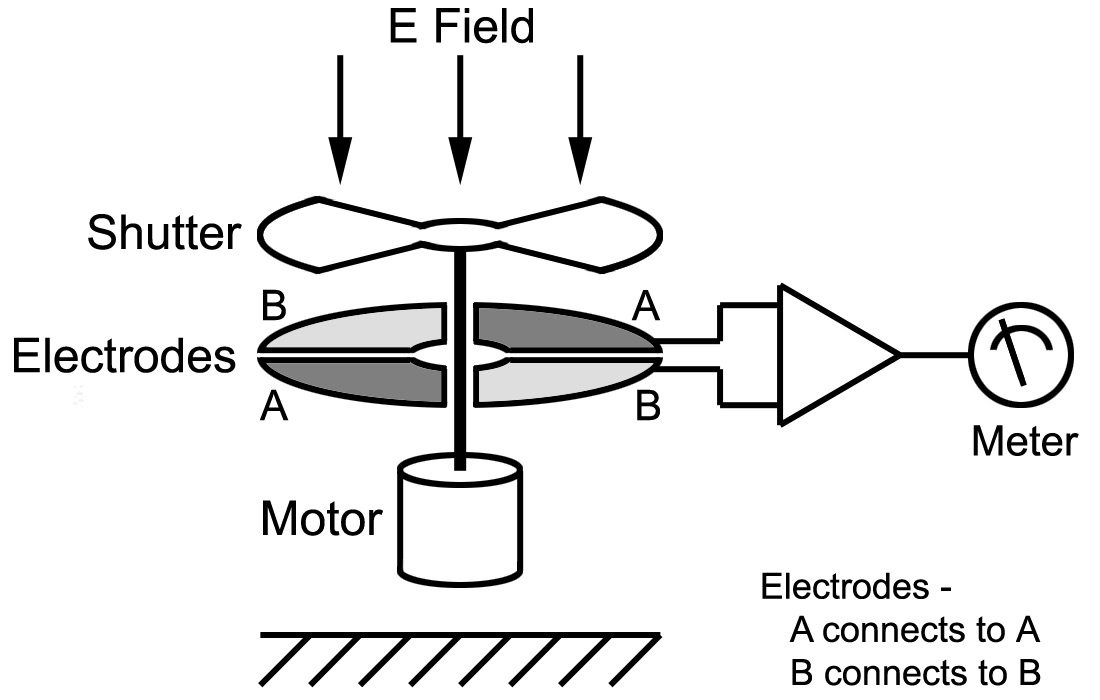}
\vspace*{10pt}
   \caption{Schematic of the working principle of the Field Mill.}
\label{fig:field-mill}
\end{center}
\end{figure}

A device commonly named Field-Mill (FM) has been adopted as voltage monitor. In the case that will be presented soon the features of the FM are unprecedented and this makes it especially interesting.
The measurement principle on which the FM is based is depicted in Fig~\ref{fig:field-mill}.
The motor in Fig~\ref{fig:field-mill} while spinning the shutter chops the E field and this induces a signal on the electrodes, also indicated as pads. The signal is measured in a differential mode between couples of opposite pads. 
The latter are read out by means of a charge amplifier whose output is calibrated as function of the voltage
that generates the E-field. 

\begin{figure}
\begin{center}
      \includegraphics[scale=0.3]{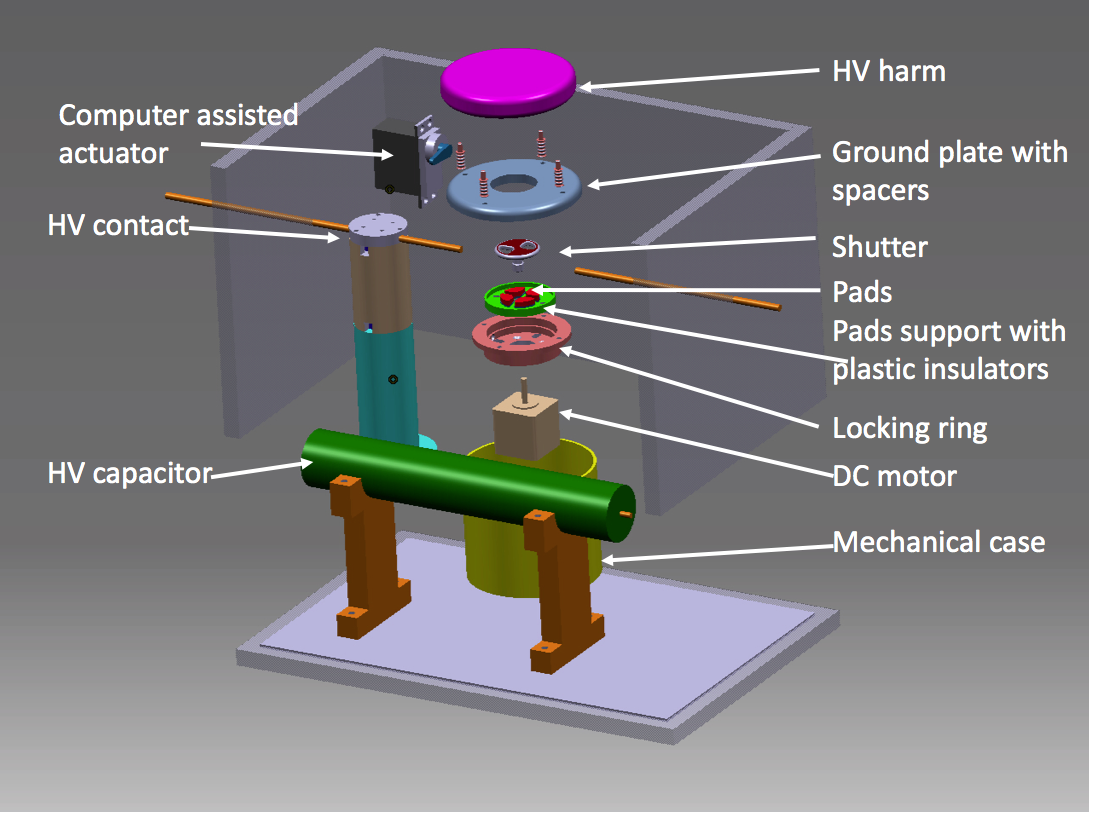}\includegraphics[scale=0.3]{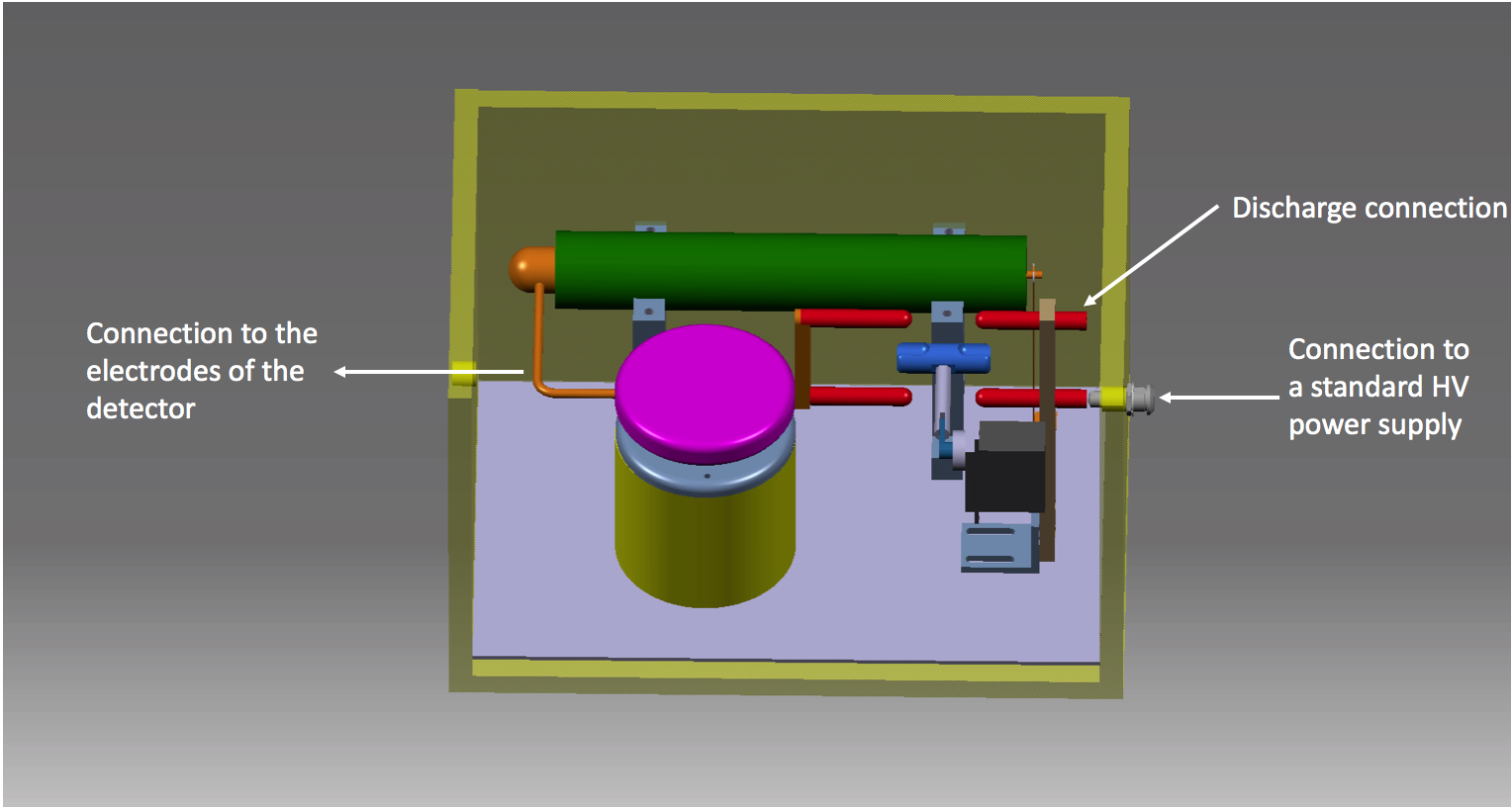}
\vspace*{10pt}
   \caption{(left) Exploded view of the HV system with the explanation of each components. (right) View of the assembly of the HV system.}
\label{fig:FieldMill_ExplodedImage}
\end{center}
\end{figure}

Fig~\ref{fig:FieldMill_ExplodedImage}, by means of an exploded view, shows all components of the HV system and of the FM.
The legend in Fig~\ref{fig:FieldMill_ExplodedImage} (top) explain all the components used for the HV and FM system and helps to understand how the voltage we aim at monitoring, on the top arm of the FM, is correlated to an E field and to the signal that is measured.
In Fig~\ref{fig:FieldMill_ExplodedImage} (bottom) an assembly of the system is visible and also the details 
of the HV connector that allow to connect a standard HV power supply to charge up the electrodes.

\begin{figure}[h!]
\begin{center}
     \includegraphics[scale=0.3]{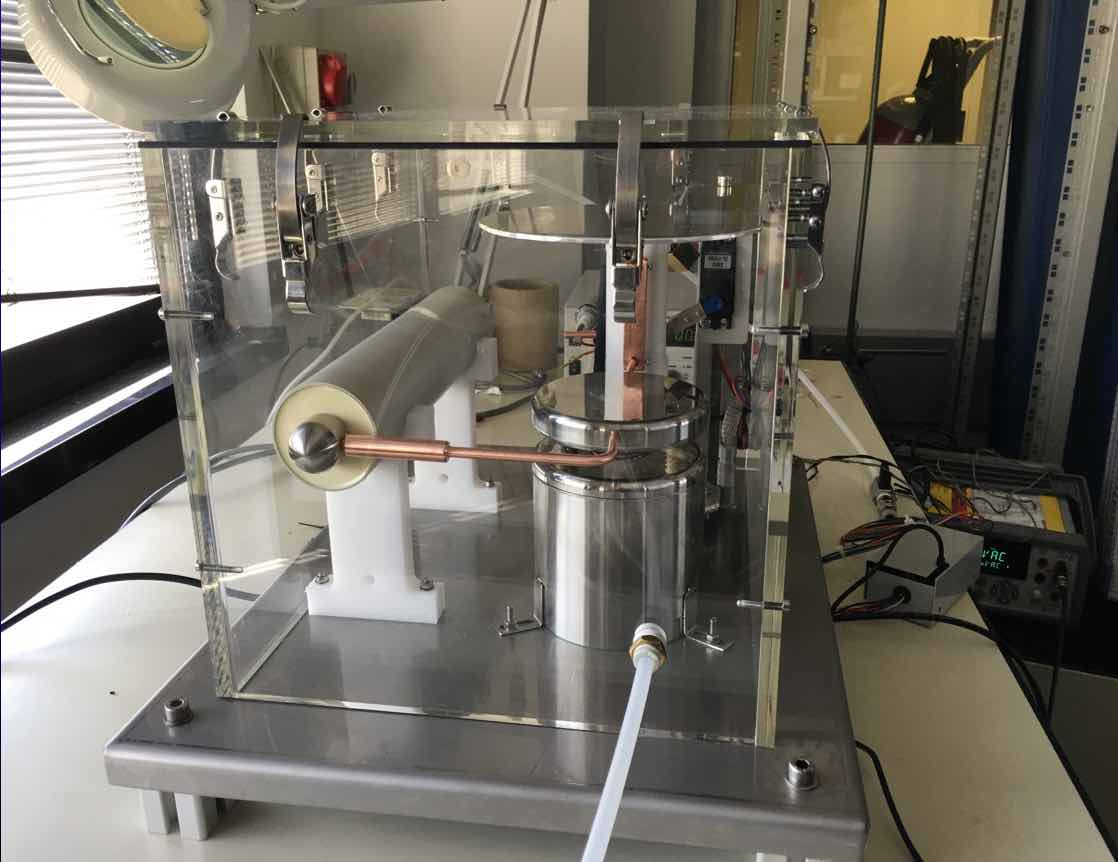}
\vspace*{10pt}
   \caption{Photograph of the prototype of the HV system together with the FM.}
\label{fig:HV_photograph}
\end{center}
\end{figure}

\begin{figure}[h!]
\begin{center}
      \includegraphics[scale=0.25]{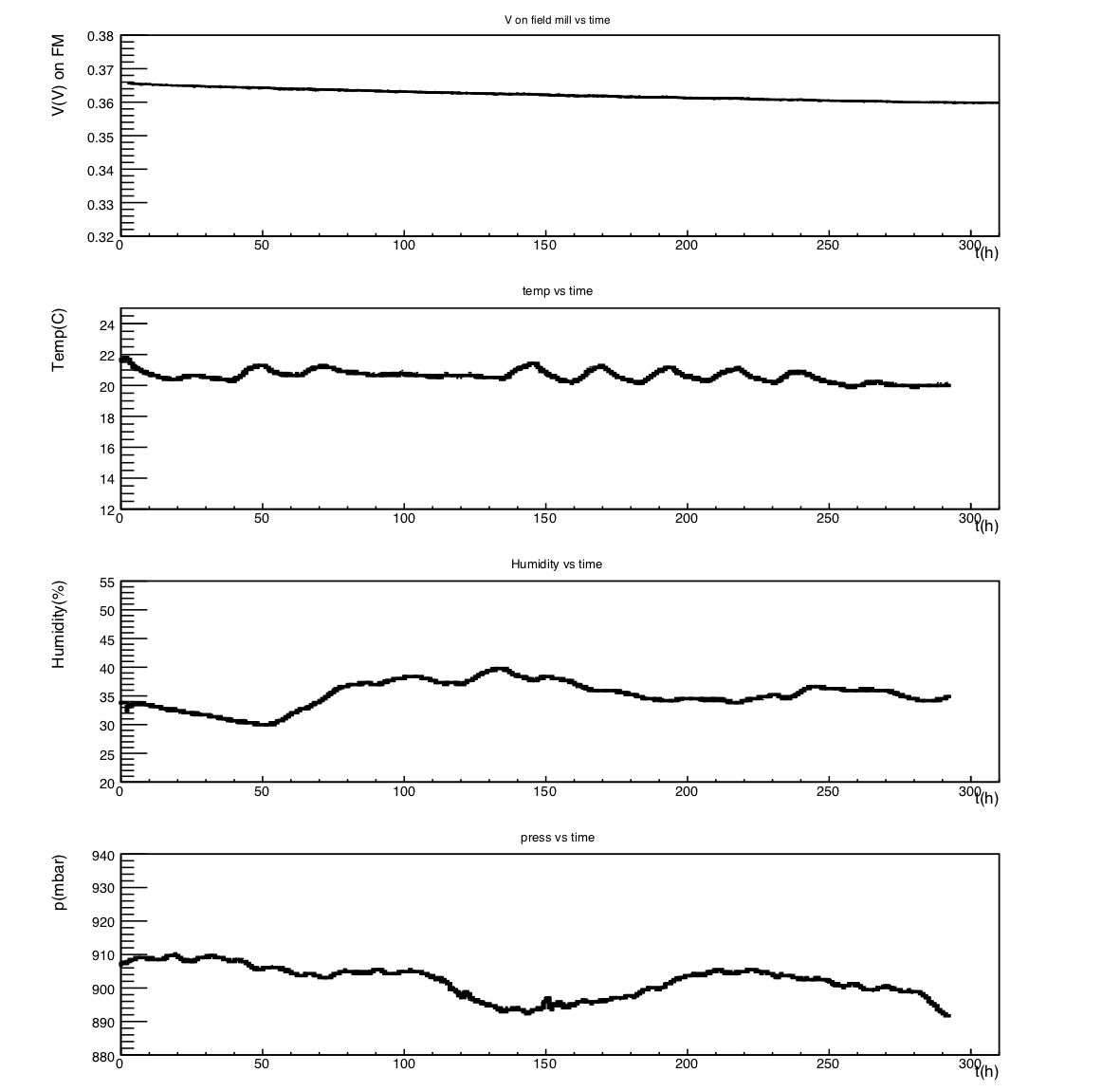}
\vspace*{10pt}
   \caption{Time series of Field-Mill measurements and environmental parameters at an above-ground location.}
\label{fig:monitored_parameters_above}
\end{center}
\end{figure}

In Fig~\ref{fig:HV_photograph} a photograph shows a prototype of the HV system with the FM built and working at the LNGS.
Some of the most relevant components highlighted in Fig~\ref{fig:FieldMill_ExplodedImage} can be recognized in the photograph reported in Fig~\ref{fig:HV_photograph}.
The plots in Fig~\ref{fig:monitored_parameters_above} show the signal measured by the FM and also the environment parameters such as temperature, humidity and 
atmospheric pressure as function of time. The environment parameters are monitored because they are expected to influence the measurement.

\begin{figure}[h!]
\begin{center}
\includegraphics[scale=0.255]{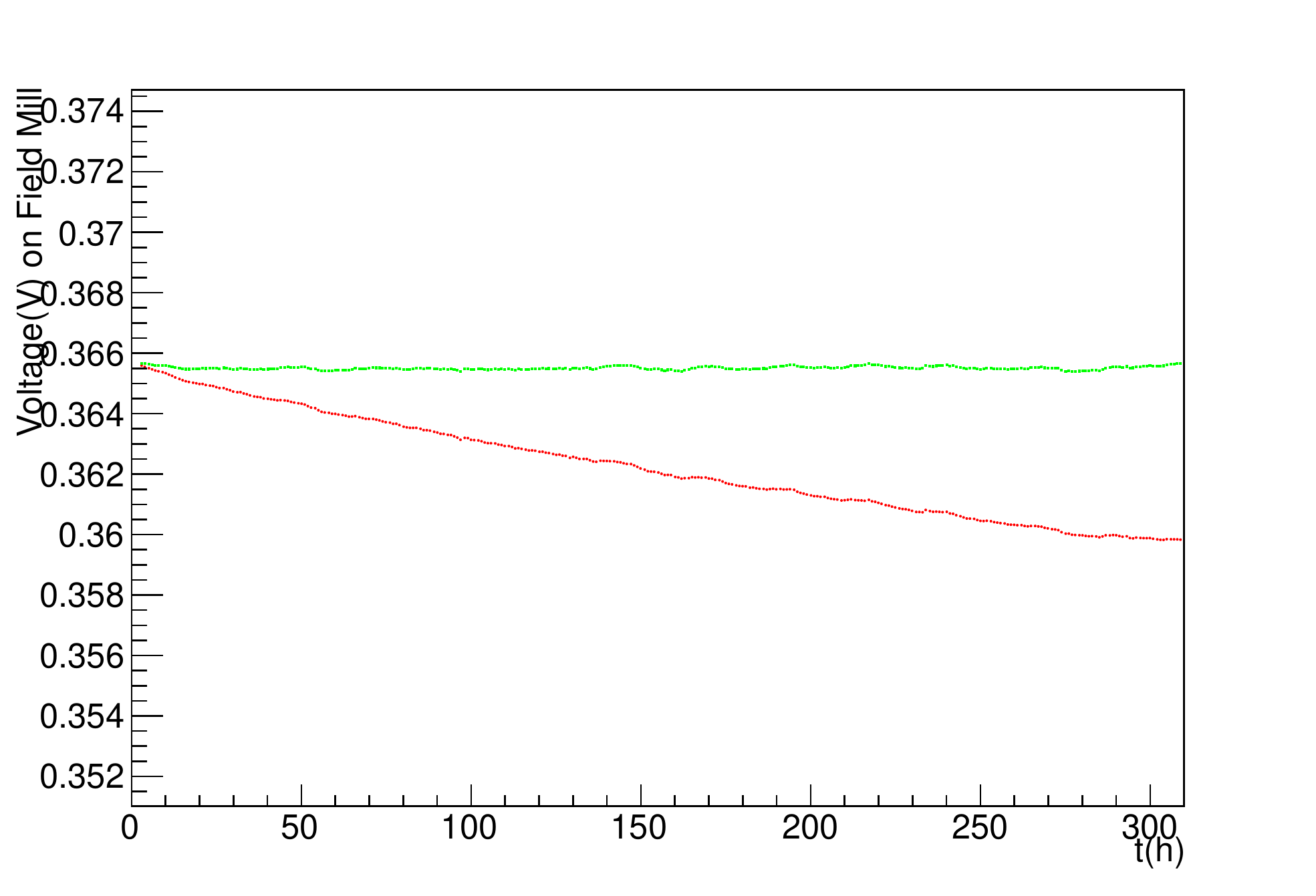}\includegraphics[scale=0.25]{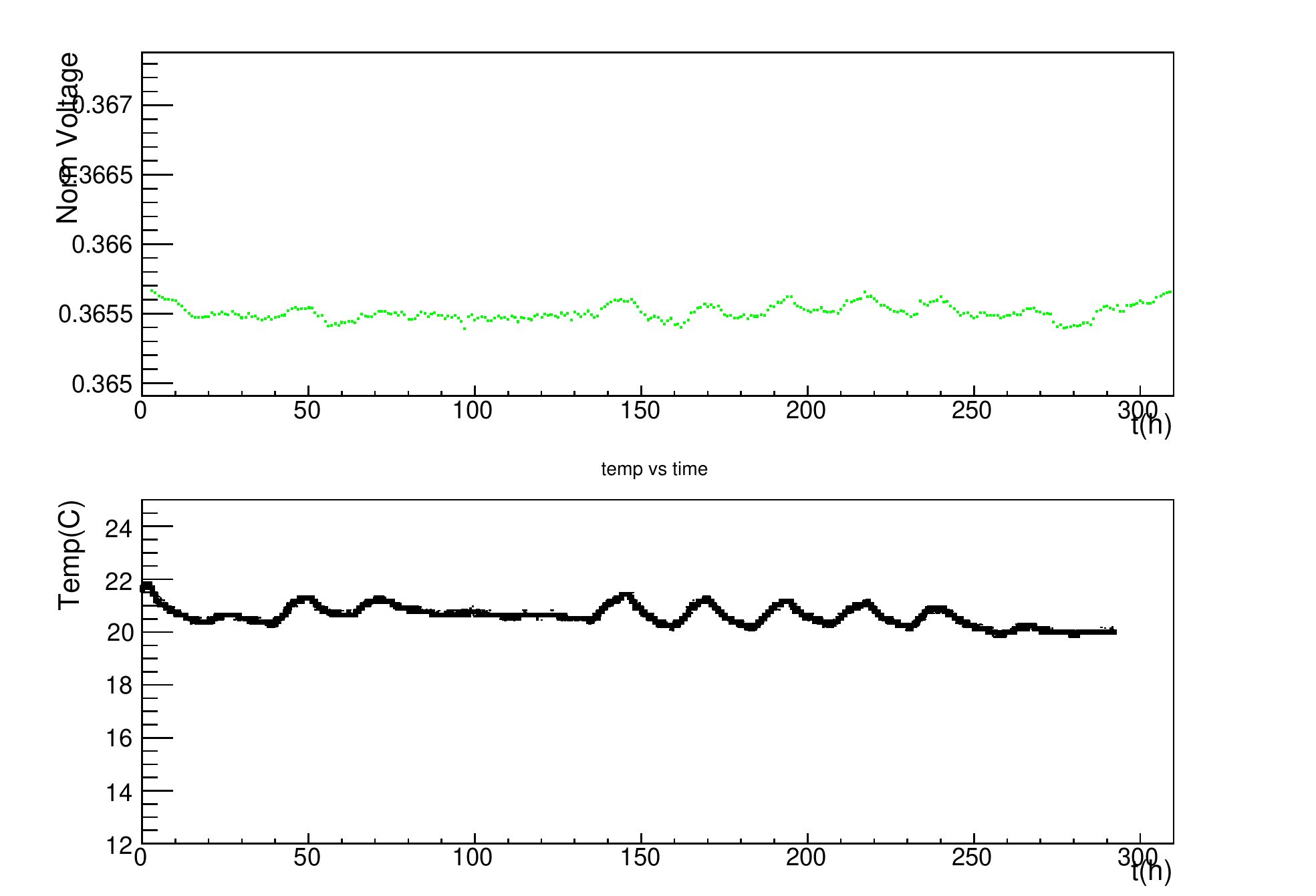}\\
\includegraphics[scale=0.45]{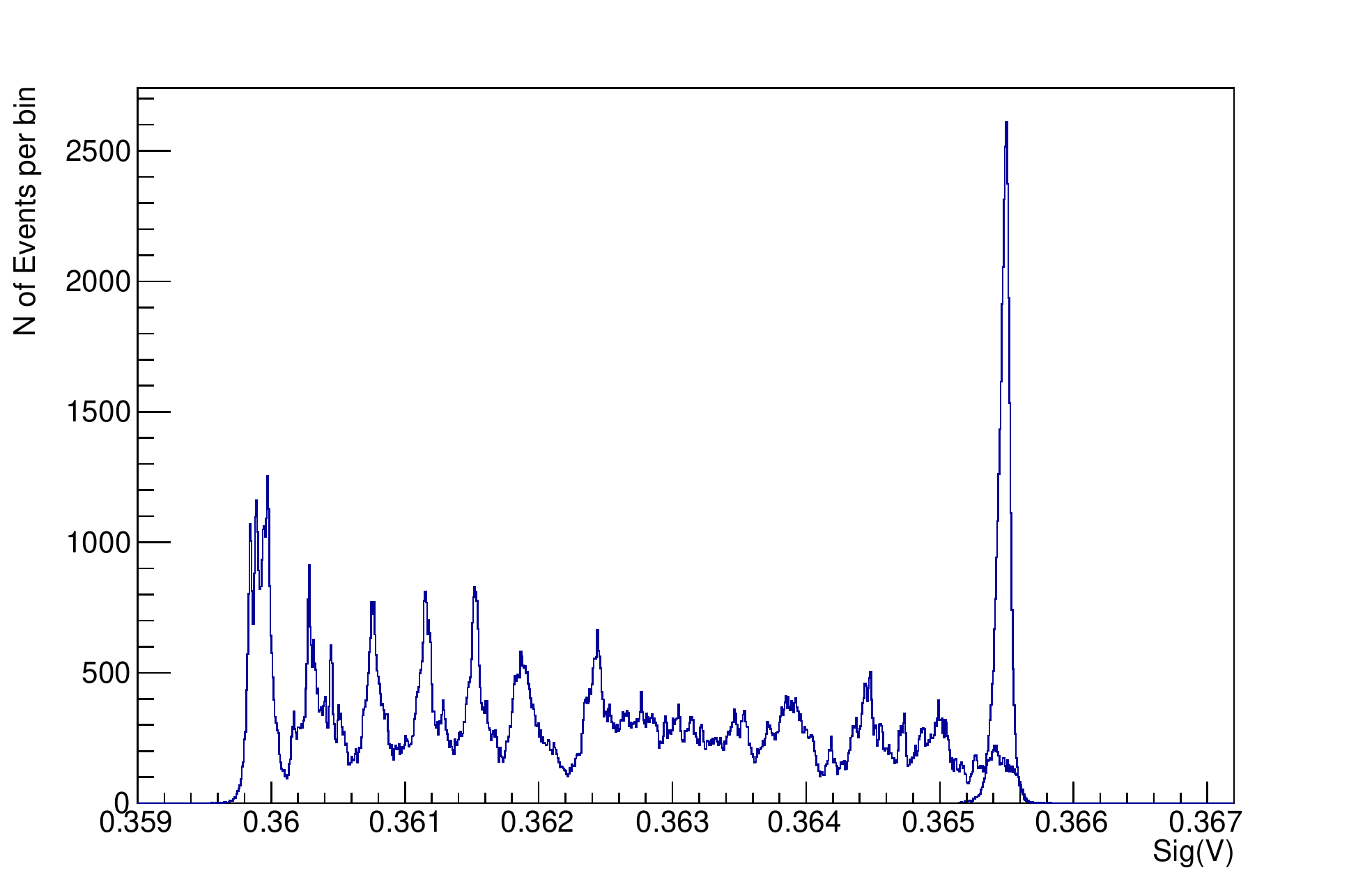}
\vspace*{10pt}
   \caption{Above-ground Measurements: (top-left) Long decay-time of voltage from 5000V and its flattened curve with a second order polynomial fit. (top-right) Temperature zoom together with the flattened curve in the same time window. (bottom) Field-Mill measurements in full time range; the narrow peak is obtained in a time interval where the temperature happened to be stable within 0.1 $^o$C.}
\label{fig:longdecay_tempcorrelated_above}
\end{center}
\end{figure}

A closer look at the measurements is shown in Fig~\ref{fig:longdecay_tempcorrelated_above}. In the top-left plot of Fig~\ref{fig:longdecay_tempcorrelated_above} the curve showing the long decay-time of the capacitor, charged up at 5000 V, is visible.
The curve is obtained with the capacitor disconnected from the standard power supply. The decay rate of the FM signal is well described by a second order polynomial function and the parameters of the first and second power term are $2.532\time 10^{-5}$~mV/h and $2.100^{-8}$~mV/h/h, respectively, both with an uncertainty of a part per mill. The decay rate observed on the FM output translate to a decay of the voltage on the capacitor, and the electrode, of $\sim 1.5$~V/h which is still one order of magnitude far from what we want to achieve.
The flattened curve of the Fig~\ref{fig:longdecay_tempcorrelated_above} (top-left) is obtained by correcting for the decay-rate. In the top-right plot
a zoom of the curve showing the temperature of the environment as function of time is shown together with zoom of the 
flattened curve just mentioned. One can clearly see the effect of the temperature on the output of the FM and the day-night effect. From this it is clear that a stabilization of the temperature is necessary. We aim at getting the temperature stable at better than $0.1^o$~C.

In the bottom plot of Fig~\ref{fig:longdecay_tempcorrelated_above} the distribution of the signal measured by the FM is shown after the correction 
for the long decay-time along the full acquisition time window and in a short period in which temperature happened to be stable at better then $0.1$~T (hours in the range $90-110$). In this case the signal distribution show a narrow peak with an RMS of $\sim 4\times10^{-5}$.

\begin{figure}[h!]
\begin{center}
      \includegraphics[scale=0.2]{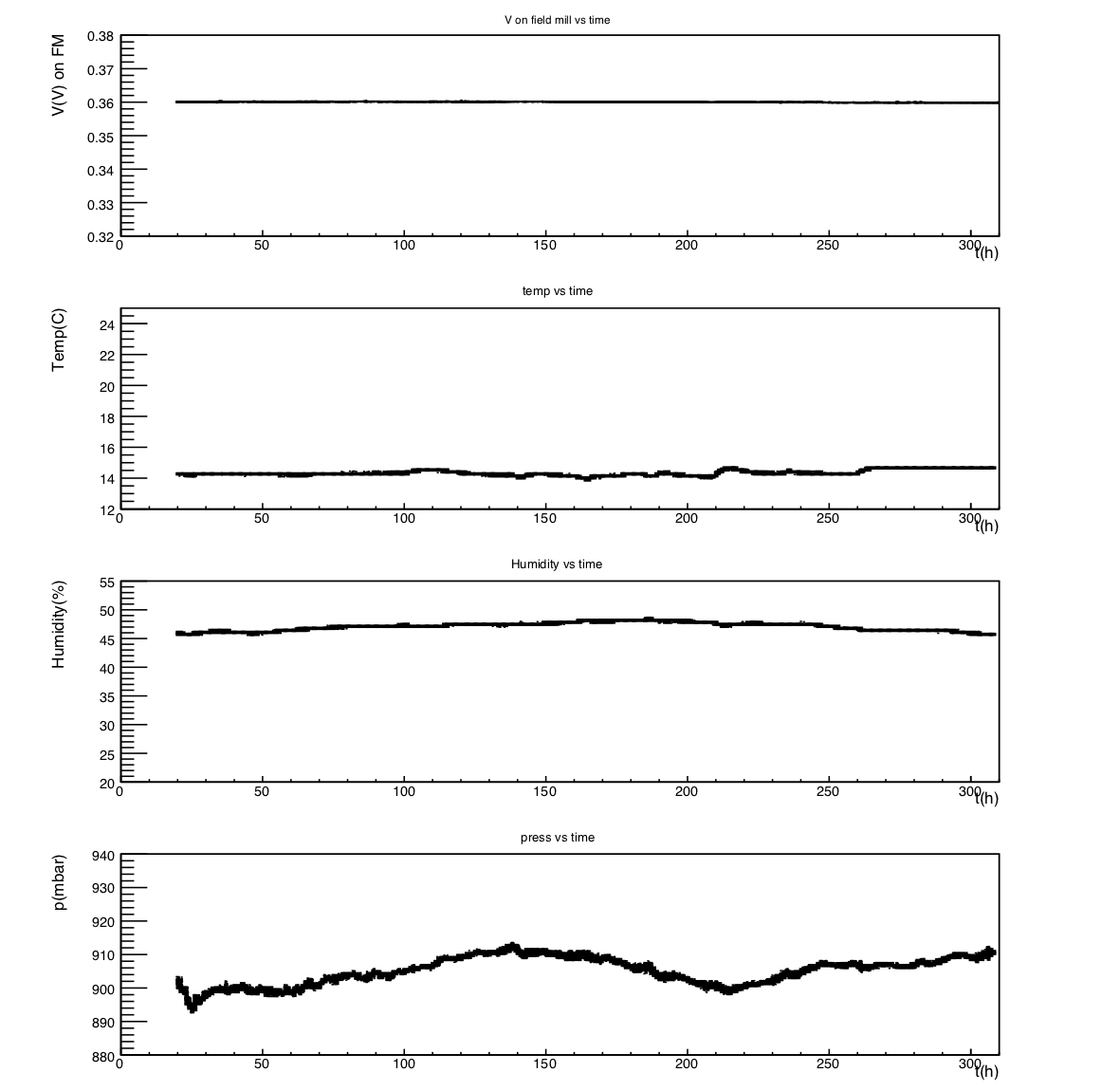}
\vspace*{10pt}
\caption{Time series of Field-Mill measurements and environmental parameters setup underground in Hall C.}
\label{fig:monitored_parameters_under}
\end{center}
\end{figure}
Thus, we moved the setup underground in Hall C and the monitored parameters as function of time in the new location are reported in the Fig~\ref{fig:monitored_parameters_under}. Also in this case the capacitor is charged up to 5000 V. The temperature 
in hall C is stable at better than $0.1^o$~C and produce soon a visible effect on the stability of the 
FM measurement.

\begin{figure}[h!]
\begin{center}
\includegraphics[scale=0.259]{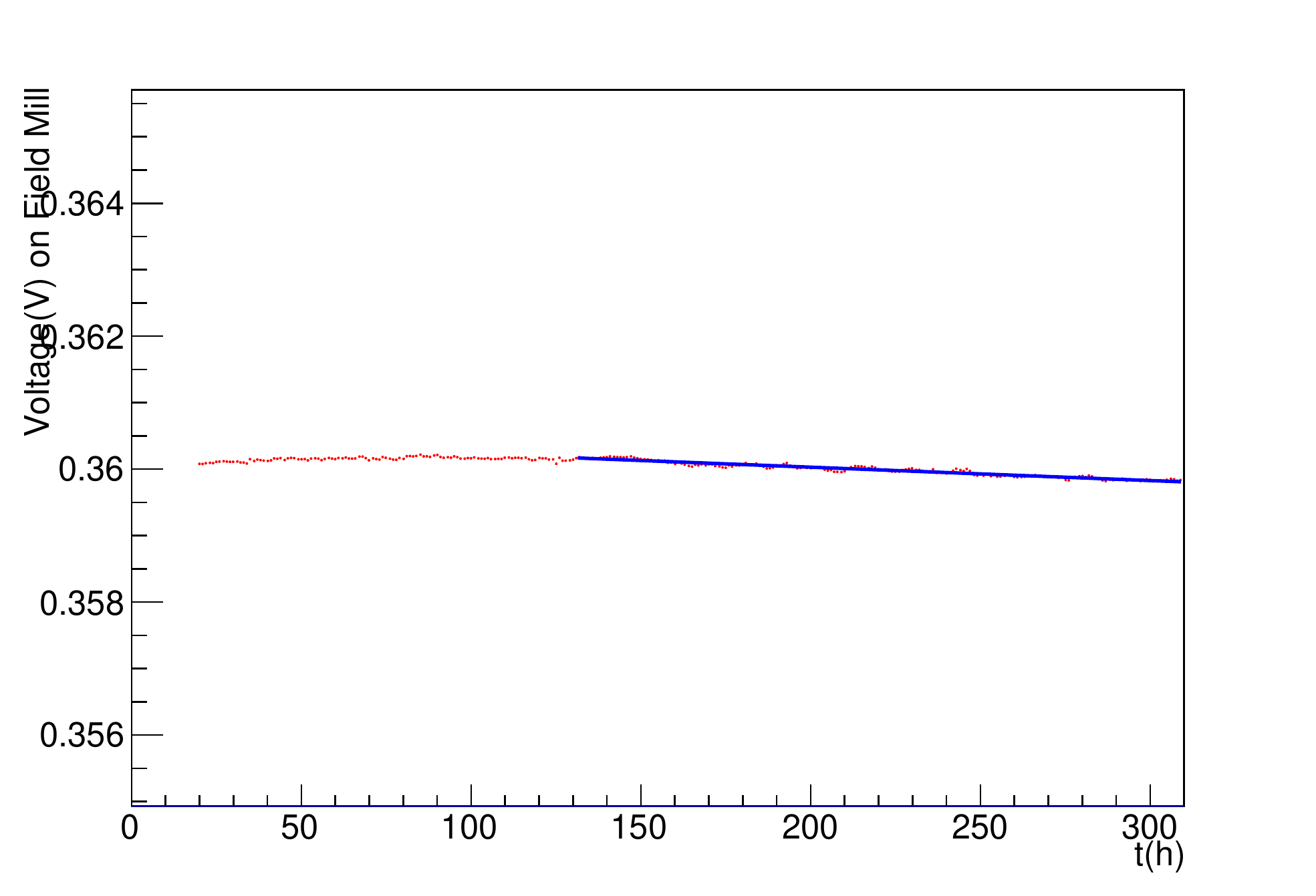}\includegraphics[scale=0.25]{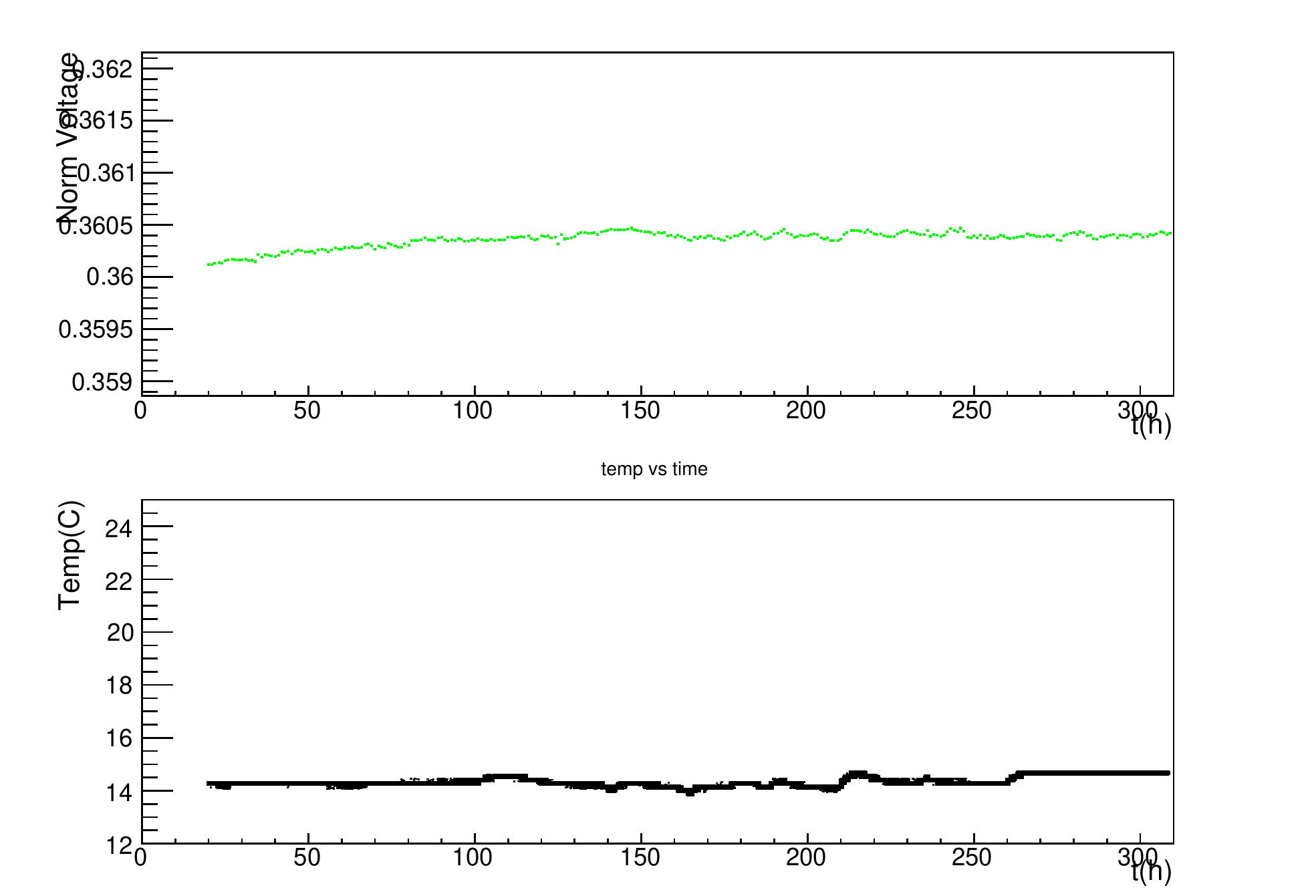}\\
\includegraphics[scale=0.45]{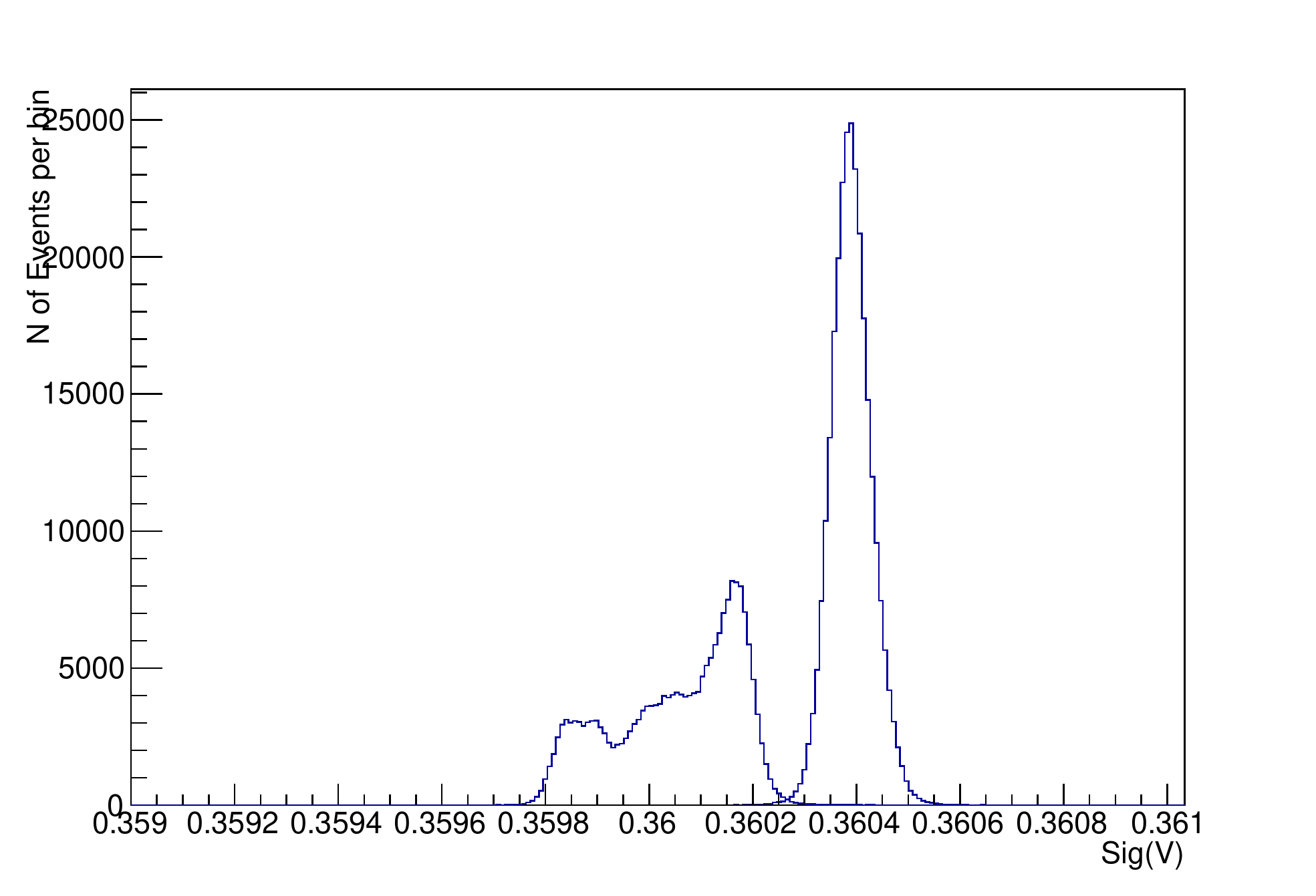}
\vspace*{10pt}
   \caption{Underground (Hall C) Measurements: (top-left) Long decay-time of voltage from 5000V and its flattened curve with a first order polynomial fit. (top-right) Temperature zoom together with the flattened curve in the same time window. (bottom) Field-Mill measurements in full time range; the narrow peak is obtained at the time above hour 140
   after the first order polynomial correction.}
\label{fig:longdecay_tempcorrelated_under}
\end{center}
\end{figure}

Fig~\ref{fig:longdecay_tempcorrelated_under} shows a data set comparable to those recorded above ground and on the top-left of Fig~\ref{fig:longdecay_tempcorrelated_under} the signal of the FM is plotted, which shows a much longer decay rate. The same plot shows the 
flattened curve after the correction for the decay-rate. In this case after a stabilization time lasting until hour 140 the decay is well described
by a first order polynomial function with a decay rate of $2.\times 10^{-6}$~mV/h with an uncertainty at per mill. 
As desired the decay rate decreased with respect to what we measured above ground by an order of magnitude. 
This improvement can be due to possible direct effects, either of cosmic rays, absent in underground site, or to the ions generated by the cosmic rays in the atmosphere around the measuring device. While the effect of the decay rate of Rn ionization even though reduced is expected to still affect the voltage stability. 
Figuring out among different hypotheses requires to install everything in a  vacuum chamber on which we are currently working. 

In the top-right plot of Fig~\ref{fig:longdecay_tempcorrelated_under} a zoom is visible on the flattened curve of the FM output as function of time and the effect of temperature stability is significantly reduced, almost not visible.
This should largely reduce the width of the peak that we can see in the bottom plot of the Fig~\ref{fig:longdecay_tempcorrelated_under}. The peak is obtained by making the distribution of the signal of the FM after the hour 140 when the FM response get stable and it can be corrected by means of a linear decay curve.

In fact, the RMS of this peak is roughly the same as the one measured above ground ($\sim 4\times10^{-5}$). We think that the reason why the width of the 
peak was not reduced below $10^{-5}$ is that the grounding of the shutter is based on two
conflicting requirements, having a brush-less motor and a good grounding contact with the shaft. The two conditions make very difficult to achieve a good ground. A sliding contact with the shaft of the shutter is realized outside of the motor and provides a grounding that is still quite rudimentary and on which we are working to improve and make it stable and the contact conditions reproducible.

\subsection*{Design and construction of a high precision electron gun}

The unprecedented precision and stability of the energy scale of the PTOLEMY detector imposes to have a calibration system with similar or better energy resolution than the part per million we aim at achieving.
\begin{figure}[h!]
\begin{center}
\includegraphics[scale=0.4]{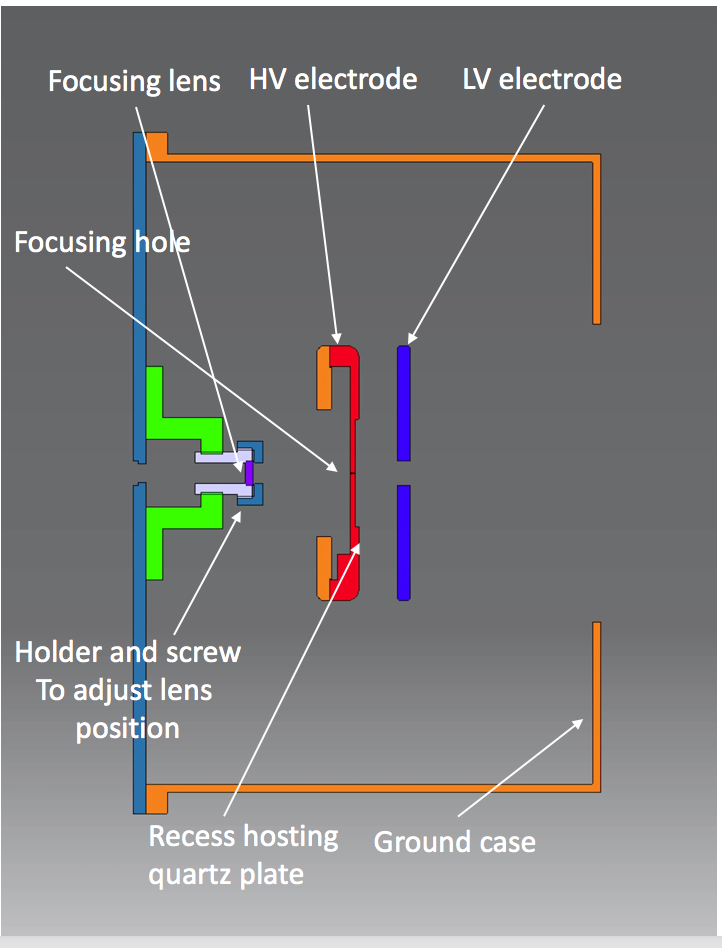}\includegraphics[scale=0.12]{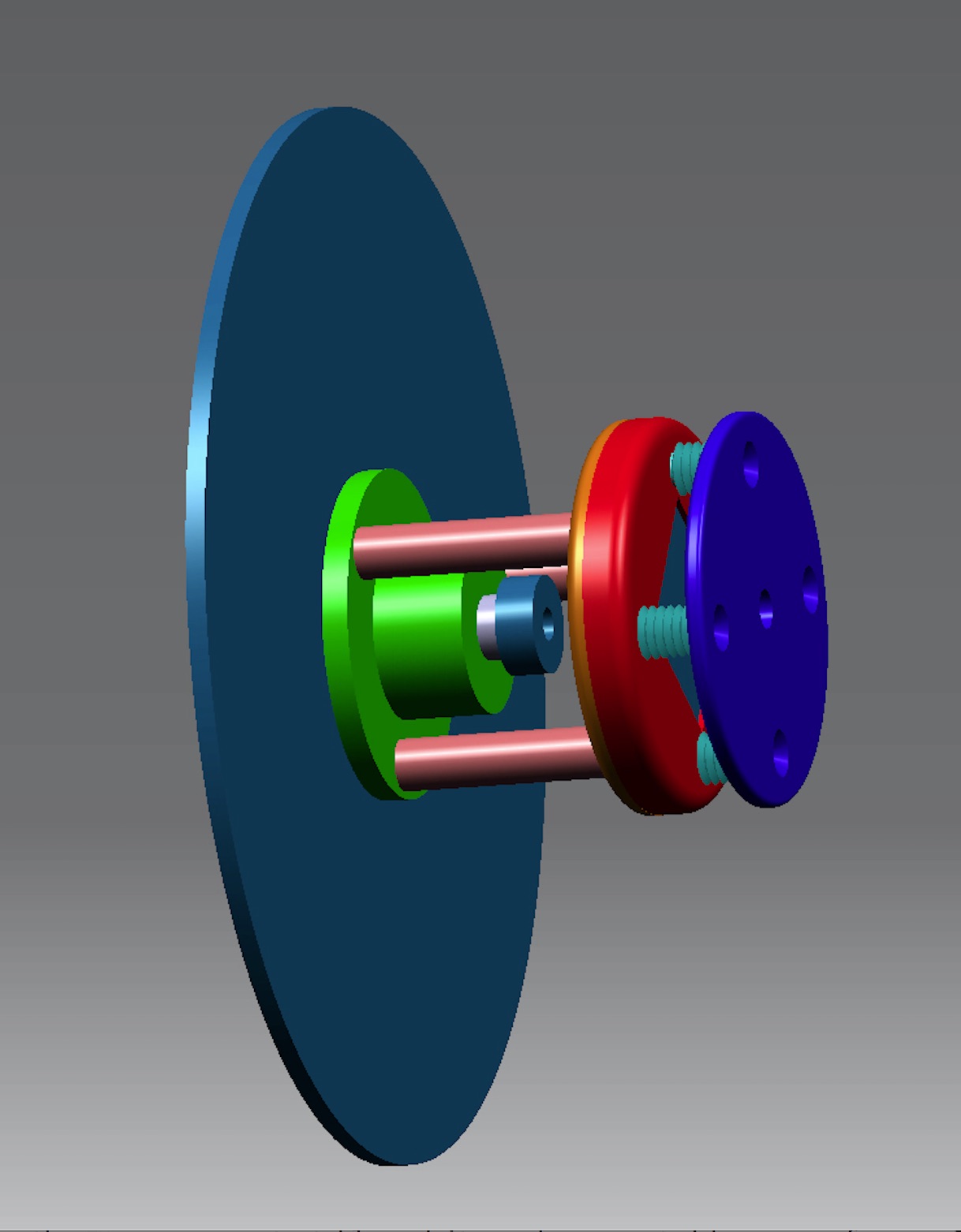}
\vspace*{10pt}
   \caption{(left) Cut view of the electron gun drawing. Most relevant components are indicated. (right) 3D view of the electron gun.}
\label{fig:E_gun_drawings}
\end{center}
\end{figure}

The design of our electron gun is based on an idea developed by the KATRIN collaboration and published in \cite{katrinegun}. In Fig.~\ref{fig:E_gun_drawings} the drawing is shown and the most important components are explained.

The principle is based on the idea that a UV light source, by means of a lens 
is focused at middle of the thickness of a quartz plate which is coated with a Au layer $5-10$~nm thick. If the UV light wavelength matches the work function of Au then an electron with very little kinetic energy can be extracted and then accelerated by means of a suited E field.  

At the level of precision of electron energy we aim at achieving ($E_{e} < 0.05$~eV) the work function depended on the actual features of the Au deposition, the impurities and the surface condition and must be measured 
on the setup. 

\begin{figure}[h!]
\begin{center}
\includegraphics[scale=0.08]{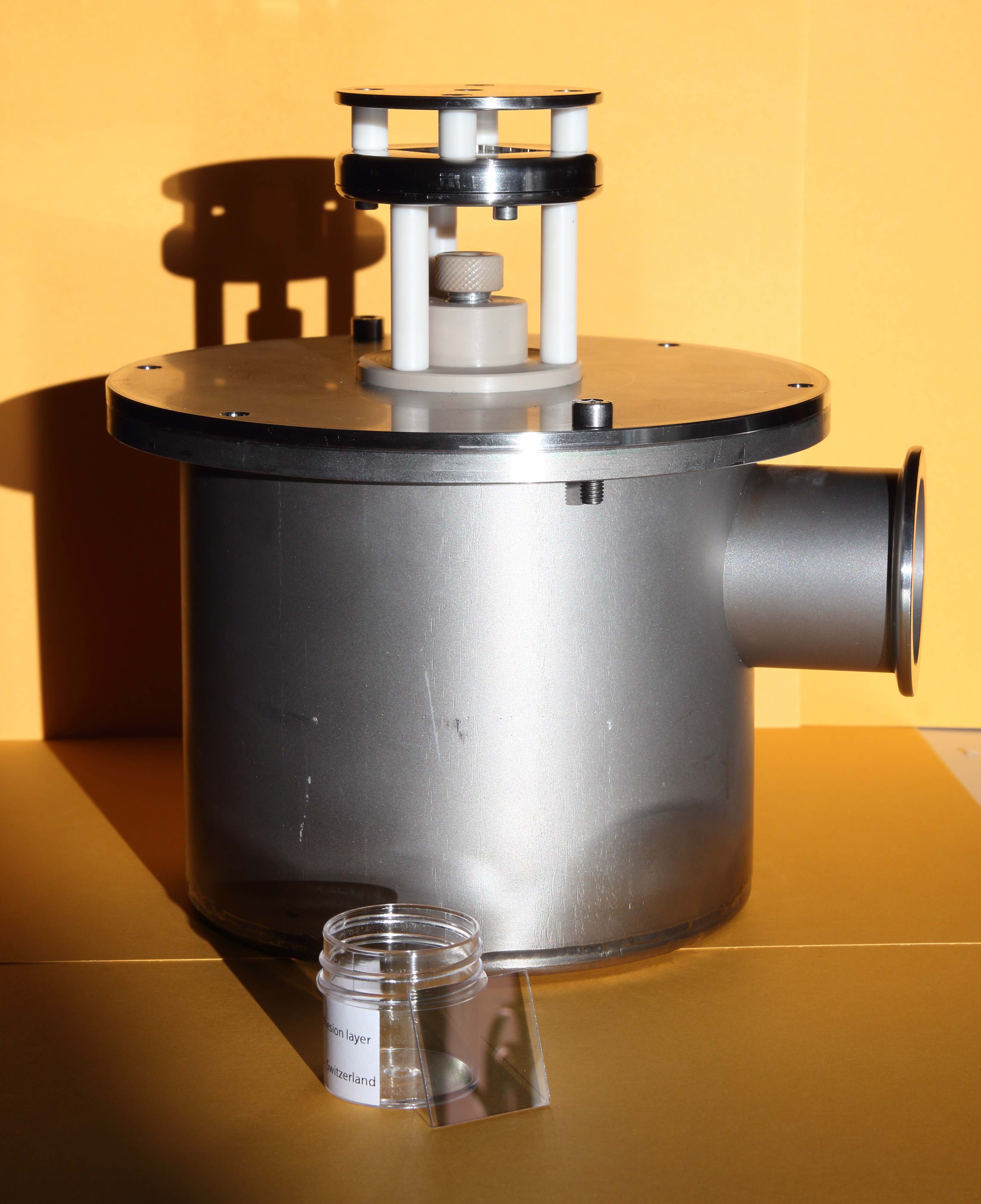}
   \caption{Photograph of the electron gun prototype built at LNGS. The main components already shown in the drawing are visible. At the bottom of the photograph the quartz plate with the Au coating is visible.}
\label{fig:E_gun_photograph}
\end{center}
\end{figure}

The photograph in the Fig~\ref{fig:E_gun_photograph} shows the prototype built at the LNGS. At the moment we still miss the UV source and the vacuum chamber in which the electron gun need to be installed. The latter is in assembly phase. The discussion is still ongoing on the detector necessary to study the electron beam features in the R\&D setup prior to install it in the 
PTOLEMY prototype.

\clearpage

\section{Cryogenics}

%\noindent \large {\bf Coordinator: K.~Schaeffner} \normalsize

%\begin{itemize}
%\item[-] Helium recuperation infrastructures
%\item[-] Run and handling of dilution Refrigerators
%\end{itemize}

The Cryogenics WP will take care of the organization and the implementation of the infrastructural and service requirements of the $^3$He/$^4$He dilution refrigerator and the superconducting magnets.

\begin{figure*}[h!]
\begin{center}
   \includegraphics[scale=1.0]{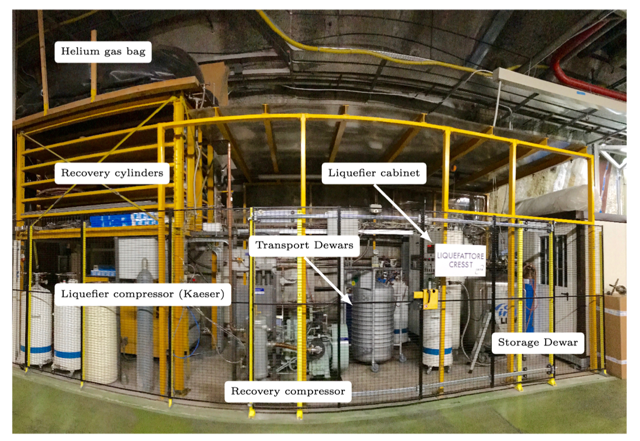}
%   \vspace*{-15pt}
   \caption{LNGS underground laboratory:
   helium gas handling facility on "Galleria TIR" (photo: courtesy of the CRESST Collaboration).}
\label{fig:cryo_helium}
\end{center}
%\vspace*{-35pt}
\end{figure*}

 The required amount of liquid helium (about 7 liters/day in operation conditions) is planned to be provided by a helium liquefier located underground (galleria TIR) at LNGS and owned by the Max-Planck-Institute for Physics in Munich, as shown in Fig.~\ref{fig:cryo_helium}.

The responsible is Dr. Federica Petricca, spokesperson of the CRESST Dark Matter search. CRESST has a written MoU with INFN-LNGS which in detail organises the conditions and the supply of liquid helium to onsite experiments. Since PTOLEMY has a very moderate consumption the supply is feasible and can be granted. Since the CRESST facility is equipped with a helium recovery system, all the helium evaporation can be stored back and re-liquefied. To do so PTOLEMY will be connected to the recovery system by means of tubes running from the magnets and the dilution refrigerator back to recovery balloon located in galleria TIR.

In exchange for the liquid helium supply consumers have to contribute to the costs for helium gas to balance the loss of helium during the liquefying process. 

The required amount of liquid nitrogen, mainly used for the superconducting magnets, makes up about 40 liters/day. At LNGS big storage tanks providing liquid nitrogen via commercial companies are already put in place. The required amount by PTOLEMY can be granted once communicated to the responsible services at LNGS.

Moreover, this WP will also provide assistance and profound knowledge in running dilution refrigerators and will take care of other cryogenic services necessary for a successful operation of PTOLEMY. 

\clearpage

\section{TES calorimeter}
%\noindent \large {\bf Coordinators: F.~Gatti and M.~Rajteri} \normalsize
%\begin{itemize}
%\item[-] Hardware developments
%\item[-] Multiplexing readout
%\end{itemize}

\subsection*{State of the art}

The present Transition-edge sensors (TESs) technology for low temperature detectors first began to develop in the 1990’s under the applications of innovative and highly performing instruments for Dark Matter direct searches \cite{Colling, Irwin95}, X-ray astrophysics \cite{Irwin96} and neutrino mass searches \cite{Orlando}, CMB measurements \cite{Irwin, Biasotti14} and microanalysis \cite{Wollman, Maurizio}.  This sensor technology, that is based on superconducting films operated at the edge of the resistive transition from the superconducting to normal state of matter in combination with very low noise SQUID readout electronics, enables very high spectral resolution capability of single quanta detection as well as very low Noise Equivalent Power in EM radiation measurements. 
\begin{figure*}[h!]
\begin{center}
   \includegraphics[scale=0.4]{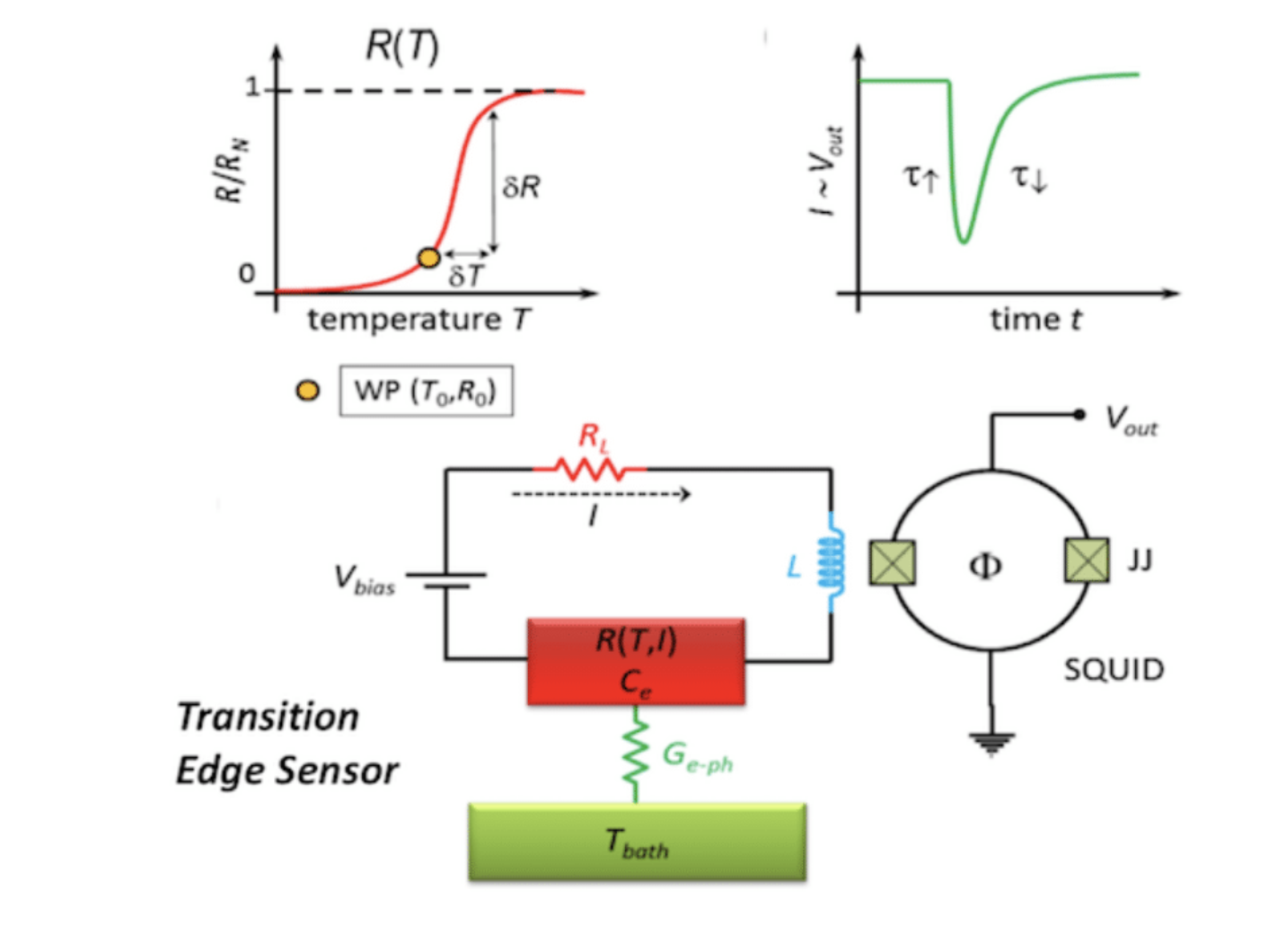}
%   \vspace*{-15pt}
   \caption{Working principle of a TES and its readout schema.}
\label{fig:tes_scheme}
\end{center}
%\vspace*{-35pt}
\end{figure*}
It has been demonstrated also the use of TESs for the detection of individual photons from the ultraviolet through the optical to the mid infrared \cite{Cabrera}. In this energy regime TESs have the capability to count individual photons in a light pulse, and show extremely low dark count rates, on the order of few mHz, a fundamental property for every low detection rate experiment.  Nowadays TESs are at the leading-edge as detectors for X-rays \cite{Ullom}, reaching an energy resolution of 1.6~eV at 5.9~keV \cite{Smith}, and 0.72~eV at 1.5~keV \cite{Lee}, but also as single photon detectors from UV to NIR \cite{Lolli} with energy resolution (FWHM) of 0.11~eV at 0.79~eV. Recently, a charged particle detector was developed based on TES, with the goal of an anti-coincidence device for X-ray detectors on the ATHENA satellite mission \cite{Biasotti16}. TES detector have been used for detecting low energy atoms, since the end of ‘60s \cite{Gallinaro} and molecules \cite{Twerembold}, but, up to now, TESs have never been used to detect electrons in vacuum in order to measure their kinetic energy.

\subsection*{TES electron detectors}

Detection of low energy electrons with 0.1~eV resolutions at very low energy (10~eV) is a complex task at the limit of the present technology of electrostatic electron spectrometers: sub-eV resolution can be achieved at expense of low transmission efficiency (5-10\%).  On the other hand, cryogenic single photon detectors have already shown stable configurations with 0.1 eV resolution and with more than 90\% quantum efficiency. Are these detectors suitable for detecting electrons with similar energy resolution and high efficiency?  In principle the physical detection processes of UV/Vis/IR and low energy have many similarities.  Photons are detected by means of a direct electron gas heating by the related electric field penetrating the metals. Being the electron-electron interactions stronger than the electron-phonon, the photon energy is promptly distributed to the whole electron gas in a time scale much shorter than the one to the phonon system. This means that the TES is sensor and detector at same time: the TES directly measures the electron gas temperature rise after the photon absorption. The consequent excess of heat in the electron gas is slowly released to the phonon system and finally to the substrate that acts as heat sink.  In case of electrons, if the track-length, that should be hundreds of angstroms, is fully enclosed in the TES metal film, a local small hot spot with hotter electrons is expected. This is quite different from the photon absorption whose wave-packet is much larger. However, thanks to the stronger electron-electron interaction, the primary electron energy should be shared among the all electrons of TES in short time, at least in the electron diffusion time over the TES film. The dynamic of the TES cooling is expected to be equal in both cases. In conclusion few differences in the primary energy absorption processes are expected. These could give rise to different features of the rising edge of the pulse, like a possible position dependence at very short timescale. At first glance we don’t expect any possible effect on the energy resolution because of the intrinsic integration time that is on order of magnitudes larger time scale. 

\begin{figure*}[h!]
\begin{center}
   \includegraphics[scale=0.6]{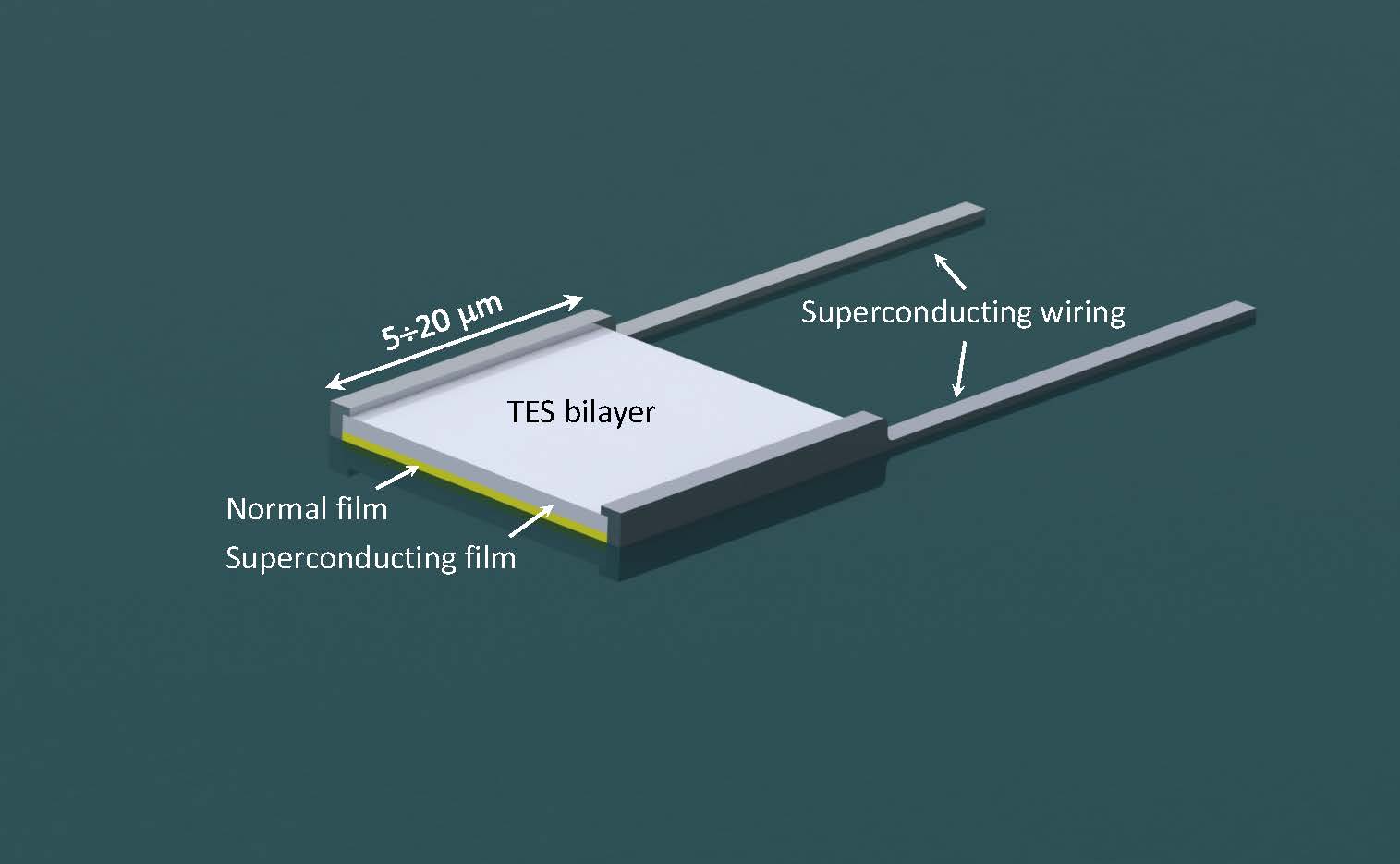}
%   \vspace*{-15pt}
   \caption{Artistic view of the bilayer of a TES setup together with its superconducting wires.}
\label{fig:tes_drawing}
\end{center}
%\vspace*{-35pt}
\end{figure*}

The possible plan of work should includes investigations of the interactions of electrons with the small metal film of TES: tracks, back reflection, secondary electrons losses, pulse shape and rise-time, energy resolution.  A trade-off study should follow in order to better match the PTOLEMY requirements. Investigations on how to improve the energy resolution, that is limited by the thermal fluctuation noise and is proportional to the transition temperature (T$_C$) and to the square root of the heat capacity, should be performed.  This means that to improve the 0.11 eV of energy resolution already obtained at INRIM \cite{Lolli}, a reduction of the transition temperature and of the sensor heat capacity is needed. The proximity effect between superconducting and normal materials can be used to tune the T$_C$ of a bilayer, while the dimensions of the active area and film thickness of the detector can be used to control the heat capacity, in order to have the highest energy resolution compatible with the energy to be detected. The material development will be carry out at INRiM, working on TiAu and TiPd bilayers, and at UniGe, on IrAu.
The TESs fabricated with T$_C < $ 100mK will be characterized trough single photon detection with the experimental setup for telecom wavelengths already present at INRiM. In this way, taking advantage of the photon-number resolution capability of this kind of detector, it will be possible to evaluate the energy resolution for the photon number corresponding to the desired electron energy. The obtained results will be used as a feedback for the TES fabrication and design process in order to obtain the expected energy resolution.

The TES sizes are on the order of tens of micrometer, this means that it is necessary to develop an array of many elements to cover a larger area. 
In order to limit the number of readout SQUID channels, three different approaches are possible: FDM multiplexing, the parallel connection of many TES or mixed configuration of both.  The FDM could be done at low frequency (10 MHz band) or at microwave (GHz band). The first has higher TRL and is used in CMB very large detector arrays in Atacama and South Pole telescopes and is under test for the readout of the GSFC-NASA 1 Kpixel array for the ATHENA X-ray space telescope. The GHz FDM electronics is under
development for the project Holmes (Milano-Bicocca Group). The readout of 100 TESs in parallel has been developed by UniGe for the cryogenic anti-coincidence of Athena. This method requires a fine tuning of the Signal to Noise: all the TESs in parallel will contribute to the full noise, while only one generate the signal.  A possible mixed configuration could be an interesting solution to be exploited.

\begin{figure*}[h!]
\begin{center}
   \includegraphics[scale=0.5]{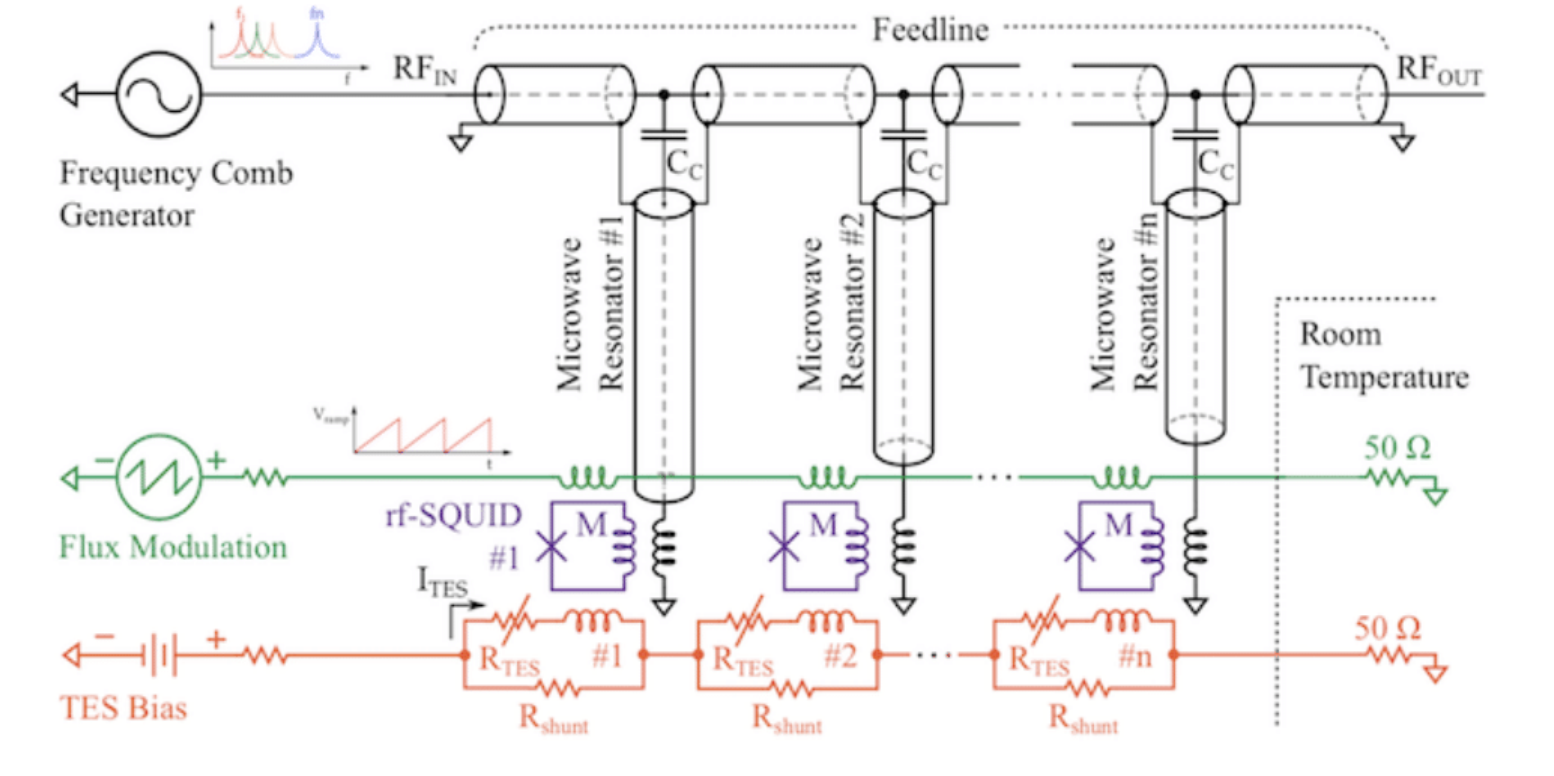}
%   \vspace*{-15pt}
   \caption{Multiplexing schema of the readout of a TES array.}
\label{fig:tes_multi}
\end{center}
%\vspace*{-35pt}
\end{figure*}

\clearpage

\section{EM filter design}

%\noindent \large {\bf Coordinators: C.~Tully and A.G.~Cocco} \normalsize

The aim of the work package is to model and simulate the MAC-E filter for the PTOLEMY prototype (phase I) and to design a new filter suitable for the PTOLEMY phase II test setup, as described in the initial part of this document.

\begin{figure*}[h!]
\begin{center}
   \includegraphics[scale=0.3]{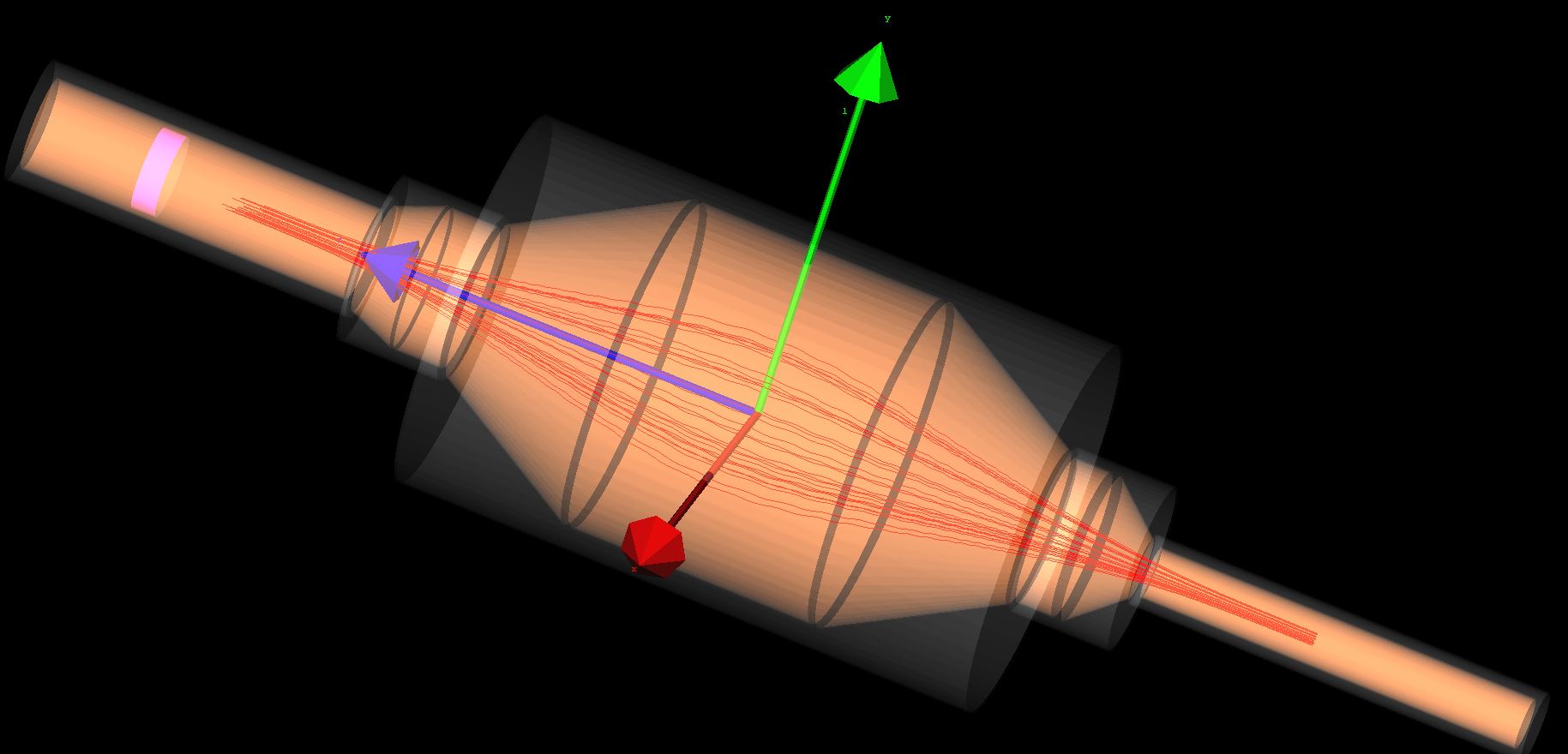}
   \vspace*{5pt}
   \caption{GEANT4 particle tracking of endpoint electrons using a computed COMSOL EM field map.
 }
\label{fig:ptolemy_geant4}
\end{center}
%\vspace*{-15pt}
\end{figure*}

\begin{figure*}[h!]
\begin{center}
   \includegraphics[scale=0.3]{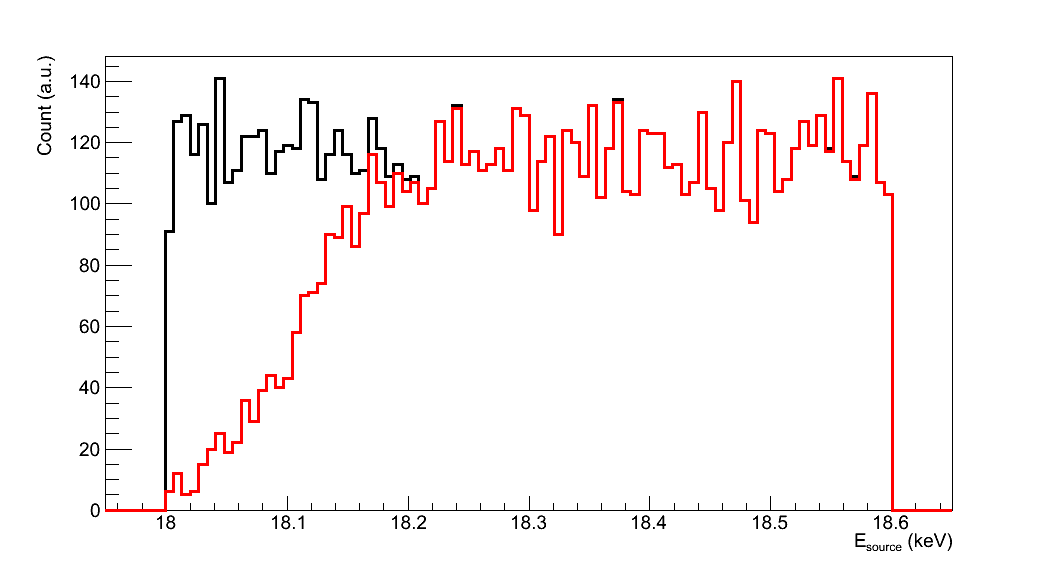}
   \vspace*{5pt}
   \caption{MAC-E filter preliminary resolution plot. Electrons arriving at the calorimeter (red) vs electrons generated at the target site (black).}
\label{fig:maceff}
\end{center}
%\vspace*{-15pt}
\end{figure*}

For what concerns the first part, numerical finite elements computation with COMSOL has been used to calculate the field map for the PTOLEMY prototype. The map has been given as input to the GEANT4 model of the detector and electrons have been traced from the inner part of the first superconducting magnet to the TES calorimeter. This is illustrated in Fig.~\ref{fig:ptolemy_geant4}.
A preliminary evaluation of the MAC-E filter resolution is shown in Fig.~\ref{fig:maceff}.

\begin{figure*}[h!]
\begin{center}
   \includegraphics[scale=0.4]{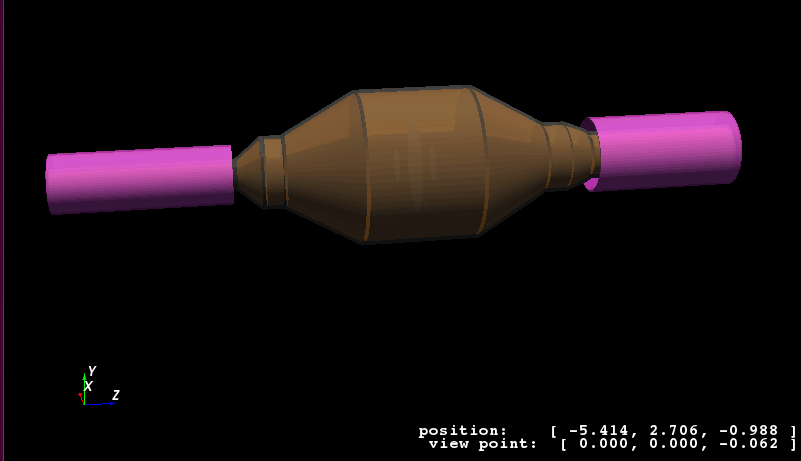}
%   \vspace*{5pt}
   \caption{Kassiopeia prototype modelization. In purple the inner bore of the superconducting magnets.}
\label{fig:kassiopeia_prototype}
\end{center}
%\vspace*{-15pt}
\end{figure*}

\begin{figure*}[h!]
\begin{center}
   \includegraphics[scale=0.25]{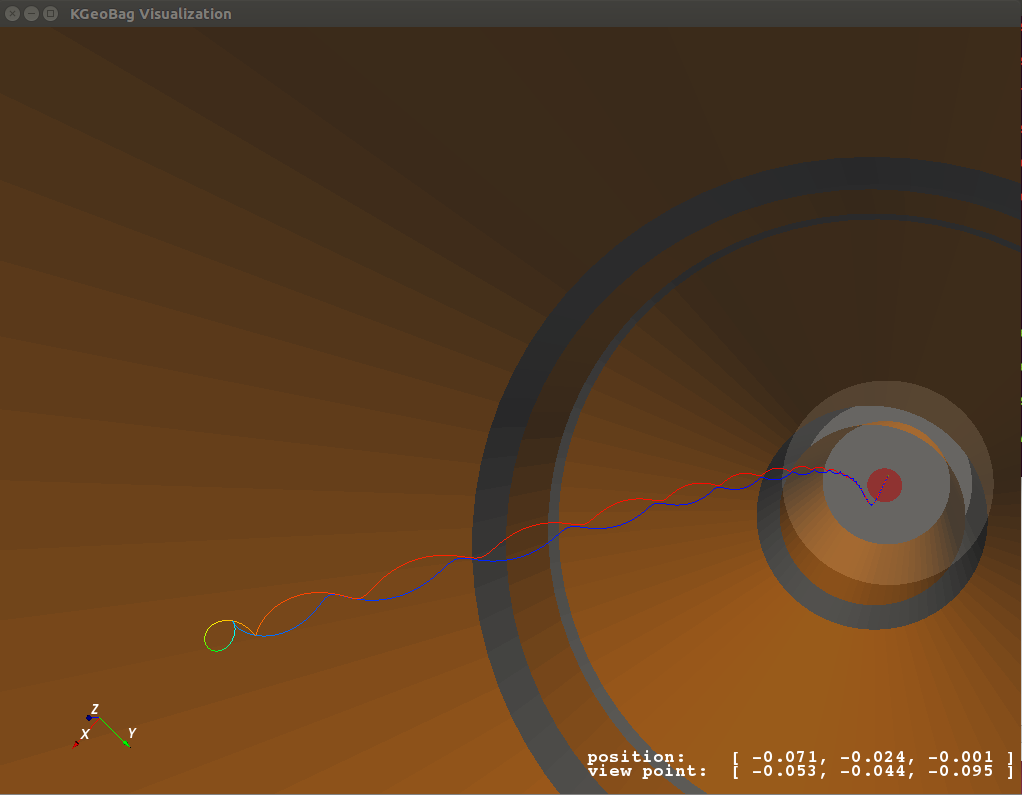}
%   \vspace*{5pt}
   \caption{Example of electron tracking in the PTOLEMY prototype using the Kassiopeia package. The electron shown inside the vacuum chamber does not have enough energy to pass the filter and is guided back to the target region (red circle on the right).}
\label{fig:kassiopeia_tracks}
\end{center}
%\vspace*{-15pt}
\end{figure*}

In order to cross check the results and to validate the filter design the PTOLEMY prototype has also been modeled using the Kassiopeia package\cite{kassiopeia}. The
Figs.~\ref{fig:kassiopeia_prototype} and \ref{fig:kassiopeia_tracks} show the prototype and electrons moving in the drift volume, respectively. Results are still under study and a complete report on the filter performances and on the GEANT4/Kassiopeia comparison are expected soon.

The second task is to develop an EM filter design for the PTOLEMY phase II. The filter must have the following characteristics:
\begin{enumerate}
\item implements the E$\times$B design in order to achieve high compactness in view of a factor $10^5 \div 10^6$ scalability,
\item couples efficiently to the graphene target achieving $2\pi$ geometrical acceptance, 
\item incorporates RF single electron sensing and frequency measurement,
\item implements self-triggering capabilities for highly efficient event reduction, and
\item couples efficiently with the TES calorimeter region.
\end{enumerate}

\begin{figure*}[h!]
\begin{center}
   \includegraphics[scale=0.26]{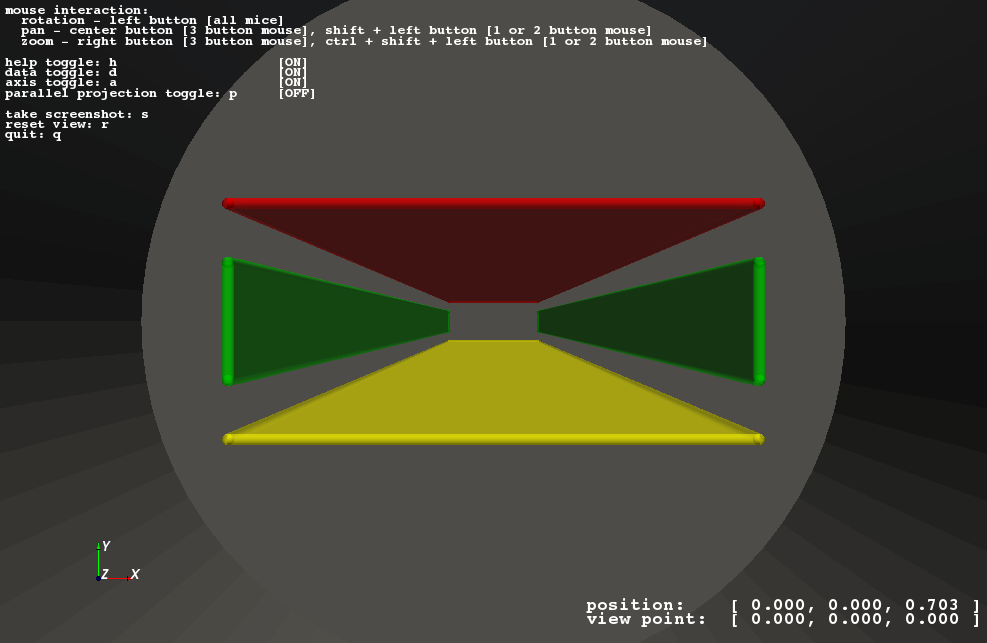}
   \vspace*{5pt}
   \caption{Kassiopeia model of E$\times$B filter electrodes. Distance between lateral (in green) and top-bottom (in red-yellow) electrodes is respectively of about 9 cm and 4 cm.}
\label{fig:kass_exb1}
\end{center}
%\vspace*{-15pt}
\end{figure*}

\begin{figure*}[h!]
\begin{center}
   \includegraphics[scale=0.2]{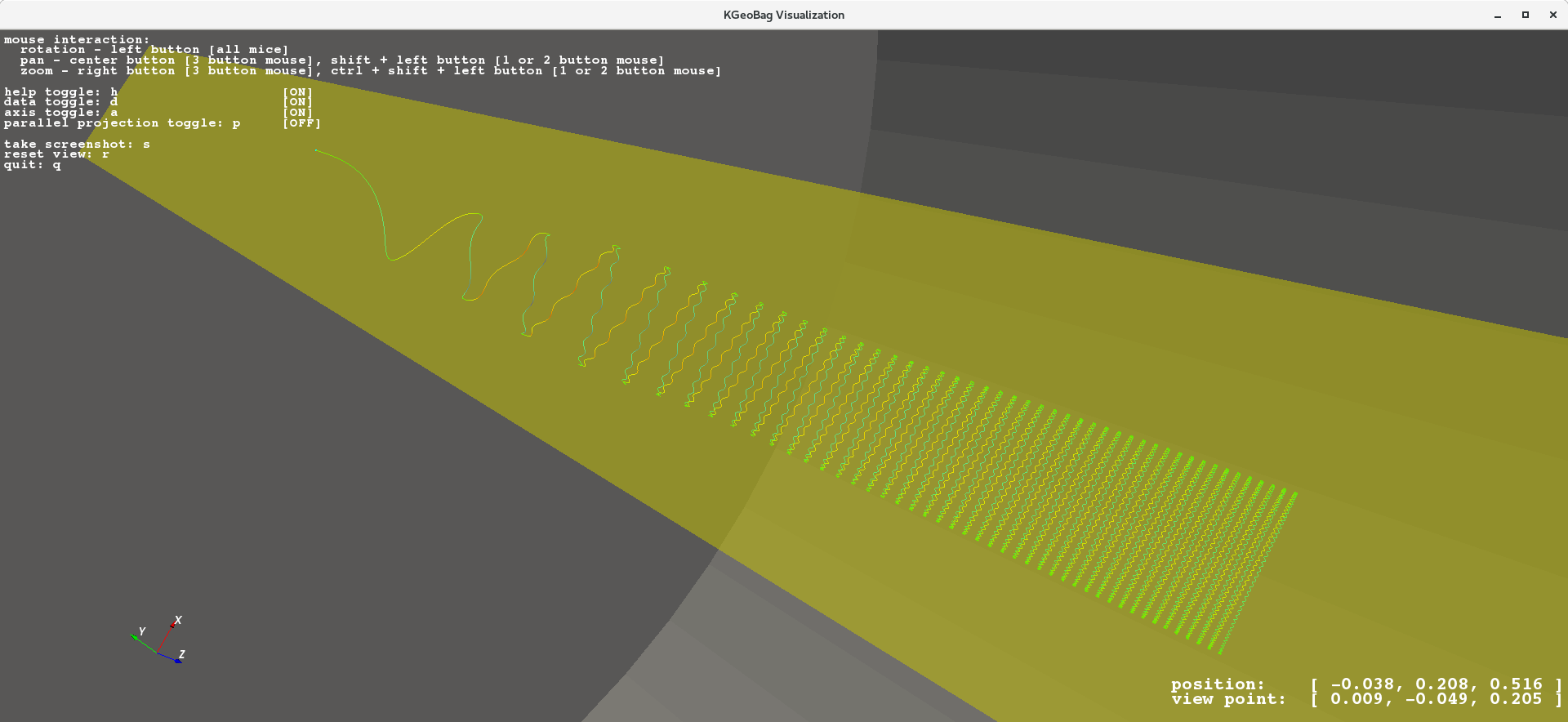}
   \vspace*{5pt}
   \caption{Electron drift in the E$\times$B filter. Total distance traveled along the Z axis is of about 50 cm.}
\label{fig:kass_exb2}
\end{center}
%\vspace*{-15pt}
\end{figure*}

\begin{figure*}[h!]
\begin{center}
   \includegraphics[scale=0.3]{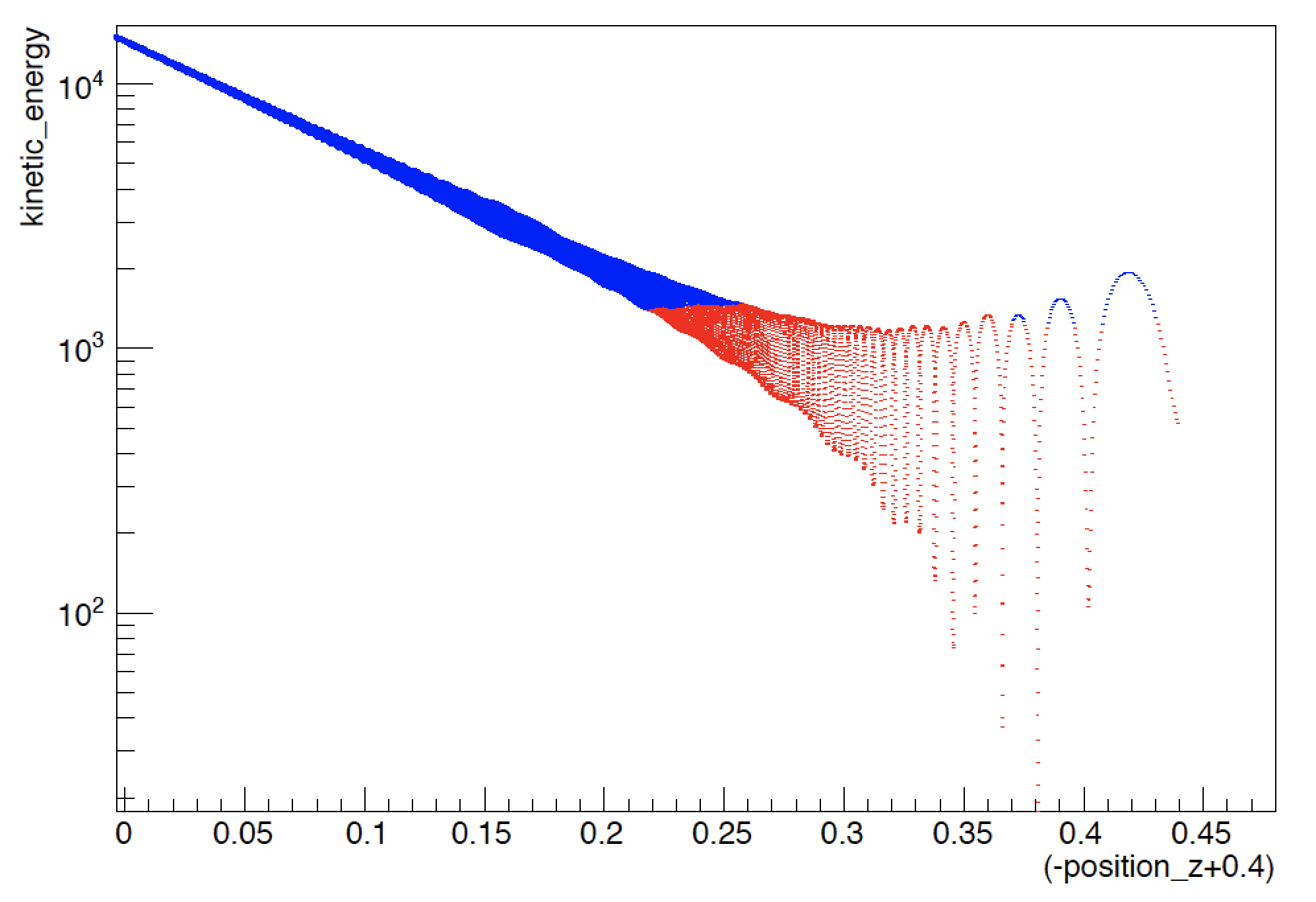}
   \vspace*{5pt}
   \caption{Electron energy as a function of the distance traveled along the E$\times$B filter. Different colors along the path are for debugging purpose only.}
\label{fig:kass_exbdrift}
\end{center}
%\vspace*{-15pt}
\end{figure*}

\begin{figure*}[h!]
\begin{center}
   \includegraphics[scale=0.4]{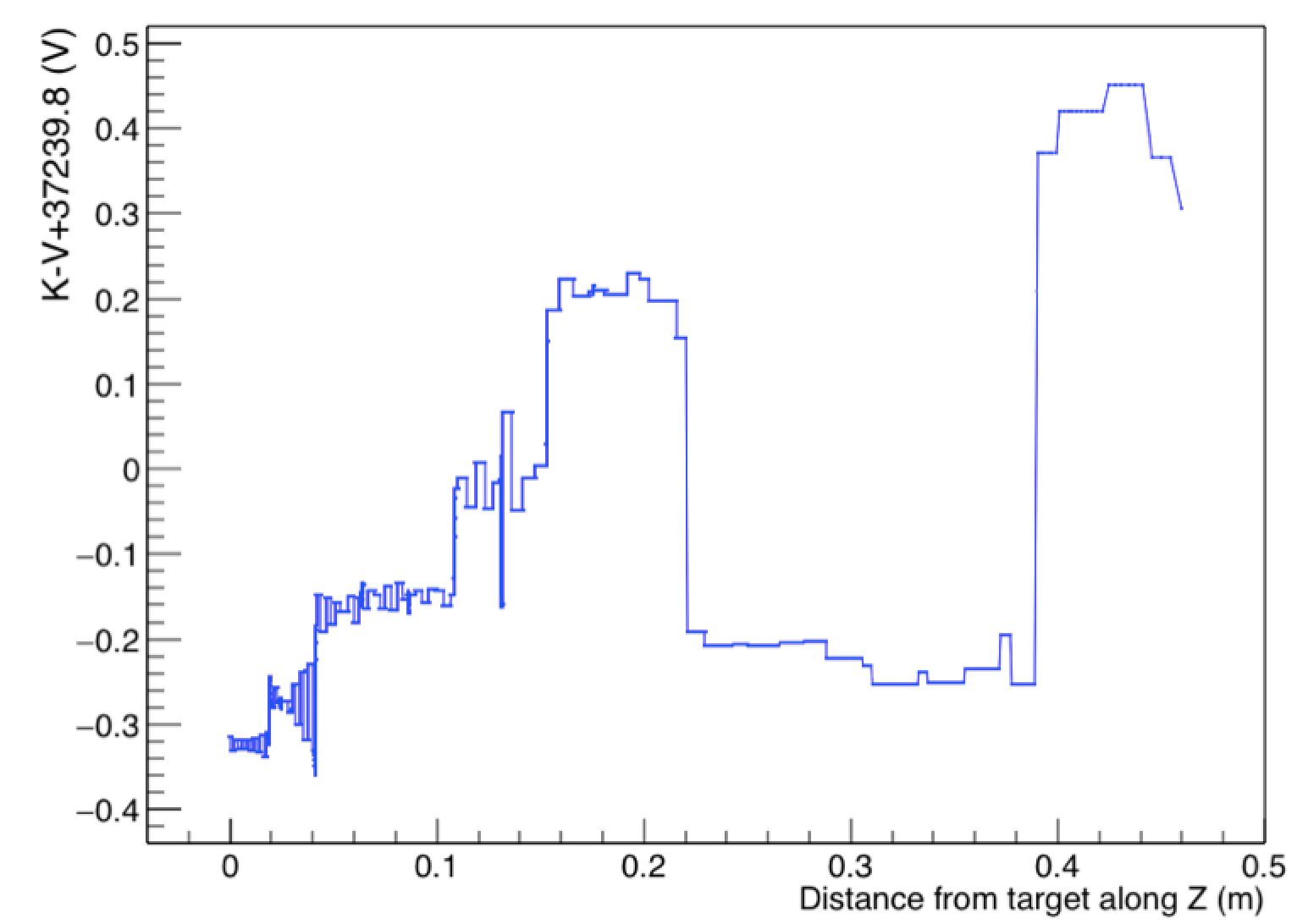}
   \vspace*{5pt}
   \caption{Electron energy balance (kinetic - potential) as a function of the distance traveled along the ExB filter. Total accuracy is below 1 eV over a 50 cm drift along the Z axis.}
\label{fig:kass_exbenergy}
\end{center}
%\vspace*{-15pt}
\end{figure*}

Work has already started within the Work Package in order to implement item 1) and preliminary results using the Kassiopeia package looks very promising. In Figs.~\ref{fig:kass_exb1} and \ref{fig:kass_exb2} an example of electron tracking in a narrow potential well is shown; electrons loose energy while drifting in the E$\times$B direction, as shown in Fig.~\ref{fig:kass_exbdrift}, by climbing the voltage potential of the drift field (orthogonal to how the PTOLEMY prototype MAC-E filter operates). Fig.~\ref{fig:kass_exbenergy} shows the tracking and field simulation accuracy.

Items 2) and 5) are currently under investigation in a joint effort with the Graphene and TES Work Packages. Items 3) and 4) are very important and depend also on the study needed to setup the RF tagging concept. This will be most efficiently developed together with the Project~8 Collaboration that has recently implemented the first measurement of single electron RF detection~\cite{project8}.

%Goal 5) will be evaluated using numerical simulation methods for EM fields.  COMSOL is used for the calculation of the EM field maps, and GEANT4 is used for the full simulation of the particle tracking.  An example of the particle tracking for the prototype setup is shown in Figure~\ref{fig:ptolemy_geant4}.
%There are two major challenges associated with scalability, one external and requires partnership with industry and one internal and requires a design for the CNB target that satisfies the needs for an underground installation.

\clearpage

\section{Modeling and Simulation} 

%\noindent \large {\bf Coordinators: A.G.~Cocco and C.~Mancini} \normalsize
%insert text here

The PTOLEMY experiment aims at the detection of a few electrons having an energy of that is a fraction of an eV above those of the $\beta$-decay endpoint. The extremely low rate of signal events and the precision needed in measuring their energy demand for very high accuracy models. The design of the Electromagnetic filter and the study of the systematic effects that may spoil the precision on the electron energy measurement are examples of items that require extreme care when dealing with numerical precision.

The PTOLEMY strategy focuses toward the use of the most performing tools in the modelization of the various aspects of the detector. Expertise has been acquired in the use of COMSOL~\cite{comsol} for finite element methods to be used in magnetic and electric field simulation as well as for thermal studies. The GEANT4~\cite{geant4} package has been used to model electrons drifting in the PTOLEMY prototype and work is being done in order to study the effect of a shallow depth underground site for the future installation of the full scale detector. Kassiopeia~\cite{kassiopeia} has been adopted for a precise field evaluation and electron tracking into the newly designed E$\times$B filter.

\begin{figure*}[h!]
\begin{center}
   \includegraphics[scale=0.3]{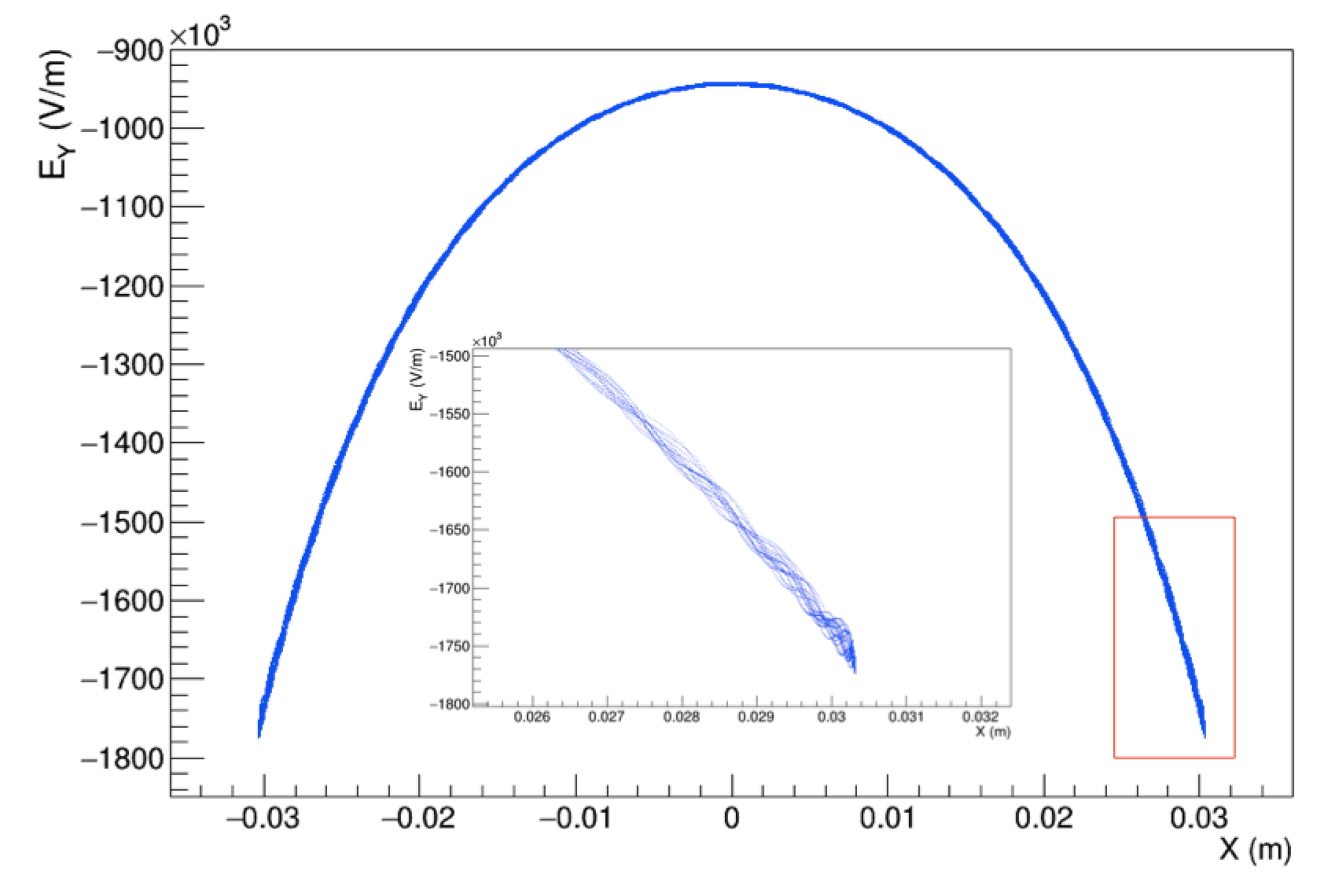}
   \vspace*{-15pt}
   \caption{Example of Kassiopeia data analysis in ROOT. Electric field (Y component) is shown as a function of the electron horizontal position. Insert represents a zoom (red box selection area) showing the cyclotron motion details.}
\label{fig:exb_analysis}
\end{center}
%\vspace*{-35pt}
\end{figure*}

Analysis tools are also very important in order to handle the amount of data expected in order to validate the detector choices. ROOT~\cite{root} has been adopted as the main analysis framework due to its high flexibility and the presence of interfaces with the most common packages. An example of analysis of preliminary data from the E$\times$B filter design is shown in Fig.~\ref{fig:exb_analysis}.

A big amount of work has been done so far but a lot remains to do in order to properly design the detector. Many more aspects of the developments described in this document require simulation tools and manpower to drive the efforts towards the right direction; this may be safely afforded by profiting of the many people in the collaboration willing to contribute to the Working Package. 

\clearpage

\section{Prototype tests}
%\noindent \large {\bf Coordinator: M.~Messina} \normalsize
The prototype tests at the Laboratori Nazionali del Gran Sasso will proceed through the following steps:
\begin{itemize}
\item[-] Agreement on space allocated on surface and underground
\item[-] Above ground area setup
\item[-] Underground area setup
\item[-] Shipment of central vacuum chamber and electrodes from Princeton
\item[-] Vacuum and electrical tests
\item[-] Setup of the HV supply with resistive divider
\item[-] Setup of the new high precision HV and Field-Mill
\item[-] Measurement of HV stability in various conditions
\item[-] Move the central chamber underground
\item[-] Underground measurement of HV stability in various conditions
\item[-] Shipment of superconducting magnets from Princeton
\item[-] Shipment of Dilution Refrigerator from Princeton
\item[-] Magnets and Cryogenic infrastructure setup underground
\end{itemize}

Initial work plan: \\
A design company, one that has done work previously for LNGS, will layout the details of the space usage, housing and services for the prototype based on the 3D CAD designs, equipment needs and in consultation with LNGS responsables.  The central vacuum tank, nine copper electrodes and vacuum pumping equipment will be shipped from Princeton University to LNGS.  The initial goal for the prototype is to demonstrate HV stability with the three primary electrodes of the central chamber, closed, under vacuum and through the HV feed-through system developed at LNGS.  The next steps will involve shipment and installation of two superconducting magnets and windowless vacuum interface to the custom dilution refrigerator dewar.
The integration of the TES calorimeter, MAC-E filter and graphene source with the prototype at LNGS will provide the first demonstration of a calorimeter-based endpoint measurement system with an energy resolution of 0.05~eV or better.  This is a fundamental step toward the detection of the cosmic neutrino background.  The translation of these technologies to a scalable underground experiment will proceed concurrently with simulation studies and build on the performance achievements of the prototype.

Status: \\
The central vacuum chamber, electrodes and HV supply are in the process of being packed for shipment to LNGS.  Approval for receiving these shipments to LNGS is still pending.  The next phase of shipments will be for the two superconducting magnets and the dilution refrigerator once the LNGS space has been prepared.

%\textcolor{red}{From old main-body: (Please check !)}

Longer-term goals: \\
To further evaluate and validate the proof-of-principle for the work package developments of the high radio-pure Carbon-12 graphene target, the EM filter and the TES cryogenic calorimeter, a low background evaluation setup is needed.  The filter, equipped with a 
precision HV system, will provide a stable high-precision voltage reference and step down the 18.6~keV kinematic energy of endpoint electrons to match the dynamical range of the TES cryogenic calorimetry.  The transport of electrons through the magnetic spectrometer with final measurement with a cryogenic calorimeter involve several challenges.  To avoid infrared heating of the cryogenic calorimeter the electron trajectory will have to be bent to absorb line-of-sight radiation with shielding at thermal stages corresponding to LN$_2$, LHe and eventually 100~mK driven by the dilution refrigerator cooling power.

The energy resolution of the cryogenic calorimeter will be evaluated with in an situ IR photon sources within the dilution refrigerator.  Evaluation of the energy scale corresponding to the electron transport through the magnetic spectrometer will be established with a single electron gun, with a conducting source photocathode that is voltage referenced to the calorimeter.  The evaluation of the graphene source and final state smearing of electrons from tritium will be achieved with a very small admixture of the tritium isotope in a hydrogen-loaded graphene substrate, at the level of 1~nanogram (370~kBq).  We will also investigate short-lived calibration EC line sources for an absolute energy scale and resolution based on line separations.

The evaluations of low background conditions will proceed in two different modes of operation.  For background in the endpoint region of tritium, the cryogenic calorimeter measurement will be made after the filtering and step down of the kinetic energy from the magnetic transport of the spectrometer.  The $^{14}$C $\beta$-spectrum is broad enough to cover the entire energy range of interest, including the tritium endpoint.  With high radio-purity Carbon-12 graphene and in the low background conditions of the LNGS, we hope to probe background sources originating from electrons liberated from the graphene at a level that is a factor of roughly one million lower than can be achieved with surface measurements.  In practical terms, we plan to extrapolate backgrounds measured across keV energy ranges in microgram quantities of substrate to a narrow 0.1eV energy window for gram-level target masses, where an assumption would be made on the shape of the background to extract a rate. We also hope to test the factor of background suppression for an RF detection of a single electron in coincidence with the cryogenic calorimeter.  This data will have an important impact on the scalability of the telescope design based on the fundamental elements of the substrate and cryogenic calorimeter components.

A complementary physics program based on high radio-purity graphene targets is proposed to further the research into backgrounds originating from interactions in the graphene target and to explore directional detection sensitivity for MeV dark matter searches.
The details of the MeV dark matter searches are described in the physics case work package.  In this mode of operation the target mass would reside within vacuum space of the custom cryostat sample space, which exceeds a volume of 10$^3$~cm$^3$.  The environmental conditions of the PTOLEMY prototype at LNGS are well-suited to further
knowledge in a broad range of areas, from high radio-pure graphene to probes of MeV dark matter through the development of high sensitivity targets and low energy detection.  We expect that the PTOLEMY prototype tests will not only provide the proof-of-principle basis for development of a cosmic neutrino telescope, but in turn will deliver a rich program of low background physics.

\clearpage

\section{Outreach}

%\noindent \large {\bf Coordinator: TBD} \normalsize

Begin discussions with Gran Sasso Science Institute (GSSI) and the Princeton University International Internship Program (IIP) to create an internship program that is roughly half local students from the surrounding LNGS region and half international or students from outside the local region.  The program would expose students to a wide range of practical training and methods in material science, cryogenics, vacuum and HV systems, low background techniques, RF measurements, data-acquisition/trigger methods and readout, simulation and data analysis.  Depending on the scope of the internships, a partnering with other programs/opportunities in the local region would also be encouraged, such as on environmental/geological studies or arts/humanities/cultural programs.

\clearpage

\bibliographystyle{h-physrev}

\bibliography{main} 

%\def\bibname{References from Individual Chapters}
%\begin{thebibliography}{99}
%\end{thebibliography}

\end{document}